\documentclass[fleqn,usenatbib]{mnras}

\usepackage{newtxtext,newtxmath}
\usepackage{subfig}
\usepackage{tablefootnote}
\usepackage{pdflscape}

\usepackage[T1]{fontenc}

\DeclareRobustCommand{\VAN}[3]{#2}
\let\VANthebibliography\thebibliography
\def\thebibliography{\DeclareRobustCommand{\VAN}[3]{##3}\VANthebibliography}

\usepackage{graphicx}	
\usepackage{amsmath}	
\usepackage{multirow}
\usepackage{color}
\usepackage{xspace}

\definecolor{sunrise}{RGB}{245, 123, 66}

\definecolor{firebrick}{RGB}{178, 34, 34}

\newcommand{\NII}{{[N\,{\sc ii}]}}
\newcommand{\NIIs}{{[N\,{\sc ii}]\,}}

\newcommand{\SII}{{[S\,{\sc ii}]}}
\newcommand{\SIIs}{{[S\,{\sc ii}]\,}}

\newcommand{\OIII}{{[O\,{\sc iii}]}}

\newcommand{\OIIIs}{{[O\,{\sc iii}]\,}}

\newcommand{\OII}{{[O\,{\sc ii}]}}
\newcommand{\OIIs}{{[O\,{\sc ii}]\,}}

\newcommand{\OI}{{[O\,{\sc i}]}}

\newcommand{\HeII}{{He\,{\sc ii}\,}}

\newcommand{\Ha}{H$\alpha$}
\newcommand{\Has}{H$\alpha$\,}
\newcommand{\Hb}{H$\beta$}
\newcommand{\Hg}{H$\gamma$}
\newcommand{\Hbs}{H$\beta$\,}

\newcommand{\beagle}{\textsc{beagle}\xspace}
\newcommand{\cigale}{\textsc{cigale}\xspace}

\title[JADES Type 1 AGN]{JADES: comprehensive census of broad-line AGN from Reionization to Cosmic Noon revealed by JWST}

\author[Juodžbalis et al.]{
Ignas Juodžbalis,$^{1, 2}$\thanks{E-mail: ij284@cam.ac.uk}
Roberto Maiolino,$^{1, 2, 3}$
William M. Baker,$^{1, 2, 4}$
Emma Curtis Lake,$^{5}$
Jan Scholtz,$^{1,2}$
\newauthor
Francesco D'Eugenio,$^{1, 2}$
Bartolomeo Trefoloni, $^{6, 7, 8}$
Yuki Isobe, $^{1, 2, 9}$
Sandro Tacchella,$^{1,2}$
Andrew J.\ Bunker,$^{10}$
\newauthor
Stefano Carniani,$^{8}$
St\'ephane Charlot,$^{11}$
Gareth C. Jones,$^{1, 2}$
Eleonora Parlanti,$^{10}$
Michele Perna,$^{12}$
\newauthor
Pierluigi Rinaldi,$^{13}$
Brant Robertson,$^{14}$
Hannah \"Ubler,$^{15}$
Giacomo Venturi,$^{10}$
Chris Willott$^{16}$
\\
$^{1}$Kavli Institute for Cosmology, University of Cambridge, Madingley Road, Cambridge, CB3 OHA, UK.\\
$^{2}$Cavendish Laboratory - Astrophysics Group, University of Cambridge,
19 JJ Thomson Avenue, Cambridge, CB3 OHE, UK.\\
$^{3}$ Department of Physics and Astronomy, University College London, Gower Street, London WC1E 6BT, UK \\
$^{4}$ DARK, Niels Bohr Institute, University of Copenhagen, Jagtvej 128, DK-2200 Copenhagen, Denmark\\
$^{5}$Centre for Astrophysics Research, Department of Physics, Astronomy and Mathematics, University of Hertfordshire, Hatfield AL10 9AB, UK\\
$^{6}$ Dipartimento di Fisica e Astronomia, Università di Firenze, via G. Sansone 1, 50019 Sesto Fiorentino, Firenze, Italy \\
$^{7}$ INAF – Osservatorio Astrofisico di Arcetri, Largo Enrico Fermi 5, I-50125 Firenze, Italy\\
$^{8}$Scuola Normale Superiore, Piazza dei Cavalieri 7, I-56126 Pisa, Italy\\
$^{9}$Waseda Research Institute for Science and Engineering, Faculty of Science and Engineering, Waseda University, 3-4-1, Okubo, Shinjuku, Tokyo 169-8555, Japan \\
$^{10}$Department of Physics, University of Oxford, Denys Wilkinson Building, Keble Road, Oxford OX1 3RH, UK\\
$^{11}$Sorbonne Universit\'e, CNRS, UMR 7095, Institut d'Astrophysique de Paris, 98 bis bd Arago, 75014 Paris, France\\
$^{12}$ Centro de Astrobiolog\'ia (CAB), CSIC–INTA, Cra. de Ajalvir Km.~4, 28850- Torrej\'on de Ardoz, Madrid, Spain\\
$^{13}$Steward Observatory, University of Arizona, 933 N. Cherry Avenue, Tucson, AZ 85721, USA\\
$^{14}$Department of Astronomy and Astrophysics University of California, Santa Cruz, 1156 High Street, Santa Cruz CA 96054, USA\\
$^{15}$ Max-Planck-Institut f\"ur extraterrestrische Physik (MPE), Gie{\ss}enbachstra{\ss}e 1, 85748 Garching, Germany\\
$^{16}$NRC Herzberg, 5071 West Saanich Rd, Victoria, BC V9E 2E7, Canada
}

\date{Accepted XXX. Received YYY; in original form ZZZ}

\pubyear{\the\year{}}

\begin{document}
\label{firstpage}
\pagerange{\pageref{firstpage}--\pageref{lastpage}}
\maketitle

\begin{abstract}
The depth and coverage of the first years of JWST observations have revealed low luminosity active galactic nuclei (AGN) across a wide redshift range, shedding light on black hole (BH) assembly and feedback. We present our spectroscopic sample of 34 Type 1 AGN obtained from JADES survey data and spanning $1.5 < z < 9$. Our sample of AGN probes a BH mass range of $10^{6-9}$~M$_{\odot}$ at bolometric luminosities down to $10^{43}$~erg~s$^{-1}$, implying generally sub-Eddington ratios of $<0.5L_{\rm Edd}$. Most of these AGN are hosted in low mass ($M_{\star}\sim10^8$~M$_{\odot}$) galaxies and are overmassive relative to the local $M_{BH}-M_{\star}$ relation, while remaining consistent with the local $M_{BH}$-$\sigma_*$ relation. The wide redshift range provided by our sample allows us to trace the emergence of local $M_{BH}$-$M_*$ scaling relation across the cosmic epoch. Additionally, we explore the capability of narrow-line diagnostics in identifying Type 2 AGN and find that a significant fraction of our AGN would be missed by them due to low metallicity or lack of high energy ionizing photons (potentially due to dust absorption, dense gas blanketing the broad and narrow line regions, or intrinsically soft ionizing spectra). We explore the UV luminosity function of AGN and their hosts and find that it is subject to significant cosmic variance and is also dependent on the AGN bolometric luminosity. Finally, we show that the electron and Balmer scattering scenarios recently proposed to explain the broad components of the Balmer lines are untenable on multiple grounds. There is no evidence that the black hole masses have been overestimated by orders of magnitude as proposed in those scenarios.
\end{abstract}




\section{Introduction}
The origin and growth of the first supermassive black holes (SMBHs) and their influence on the galaxies hosting them have long remained an open problem in astrophysics. Tight scaling relations connecting the black hole (BH) mass to host galaxy properties found in local active galactic nuclei (AGN) \citep{VolonteriBHmass, KormendyHo2013, Greene2020} suggest a strong coupling between BH and galaxy evolution. However, while various surveys have greatly expanded the redshift frontier through the years \citep{Merloni2010, Bongiorno2014, Lyu2022}, until recently rest-frame optical studies in the $z > 4$ regime have remained limited to only the brightest and most massive quasars (QSOs) \citep{XQR30, Fan2023}, missing the low mass, low luminosity AGN and complicating investigation into BH seeding and emergence of local scaling relations. Nevertheless, the presence of $\sim 10^9$~M$_{\odot}$ BHs at redshifts as high as 7 \citep{Banados2018, Wang2020} had already reinvigorated interest in SMBH seeding models. These models invoke direct collapse BHs (DCBHs) \citep{Bromm2003, Luo2018}, seeds produced by runaway collisions in dense clusters \citep{Davies2011, Kroupa2020}, or Population III stellar remnants growing at super-Eddington rates \citep{Trinca2022, Schneider2023}, among other scenarios, in attempt to explain these discoveries. The need to test these models has presented an even stronger impetus to search for smaller and less luminous BHs at high redshifts.

Feedback from AGN has also been invoked as a potential quenching mechanism which prevents high mass galaxies from forming stars, bridging the discrepancy between galaxy luminosity and dark matter halo mass functions \citep{Man2018}. However, the exact shape this feedback takes and its role in early galaxy formation is still an open problem, necessitating population studies probing AGN and their hosts at high redshifts \citep{Harrison2024}.

The launch of JWST has greatly expanded the scope of AGN research possible in the high redshift regimes with results from first spectroscopic surveys revealing abundant populations of faint Type 1 AGN out to well before the epoch of reionization \citep{Harikane_AGN, Maiolino_AGN, Matthee2024, Taylor2024, Greene2024}, exhibiting broad features in their Balmer emission lines. However, these discoveries soon posed more questions than answers, with the newfound objects displaying considerable X-ray weakness \citep{Maiolino_xray_weak, Yue2024, Ananna2024}, radio weakness \citep{Mazzolari2024CEERS,Mazzolari2024radio}, signatures of metal poor BLR \citep{Trefoloni2024} and being significantly overmassive with respect to the local BH mass - stellar mass relations \citep{Ubler2023, Harikane_AGN, Maiolino_AGN, Juodzbalis2024}. These overmassive, yet in some cases dormant ($\lambda_{Edd} < 0.05$), BHs have presented compelling evidence for having experienced phases of super-Eddington growth  \citep{Juodzbalis2024}. However, the number statistics of AGN with well constrained host properties remain small. Therefore, further exploration is needed to fully characterize the high redshift AGN population, constrain BH seeding models, and trace the emergence of local scaling relations.

In addition to the large population of Type 1 AGN, first results from JWST spectroscopy have revealed populations of ``mini'' or rapidly quenched low to intermediate mass ($M < 10^{10}$~M$_{\odot}$) galaxies \citep{Strait2023,Looser2024, Baker2025}, alongside significant populations of the more massive ($M > 10^{10}$~M$_{\odot}$) traditionally quiescent systems \citep{Carnall2024,Baker2024, Nanayakkara2024,Park2024, DEugenio2024_pablo}. The high-z massive quiescent galaxies are often found to host AGN \citep{Carnall2023, Baker2024, Bugiani2025, DEugenio2024_pablo} with an AGN fraction of $\gtrsim 45\%$ \citep{Baker2024} and frequently hosting outflows likely requiring AGN feedback to explain \citep{Belli2024, Valentino2025}. It remains an open question as to the potential for AGN feedback to quench the lower-mass, $M\sim 10^9$~M$_{\odot}$, systems at $z>4$ with none to-date showing signs of current AGN activity \citep[however, any AGN quenching episode could have occurred prior to observation, with typical quenching lookback times of $10-30$ Myr in rapidly quenched galaxies,][]{Looser2024, Baker2025}.  These results suggest that AGN feedback is likely to be a crucial component for understanding high-z galaxy quenching, at least at the most massive end.  Additionally, further investigation into the frequency, scope and properties of AGN in lower-mass galaxies is urgently needed.

An additional problem complicating the studies of AGN at the current redshift frontier is the difficulty of selecting Type 2 AGN, which are thought to comprise the majority of AGN population \citep{Lusso2013}. Standard narrow emission line diagnostics, such as the BPT diagram \citep{bpt}, tend to struggle at higher redshifts due to the sensitivity of their lines to metallicity \citep[see][]{Ubler2023}, which is considerably lower in the young Universe \citep{Masters2014, Coil2015}. While several alternative narrow line diagnostic diagrams have been proposed \citep{Nakajima_metallicity,  Scholtz2023, Mazzolari2024, Shirazi2012, Feltre2016}, larger samples of robustly confirmed AGN are instrumental to testing their validity and capabilities in providing a pure and complete sample of Type 2 AGN. As Type 1 AGN are selected through a direct detection of the BLR, a large sample of such objects would provide a robust benchmark in validating high redshift Type 2 diagnostics.

In this paper, we present a solid sample of 34 broad line AGN at $1.5 < z < 9$, from which 20 are newly discovered, obtained from the data collected across the entirety of the JWST Advanced Deep Extragalactic Survey  (JADES) spectroscopic survey data \citep{Bunker2020, Rieke2020, JADES_desc}. These observations offer considerable depth and a range in spectral resolution allowing for robust constraints of broad line region (BLR) and narrow line (NLR) region properties for Type 1 AGN down to $L_{bol} \sim 10^{43}$~erg~s$^{-1}$. Deep imaging, accompanying the JADES spectroscopy \citep{DEugenio2024}, allows for detailed investigations into the properties of galaxies hosting faint AGN in the early Universe. Additionally, the redshift range probed by our sample enables us to probe the redshift evolution of AGN and their host properties, tracing the emergence of local scaling relations. We will also show how the narrow line emission properties displayed by our sample AGN imply the likely presence of a Type 2 population indistinguishable from star forming galaxies by current diagnostics.

The paper is organized as follows - in Section \ref{sec:JADES} we introduce the JADES survey, its photometric and spectroscopic data used in our analysis. Section \ref{sec:fitting} describes our methods for selecting broad line AGN while Section \ref{sec:bh_prop} lays out our methods of estimating their BH masses and accretion rates. Section \ref{sec:host_prop} describes our methods of constraining the host properties of our sample BHs. The BH - host scaling relations of our sample, together with host morphologies are discussed in Section \ref{sec:scaling}. Section \ref{sec:stacks} describes spectral stacking of our sample sources and assesses the viability of searching for Type 2 AGN via narrow line diagnostics. In Section \ref{sec:BLR_shape} we present an initial exploration into the shapes of the broad lines of our sources and rule out scattering scenarios that would imply $\sim2$~dex overestimation of BH masses. The contribution of Type 1 AGN hosts to the high redshift UV Luminosity Function is discussed in Section \ref{sec:uvlf}. Lastly, Section \ref{sec:conclusion} provides summary and conclusions.

Throughout this work we assume a flat $\Lambda$CDM cosmology with $\Omega_m = 0.315$, $H_0 = 67.4$~km~s$^{-1}$~Mpc$^{-1}$ \citep{Planck2020}. All reported magnitudes are in the AB system.

\section{Data description}
\label{sec:JADES}
The data used in this study has been obtained as part of the first three years of the JADES survey. This survey consists of deep JWST imaging and spectroscopy in GOODS-N and GOODS-S fields. The full description of the survey can be found in \cite{JADES_desc}, however a summary will be provided here to add necessary context. 

The JADES photometry largely consists of near-infrared (NIRCam) imaging utilizing the seven of the wide (F090W, F115W, F150W, F200W, F277W, F356W and F444W) together with two medium (F335M and F410M) filters. The imaging consists of three tiers - Medium, covering 175 arcmin$^{2}$, Deep - covering 36 arcmin$^{2}$, and Deepest - produced by overlapping Deep pointings and covering a 9 arcmin$^{2}$ area. The total area covered by JADES imaging is thus of order 220 arcmin$^{2}$. In terms of depth, the limiting AB magnitude of the Medium pointings is 28.6 to 29.0, depending on the filter. Deep pointings range from 29.75 to 30.14, while Deepest - from 30.02 to 30.52.

The photometric data were reduced according to procedure presented in \cite{Robertson2023}, \cite{ Tacchella2023}, and \cite{Baker2023}. In summary, v1.9.2 of the JWST calibration pipeline \citep{bushouse_2022_7229890} was used together with the CRDS pipeline mapping context 1039. Stages 1 and 2 of the pipeline, meant for applying the detector corrections and flat-fielding respectively, were run with JADES' own sky-flat provided for the flat-fielding, otherwise keeping to the default parameters. After Stage 2, custom procedures were performed to account for $1/f$ noise and subtract scattered light artifacts, "wisps", along with the large scale background. Astrometric alignment was performed using a custom version of JWST {\tt TweakReg}\footnote{https://drizzlepac.readthedocs.io/en/deployment/tweakreg.html}, with corrections derived from HST F814W and F160W mosaics along with Gaia Early Data Release 3 astrometry. The images of individual exposures were then stacked in Stage 3 of the pipeline with the final pixel scale being 0.03" per pixel. 

The spectroscopic part of the survey consists of several tiers, characterized by their depth, `Medium' or `Deep' along with photometry from which targets were selected, HST for targets selected with the Hubble Space Telescope or JWST for Webb-selected objects (PIDs 1210, 1287, 1286, 1181 and 1180, PIs Eisenstain and Luetzgendorf). The first two data releases were presented in \cite{Bunker2024, DEugenio2024} with the entire data set to be presented in the upcoming Data Release 4. The NIRSpec dispersers used in the survey were prism, the R1000 and G395H/F290LP (R2700) gratings. However, the high resolution, R2700, grating had less coverage than prism and R1000; we thus used them for sample selection, while employing R2700 data for more accurate kinematics measurements for sources where it was available.

The medium resolution grating (R1000) consists of three individual gratings - G140M, G235M and G395M that can be used with four separate filters - F070LP, F100LP, F170LP and F290LP. The available combinations of these used by JADES are presented in \autoref{tab:R1000}.

Despite not using the G140M/100LP configuration, the survey data does access the 1.27 - 1.66$\mu$m range as F070LP retains data beyond the nominal cutoff at the cost of it overlapping with lower orders of dispersion. The data reduction pipeline employed by JADES is able to disentangle the different orders and give access to the full range of wavelengths covered by F070LP. The precise exposure time for each disperser is tier dependent with the HST/Medium tiers receiving a nominal 1.7 hours of exposure per source in the prism and 0.8 hours per source in each of the medium gratings, JWST/Medium tiers had 2.6 hours of exposure in each configuration. The deep tiers had 7 hour exposures in R1000 gratings and 28 hours in prism. It should be noted that not all objects within a tier received the same exposure hours. In some cases the exposure time was 1/3 to 2/3 of the nominal as the source ended up in fewer dithers while in others the exposure times were longer due to sources ending up in multiple pointings. The exposure times for each individual source are listed in JADES Data Release 3 (DR3) catalogue \citep{DEugenio2024}. In addition to the main Medium/Deep tiers an additional `Ultra Deep' set of observations was carried out in GOODS-S as part of a GO `Large Programme' (PID:3215, PIs Eisenstein and Maiolino) consisting of 7 to 47 hour exposures for prism and G395M/F290LP, G140M/F070LP configurations of R1000, the G235M/F170LP being unused. Complementing the R1000 and prism, some tiers also made use of the high resolution grating G395H, with resolution $R=2700$. However, the R2700's high resolution can reduce the S/N of broad wings and the spectra may end up truncated on the red side, thus this grating has relatively little coverage in the survey. In total, the survey contains spectra of $\sim$5000 objects across the GOODS-N and GOODS-S fields spanning redshifts from 0.6 to 14.

The NIRSpec data was processed according to procedures laid out in \cite{Bunker2023} and other similar JADES papers, such as \cite{Carniani2023}. A full description of the data reduction procedure will be presented in Carniani et al. (in prep), in summary - the spectral data was reduced using the pipeline developed by the NIRSpec GTO team and the ESA NIRSpec Science
Operations Team. The first stage of this pipeline involved removing cosmic rays and snowball artifacts and the second stage was where background subtraction was performed. Afterwards 2D cutouts of each spectrum were generated and optics corrections performed. Path-loss corrections were calculated for each observation, taking into account the intra-shutter position, assuming a point-source geometry and a 5-pixels extraction box (i.e. the full shutter height). Due the compact nature of the objects, to maximize the S/N, complementary 3-pixel extractions were used for broad line detection. Even though a 3-pixel box is not the extraction box the path-loss corrections were optimized for, when compared, the two extractions do not appear to have significant differences in the measured line fluxes. The 5 pixel extractions were utilized in constraining the host properties from the spectra. As a final step, the extracted individual exposures are stacked using a weighted sum and sigma-clipped to remove outlier regions resulting from residual bad pixels or cosmic rays. It should be noted that the sigma clipping procedure employed by the pipeline can erroneously remove bright spectral features, thus the final product contains two stacks - one with sigma clipping, the other - without.

\section{Spectral fitting and sample selection}
\label{sec:fitting}

The core method used to search for AGN in this study is emission line decomposition into multiple Gaussian components, following the methodology of \cite{Maiolino_AGN}. This is done to separate the narrow line emission, coming from low density gas ionized by either star formation or AGN activity, from the broad components indicating the presence of high velocity gas characteristic of the broad line regions surrounding an accreting BH. The main emission line chosen for the analysis was the H$\alpha$ line as this emission feature tends to be bright, only weakly absorbed by neutral hydrogen and, owing to its rest frame wavelength of $\lambda = 6563$~\AA, detectable by JWST between $z = 0.5$ and $z = 7.0$ allowing for self-consistent probing of the AGN population in the low and high redshift regimes. Additionally, results by \cite{Gravity24} have shown that, while single epoch virial estimators can significantly overestimate BH masses, this bias is smallest for \Has, for which it is consistent with the scatter on the relation. 

The procedure used to identify broad \Has line emission consisted of initially fitting two distinct models (with outflow fitting in the \OIIIs carried out at a later stage) to each R1000 spectrum. The first model contained just the narrow line emission from the H$\alpha$ line and the [NII]$\lambda\lambda$6550,6585 doublet. The narrow lines were constrained to have the same widths and the same systemic velocity. The ratio between the peaks of the [NII] lines was constrained to 3 as it is set by the atomic physics and insensitive to density and temperature \citep{NII_ratio2023}. The second model then simply contained an additional broad line component of H$\alpha$ added on top of the narrow lines. The kinematics of broad H$\alpha$ was allowed to vary with respect to the narrow lines. Fitting the two models was carried out using Bayesian methods with mostly uniform priors on the line peak height and FWHM, with the latter allowed to vary between 100 and 800~km~s$^{-1}$ for the narrow and 800 to 10000~km~s$^{-1}$ for the broad lines. The prior on line centroid velocities was Gaussian, centred on the measured redshift of the source, with a standard deviation of about 400~km~s$^{-1}$. The posterior distributions were then estimated from the priors and the data using a Markov Chain Monte Carlo (MCMC) integrator \citep{emcee}. The relative performance of the two models was quantified using the Bayesian information criterion (BIC):
\begin{equation}
    BIC = \chi^2 + k\ln{n},
    \label{eq:bic}
\end{equation}
where $\chi^2$ is the chi-square of the fit, $k$ - the number of model parameters and $n$ - the number of data points. The difference in BIC between the narrow line only model and the model containing the broad line was then computed as $\Delta BIC = BIC_{\rm Narrow} - BIC_{\rm Narrow+Broad}$. The lower floor for accepting the BLR model was $\Delta BIC > 5$. In addition, the broad component was required to have a signal-to-noise ratio\footnote{We define S/N as the ratio of the total line flux integrated over the FWHM to the quadrature-combined error in the same range.} of at least 5. In addition we inspected and removed sources for which the significance was driven by 1-2 pixels as those were likely artifacts of correlated NIRSpec uncertainties \citep{dorner2012}. This makes our selection conservative and ensures that only solid detections are included in the main sample.

An additional complication in measuring intrinsic line kinematics from a given fit is the instrumental broadening effect encapsulated in a line spread function (LSF). Any spectrum observed in a detector is a convolution between the object's intrinsic spectrum and the LSF. However, this LSF can be readily approximated as a Gaussian profile of some FWHM. Therefore, the deconvolution of measured quantities can be performed algebraically, via ${\rm FWHM_{\rm int}} = \sqrt{{\rm FWHM_{\rm meas}}^2 - {\rm FWHM_{\rm LSF}}^2}$ relation, where ${\rm FWHM_{\rm meas}}$ and ${\rm FWHM_{\rm LSF}}$ are the measured and instrumental FWHM respectively. The full description of the JWST LSF for NIRSpec-MSA observations is presented in \cite{DeGraaf2024}, however, the LSF corrections were negligible for the broad lines of our AGN as they were $>10$ times wider than ${\rm FWHM_{\rm LSF}}$.

In addition to the \Has line we also fit the \OIII$\lambda\lambda$4959,5007 doublet together with \Hbs in order to constrain the presence of ionized outflows, to disentangle BLR components from outflows, and also to derive dust attenuation utilizing the Balmer decrement. The outflow component fitted to the \OIIIs doublet used a flat FWHM prior constrained to between 600 and 2500 km~s$^{-1}$. In cases where the outflow contribution was found to be significant, we refit the \Has with an added outflow component, utilizing an informative Gaussian prior constrained by the fits to the \OIIIs lines, in order to reassess the significance of the BLR. The rationale of this step is that a BLR component does not have a counterpart in \OIIIs, hence broad wings in H$\alpha$ that are not seen in \OIIIs cannot be ascribed to outflows but to a BLR component. In the case of GN-200679, the spectrum lacked R1000 coverage of \OIIIs, thus we consider this source a tentative Type 1 AGN as the possibility that the broad \Has profile is an outflow could not be discounted. In addition, we find that, after accounting for outflows, the fitted BLR component for GN-23924 is heavily subdominant, despite remaining formally significant and passing visual inspection. We thus conservatively mark this object as tentative.

We also note that after a careful inspection of the single-Gaussian BLR fit residuals two of our sample sources (GS-49729 and GS-159717) presented significant broad wings of the broad line profile, not reproduced by a single Gaussian fit. We refit these sources with a 2-Gaussian BLR model, constraining one Gaussian to be narrower than the other and find that such a fit reproduces the BLR shape considerably better (Appendix \ref{sec:all_fits}), while such a double-Gaussian structure may be indicative of merging BHs with separate BLRs, as discussed in \cite{Maiolino_AGN}, the two components being located at the same systematic velocity suggests that both profiles originate from the same BLR and the double-Gaussian nature is due to intrinsic non-Gaussianity found in the BLR of some AGN \citep{Kollatschny2013}. We thus treat both Gaussian components as modelling the same line profile and re-estimate the broad \Has luminosity and FWHM from the combined profile. The presence of merging BHs in our sample is still a possibility, as confirmed BH mergers do sometimes exhibit matching systemic velocities of the broad lines \citep{Capelo2017}. However, access to high resolution IFU data is required for a robust survey of BH mergers in our sample AGN.

\subsection{Comparison with previous AGN samples in JADES}
The previous procedure has yielded 28 robust and 2 tentative broad line AGN sources spanning redshifts from 1.7 to 7 (\autoref{tab:all_objecs}), out of which 11 overlap with the \cite{Maiolino_AGN} sample as it originates from earlier JADES data. We do not recover all 12 BLR presented in \cite{Maiolino_AGN} as one of their tentative sources (ID-3608) does not pass our BIC criterion. In addition, one of our objects (GS-204851) has been identified as a tentative AGN candidate in \cite{Matthee2024}, as GOODS-S-13971. Our higher signal-to-noise data allows to robustly confirm the aforementioned object as an AGN. Recent work by \cite{Kokorev2024b} has identified a candidate broad line AGN in a quenched galaxy from JADES DR3 spectroscopy \citep{DEugenio2024}. While this source is in our parent catalogue, our fits fail to detect a significant broad component with $\Delta BIC < 5$, we thus do not include this source in our sample. This galaxy is included as ID 72127 within the quiescent galaxy sample of \citet{Baker2024}. The BHs apparently following the local scaling relations recently reported by \cite{Li2025} likewise do not pass our BIC and S/N selection criteria. Finally, we note that two additional objects (GN-38509 and GN-28074) have been identified as particularly interesting and published separately from the main sample \citep{Juodzbalis2024, Juodzbalis2024b}. Therefore our full sample of 30 Type 1 AGN contains 16 new sources. Recent work by \cite{Hainline2025} has shown that up to 30\% of our sample sources could be classified as Little Red Dots (LRDs), a sub population of red AGN identified in previous JWST surveys \citep{Matthee2024}, however, the majority of our AGN do not fall into this category.

In addition to the main sample above, visual inspection of the prism spectra revealed four additional objects at $7 < z < 9$ that appeared to have a broad component in the \Hbs line. Individually, these proved to be statistically insignificant, however, a stack of them, weighted by inverse root mean square (rms) error, revealed a significant broad \Hbs component as shown in Appendix \ref{sec:tent_hbeta}. The full description of our stacking procedure is provided in Section \ref{sec:stacks}, while here we summarize that the significance of the broad \Hbs feature remains even when weighting by \OIII$\lambda$5007 luminosity. Additionally, we check through jackknife resampling that the significance is not driven by one particular object. We thus place these four objects in our sample, bringing the total to 34 AGN, with 20 of them being new discoveries.

\begin{figure*}
    \centering
    \subfloat{\includegraphics[width=0.78\textwidth]{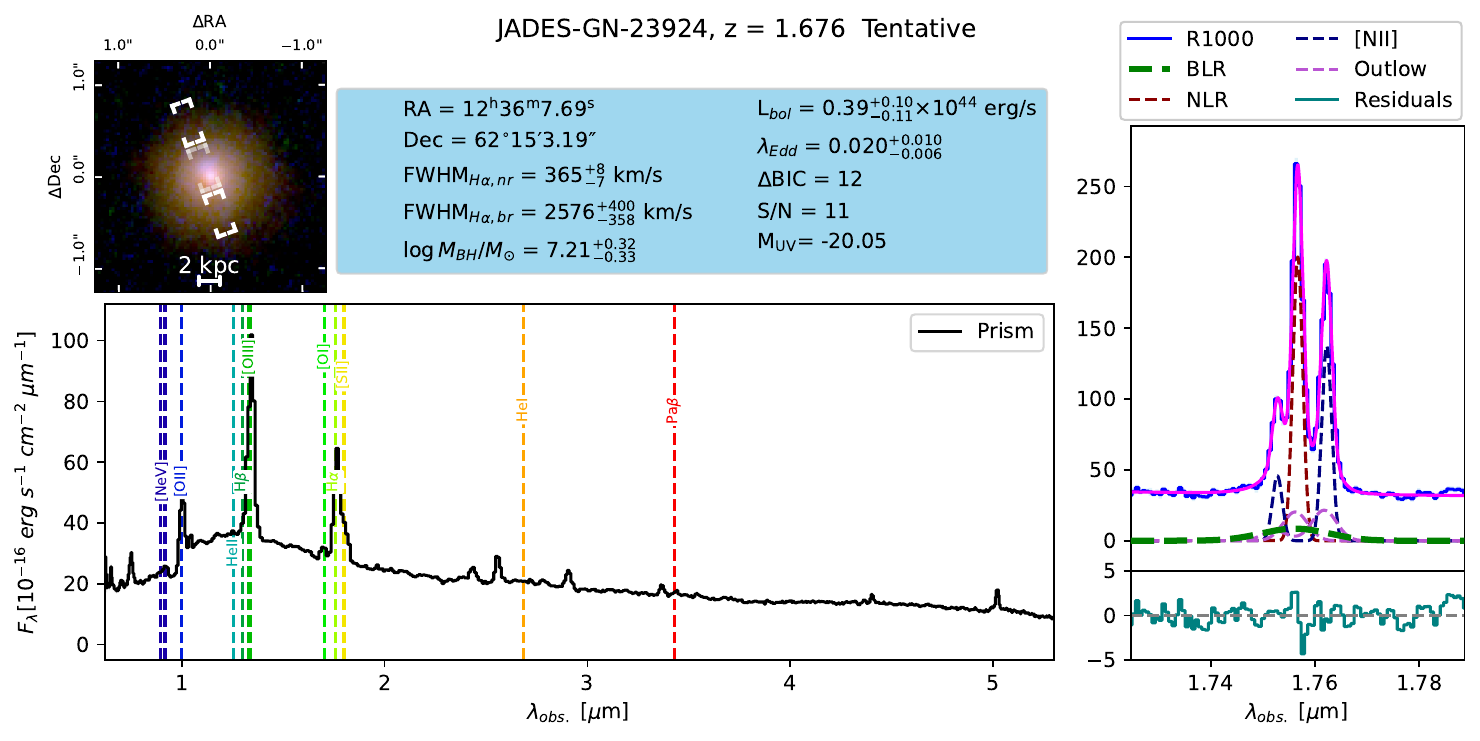}}
    \hfill
    \subfloat{\includegraphics[width=0.78\textwidth]{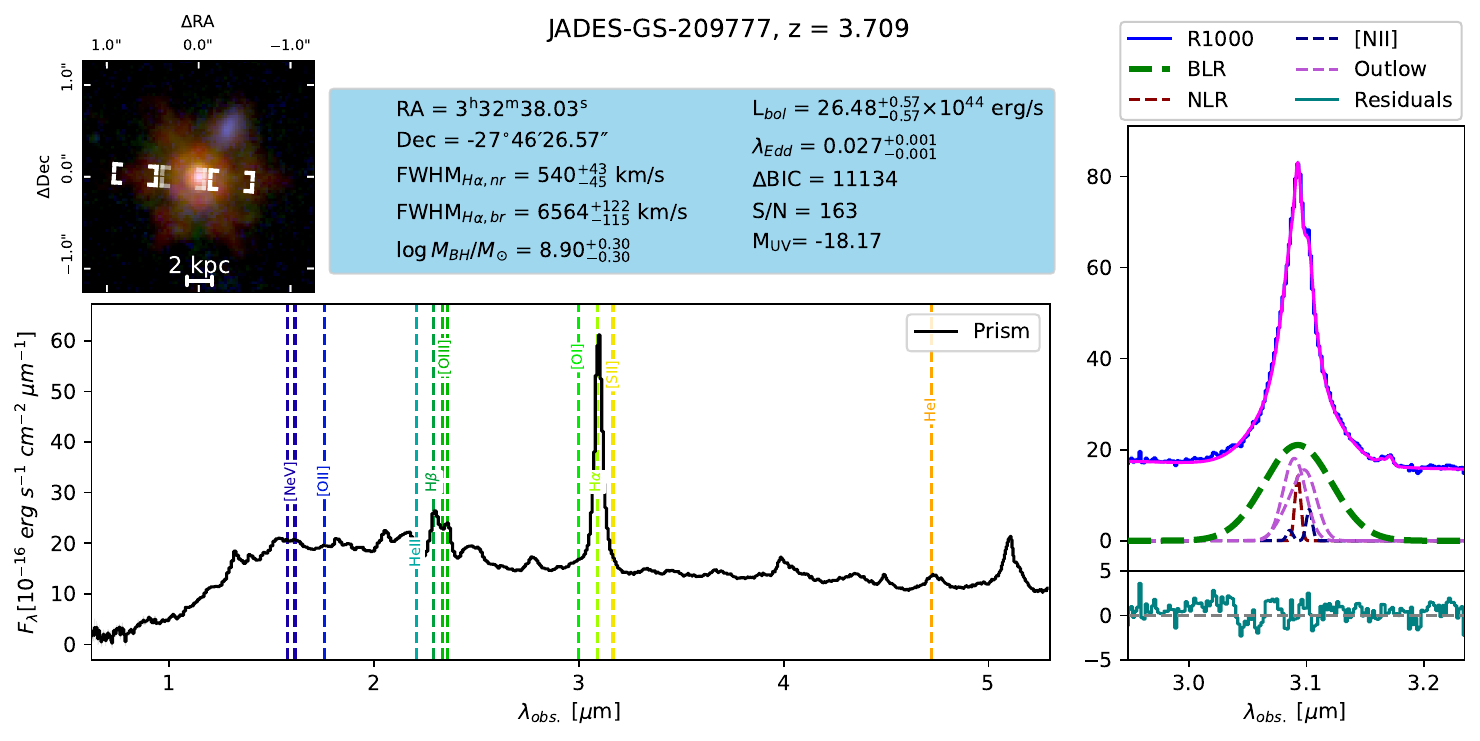}}
    \hfill
    \subfloat{\includegraphics[width=0.78\textwidth]{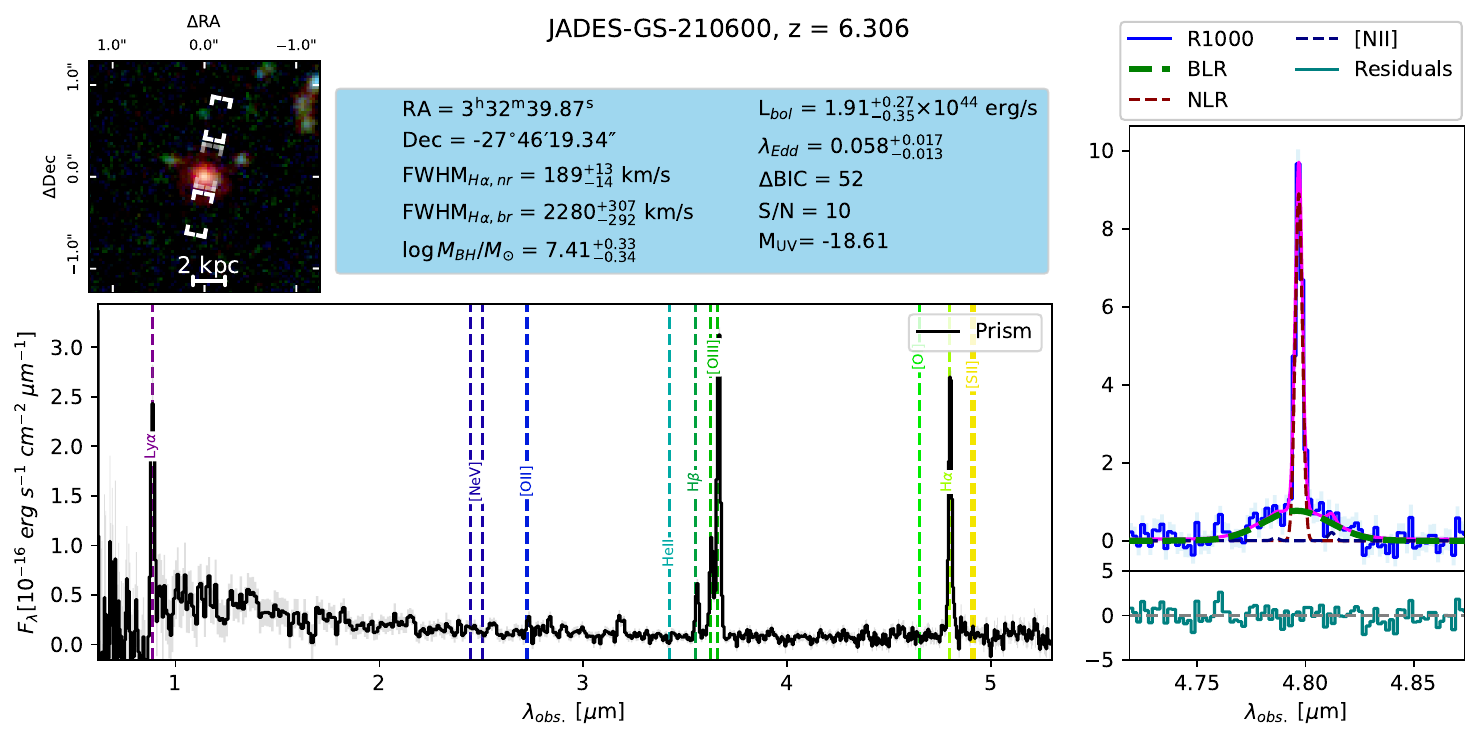}}
    \caption{Fits of objects representative of our sample sources across the redshift range. Each figure showcases the RGB NIRCam stamps at the top together with the plotted MSA slit position. The stamps are made with F090W, F200W and F400W as blue, green and red channels respectively. The prism spectrum is shown in the bottom left and the fits to the \Has line region in R1000 together with the corresponding residuals are located in the bottom right. Quantities derived from R1000 fitting are shown in the blue box.  {\bf Top: }The lowest redshift ($z = 1.676$) source of our sample, located in a massive elliptical host with a clear stellar continuum and a Balmer break. This object has been marked as tentative due to the presence of \OIIIs outflows. {\bf Middle:} A massive $z = 3.709$ quasar with a red optical continuum and \Has emission dominated by BLR and outflows with very weak narrow lines. {\bf Bottom: }One of the highest redshift sources in our sample ($z=6.306$), a low luminosity AGN.}
    \label{fig:representative_fits}
\end{figure*}

The prism spectra of all objects together with fits to the \Has line are presented in \autoref{fig:representative_fits} and Appendix \ref{sec:all_fits}. We have also carried out a search for broad \Hbs emitters across both R1000 and prism in order to push our upper redshift limit to $z = 9$, however, no significant sources were detected. Additionally, we fitted all prism spectra available in order to look for more massive dormant BHs like GN-38509, however, no additional BLR candidates were uncovered that were not already selected by our fits to R1000.

\section{Constraining BH properties}
\label{sec:bh_prop}
We utilize our fits to the broad line region (BLR) emission to constrain the black hole mass using single-epoch virial relations. Recent results by \cite{Gravity24} measured the BLR size via interferometric data, and have put into question the usage of BH virial relations using transitions such as CIV, MgII or \Hbs \citep{Netzer2007, Negrete2012}; however, their finding that the discrepancy is reduced to only a factor 2.5 when using the \Has line is reassuring, as this is within the 0.3~dex calibration uncertainties, this is verified further by \cite{Bertemes2025} testing of different BH mass constraints for a bright QSO. Furthermore, the deviations are associated with super-Eddington accretion influencing the size of the BLR, while most of our sample is likely in a sub-Eddington regime. We thus obtain our BH masses via the virial relation \citep{VolonteriBHmass}:
\begin{multline}
\label{eq:virial_mass}
    \log{\frac{M_{\rm BH}}{M_{\odot}}} = \\
    6.60 + 0.47\log{\left(\frac{L_{H\alpha}}{10^{42}\ {\rm erg\,s^{-1}}}\right)+2.06\log\left(\frac{{\rm FWHM}_{H\alpha}}{1000~{\rm km\,s^{-1}}}\right)}, 
\end{multline}
where $L_{H\alpha}$ is the luminosity of the broad \Has line and ${\rm FWHM}_{H\alpha}$ its width. We estimate the bolometric luminosities of our AGN following the calibrations of \cite{SternLbol}, which give $L_{bol} = 130L_{H\alpha}$.

Estimating the extinction correction to the BLR is not trivial even when a broad component in both \Hbs and \Has lines is seen, as the intrinsic Balmer decrement can be significantly steeper than the case B scenario, due to collisional excitation enhancing \Has relative to \Hb, and can potentially reach values up to ten \citep{Ilic2012}. In our analysis, we measure the $A_V$ utilizing fluxes of the narrow \Hbs and \Has lines and assume a standard case B decrement of 2.86 for the narrow lines along with the SMC extinction curve \citep{smc_gordon}. We use the same $A_V$ values to correct the BLR emission, assuming that the host's ISM is the dominant contributor to the extinction there \citep{Gilli_dust}.

The BH properties for the four sources with tentative broad \Hbs detections were estimated using the \Hbs based mass scaling relation from \cite{Vestergaard2006}:
\begin{equation}
    \log{\frac{M_{\rm BH}}{M_{\odot}}} = \\
    6.67 + 0.63\log{\left(\frac{L_{H\beta}}{10^{42}\ {\rm erg\,s^{-1}}}\right)+2\log\left(\frac{{\rm FWHM}_{H\beta}}{1000 ~{\rm km\,s^{-1}}}\right)}.
\end{equation}
Due to a lack of \Has coverage constraining the $A_V$, we assume $A_V = 0$ for these sources. While dust attenuation on average can be expected to decrease with increasing redshift \citep{Zhao2024}, this is not necessarily true for individual sources, thus the BH masses and luminosities may be somewhat underestimated for these objects.

For the bolometric luminosity estimates, we use the same \Ha-based calibrations as for the main sample sources \citep{SternLbol}, assuming a case B decrement. As the Balmer decrement in BLR is generally steeper than the case B value, our luminosity estimates, and consequently Eddington ratios, may be lower limits.

\subsection{Properties of the final BH sample}
\begin{figure}
    \centering
    \includegraphics[width=\columnwidth]{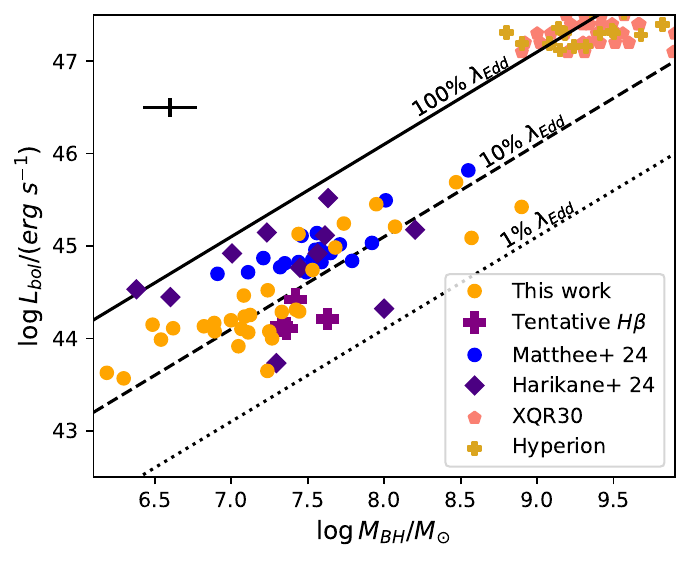}
    \caption{$L_{bol} - M_{BH}$ diagram of our sample sources (orange points) compared to previous samples by \protect\cite{Harikane_AGN, Matthee2024} (blue points and diamonds) as well as high redshift quasars from XQR30 and Hyperion, \protect\cite{XQR30, Hyperion} (light red pentagons and yellow '+' respectively). Constant Eddington ratios are shown by black lines. The average error on each of our sample points is indicated by the black bars in the upper left. The apparent correlation between $L_{bol}$ and $M_{BH}$ is likely overstated in the plot as both axes depend on the \Has luminosity through the scaling relations.}
    \label{fig:Edd_rate_mbh}
\end{figure}

Our final sample of accreting BHs and their hosts contains 34 sources with BH masses ranging from $10^{6}$ to $10^{9}$~M$_{\odot}$, with a mean $M_{\rm BH} \approx 10^{7.2}$~M$_{\odot}$, corresponding to the knee of the BH mass function at $z = 4$ \citep{Kelly2012} and aligning with other BHs discovered with JWST from other studies (\autoref{fig:Edd_rate_mbh}).  The $L_{bol}$ of our sample sources ranges from $10^{43.5}$ to $10^{45.5}$~erg~s$^{-1}$, with the majority concentrated at $\sim 10^{44}$~erg~s$^{-1}$. \autoref{fig:Edd_rate_mbh} showcases the comparison of our sample to other JWST surveys \citep{Harikane_AGN, Matthee2024} as well as high redshift QSOs \citep{XQR30} in terms of $L_{bol}$ and $M_{BH}$. As can be seen, our sample goes $\sim 1$~dex fainter in luminosity than other JWST surveys. We thus provide a deep view into the low-mass, low-luminosity regime of AGN activity that was completely inaccessible in the pre-JWST era as previous all-sky quasar surveys could only explore sources at $L_{bol} > 10^{47}$~erg~s$^{-1}$ and $M_{\rm BH} > 10^{8.5}$~M$_{\odot}$ - close to the upper BH mass range of our sample. As the upper BH mass bound ($M_{BH} \sim 10^9$~M$_{\odot}$) of our sample sources approaches the masses of typical $5 < z < 6$ QSOs, our observations have the potential to probe the `post-quasar' regime of SMBH activity (i.e. faded quasar or ``dormant'' BHs),  as indeed illustrated by GN-38509 \citep{Juodzbalis2024}.


\section{Constraining host properties}
\label{sec:host_prop}
As a core goal of this paper is investigating the scaling relations between BHs and their host galaxies, we conduct an analysis of the properties of the galaxies hosting our sample AGN. We focus on inferring their stellar velocity dispersions ($\sigma_*$) and measuring stellar masses ($M_*$). In addition, many BHs found by JWST were discovered to be overmassive relative to the stellar masses of their hosts \citep{Maiolino_AGN, Ubler2023, Furtak2023_AGN, Harikane_AGN, Kocevski2024}, while remaining consistent with the local $M_{BH} - \sigma_*$ relation \citep{Maiolino_AGN}. The enhanced statistics and redshift coverage afforded by our sample allows us to look for signs of redshift evolution of these scaling relations. We thus endeavor to robustly establish the host galaxy properties of our sample AGN.

\subsection{Velocity dispersion}

\begin{figure}
    \centering
    \includegraphics[width=\columnwidth]{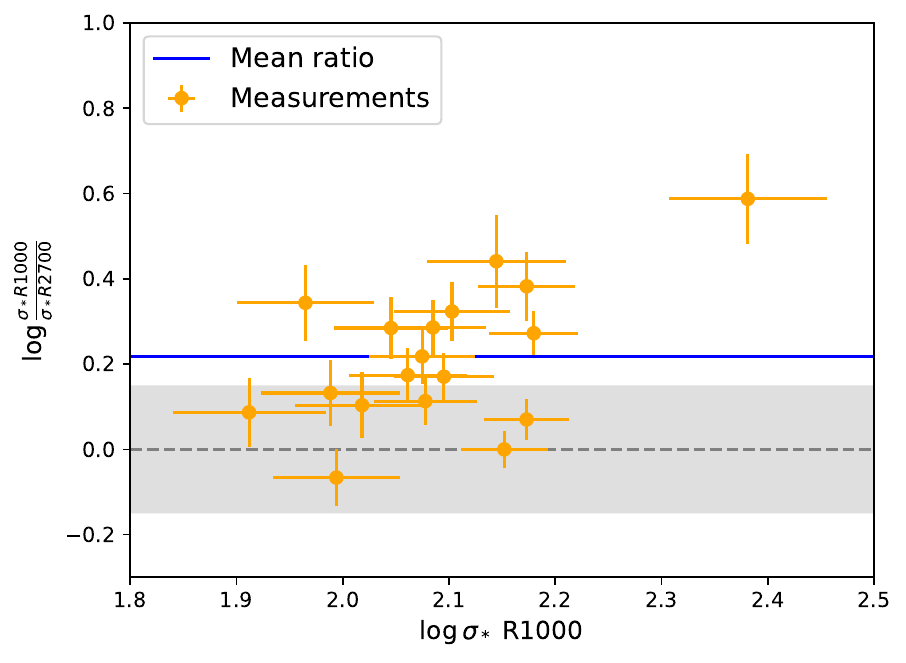}
    \caption{Comparison of inferred $\sigma_*$ values, corrected for LSF, obtained by fitting the medium and high resolution emission line spectra of ionised gas. The y-axis shows the inferred $\sigma_*$ ratio, while the x-axis shows the $\sigma_*$ inferred from R1000. The gray shaded region indicates the 0.3~dex scatter on the \protect\cite{Bezanson2018} calibration relation. The mean ratio is indicated by the blue line and shows that R1000 data on average overestimates $\sigma_*$ by about 80\%.}
    \label{fig:1000_2700}
\end{figure}

We measure the host galaxies' ionised gas velocity dispersions of our sources by fitting the nebular emission lines in the highest resolution data available (either R1000 or R2700) and using the FWHM of the narrow \Has or, if unavailable, narrow \OIIIs lines and applying the LSF correction. While the JADES survey has prism + R1000 coverage for all objects, only some tiers were observed in R2700 G395H/F290LP configuration. Additionally, since about 2/3 of the R2700 spectra are truncated to some degree (because the red part of the long R2700 possibly falling outside the detector, depending on the location on the MSA), many of the R2700 spectra do not have strong nebular lines in the available range. As a result, only 18 of our sources have R2700 coverage with strong nebular lines available for measuring the velocity dispersion. The stellar velocity dispersions were inferred from the measured ionised gas integrated line widths using the average gas-to-stellar measurements from \cite{Bezanson2018}, assuming a scaling factor of 1.3.  For the sources lacking R2700 coverage, we were forced to rely on the R1000 data. In order to assess and correct overestimation coming from using the R1000 data we perform a comparison of the inferred $\sigma_*$ values given by both gratings. The comparison is shown in \autoref{fig:1000_2700} and illustrates that, while R1000 measurements on average overestimate the $\sigma_*$ value by around 0.26~dex with a standard deviation (STD) of 0.16~dex, around half of measurements are consistent within the $\sim$0.15~dex scatter on the \cite{Bezanson2018} calibration. While \autoref{fig:1000_2700} does appear to display a linear trend, the apparent correlation is largely driven by the outlying point at $\log{\sigma_*} = 2.4$. Based on these considerations, we adopt a constant correction factor of 0.26~dex and add the STD of this factor to the fit uncertainties in quadrature.

We note that $\sigma_*$  could not be reliably measured for the tentative \Hbs sources due to low S/N of the \OIIIs doublet and narrow \Hbs in R1000, while the prism resolution of $\sim 1000$~km~s$^{-1}$ is inadequate for accurate narrow line width measurements.

\subsection{Stellar masses from spectroscopic decomposition}

We use two different methods to derive stellar masses for our sources: fits to the NIRSpec prism spectra using \beagle, and fits to the NIRCam photometry using \cigale.  The comparison of the two methods indicates a scatter of 0.4 dex but no strong systematic disagreements.  The two methods are described here.  

We use a version of \beagle\ \citep{BEAGLE} to constrain host stellar masses from NIRSpec prism spectroscopy. 
The method used is similar to the one laid out in \cite{Maiolino_AGN} and we summarise the key features here.  
We fit the prism spectroscopy extracted from a 5 pixel aperture corrected assuming a point source spatial distribution.  
In cases where the source is clearly extended we instead use \cigale\ to fit to the NIRCam photometry as described further in the text.
The continuum from the AGN is modelled as a simple power-law.  We perform two fits to  each source modeling the AGN component with a frequency power law slope of -2.33 \citep{ShakuraSunyaev1973} and -1.54 \citep{Berk2001}, with the level of dust attenuation of the power-law component allowed to vary freely and modelled by the SMC extinction law \citep{Pei1992}. The stellar component was modelled using the delayed exponential star formation history (SFH) with a 10~Myr starburst and a 300~M$_{\odot}$ IMF cutoff.  This star-formation history allows the recent star formation rate to vary freely and independently from the past star formation, reducing the biases inherent in stricter parametric forms that tie together the current star formation to the integrated stellar mass.  The dust attenuation of the stellar component was modeled with the \cite{CharlotandFall2000} attenuation curve. Before fitting, regions of the spectra containing strong emission lines were masked as \beagle\ does not contain models for BLR emission.  Additionally to the fitting performed in \cite{Maiolino_AGN}, the measured narrow \Ha\ flux was used as an upper limit to the model \Ha\ line flux from the star-forming component.  The variable parameters and priors are given in Table~\ref{tab:beagleparams}.

We note that the above procedure takes into account only spectroscopy. In principle, this could cause the stellar masses to be significantly underestimated for extended hosts due to some of their light ending up outside the shutter. To mitigate this issue and provide an additional check on \beagle stellar mass estimates, we carry out photometric fitting of our sample sources using \cigale \citep{Boquien2019}. The photometry used came from the JADES DR3 catalog \citep{DEugenio2024} and was available for 28 of our sources. The star formation history (SFH) was modeled using a delayed SFH model with an exponential burst and BC03 stellar population models from \cite{BC2003} were used to model stellar emission. Stellar dust attenuation was modeled using the \cite{Calzetti2000} dust attenuation law, while \cite{Dale2014} models were used to compute dust emission. The AGN accretion disk emission was modeled using an attenuated power law from \cite{Stalevski2016}. The parameters used for each model are summarized in \autoref{tab:cigale_params}.

After fitting each source according to the procedures above, we compare the $M_*$ values given by \beagle to those of \cigale and find that they are generally consistent within $\sim0.4$~dex. We obtain our final $M_*$ estimates for sources for which either \beagle or \cigale give reduced-chi-square $\chi^2_R < 3$, preferring the \beagle result if both fits are equally good and the source does not contain significant extended components. For hosts with extended morphology (a total of 5), we use the \cigale estimates. Following these procedures we were able to estimate stellar masses for most of our sources, except GS-209777 for which the stellar component is likely completely subsumed by AGN emission (\autoref{fig:representative_fits}) and thus both \beagle and \cigale fail to give satisfactory fits. In addition, the photometric stellar mass estimates for GN-28074 and GS-159717 are higher than the upper limit on $M_{dyn}$ \citep[D'Eugenio (in prep.)]{Juodzbalis2024b}, while \beagle fitting fails to reproduce the spectral shape well, thus we are unable to reliably estimate their host's stellar mass. The stellar mass and $\sigma_*$ estimates for our sample sources are given in \autoref{tab:host_prop}.

\subsection{AGN - host galaxy imaging decomposition and stellar masses from resolved SEDs}
\label{sec:fpho}
The stellar mass estimation methods described in the previous section are prone to degeneracies stemming from the difficulty in disentangling AGN and host galaxy emission. In order to investigate their impact on the stellar mass estimates and obtain more unbiased measurements, we use the tool \texttt{ForcePho} (Johnson B., in prep) to decompose each of our sample sources into galaxy and AGN components and perform fractional SED fitting to reexamine their stellar masses.

The decomposition procedure followed the methodology of \cite{Baker2023,Tacchella2023} and \cite{Juodzbalis2024} in which the AGN emission was modeled as a point source while the underlying galaxy was assumed to follow a S\'ersic light profile. Within \texttt{ForcePho}, the point source component has its half-light radius limited to $r_{e}<0.01''$ (making it completely unresolved by NIRCam), while the S\'ersic index is fixed to one to avoid any significant extended flux resulting from higher order profiles. The host galaxy component has a freely varying S\'ersic index from 0.8 to 6 and varying effective radius. \texttt{ForcePho} works by fitting the individual exposures simultaneously enabling it to extract sub-pixel information due to dithering \citep{Baker2023}. It enables the flux to vary freely between the bands, whilst  the structural information (e.g. position, S\'ersic index, half-light radius, and ellipticity) of the profile are based on the combined information of all the bands. This enables it to fit varying colour gradients as multiple distinct components \citep[e.g. central cores and discs,][]{Baker2023}.
The PSF is modelled as a combination of Gaussians using a Gaussian Mixture Model enabling straightforward convolutions with the S\'ersic profiles used to model the light distribution. Recovery tests of the PSF approximations in the case of multiple component fits have been performed previously in \citet{Baker2023}. 

Each fit was visually inspected and those with significant residuals were discarded. This mostly included bright, point-source dominated objects such as GN-28074, GS-209777 and GS-49729 as the Gaussian Mixture Model (GMM) approximation used by \texttt{ForcePho} breaks down for bright point sources which illuminate the non-Gaussian wings of the PSF. The decomposition also failed for GS-159717 due to significant foreground contamination by a $z\sim1$ galaxy D'Eugenio (in prep.). In addition, \texttt{ForcePho} struggled with objects exhibiting complex morphology, such as GS-204851, which consists of several distinct clumps (\autoref{fig:all_fits4}). It should be noted, however, that such clumpy hosts are likely interacting merging systems and present a general problem of which clumps should be attributed to the AGN host.

We also discard fits that appear to be well-modelled by a single S\'ersic profile. In these cases, the point source component absorbs all the flux whilst the radius of the host galaxy expands to unfeasible sizes and records near zero fluxes by fitting any remaining background.

\begin{figure}
    \centering
    \includegraphics[width=\linewidth]{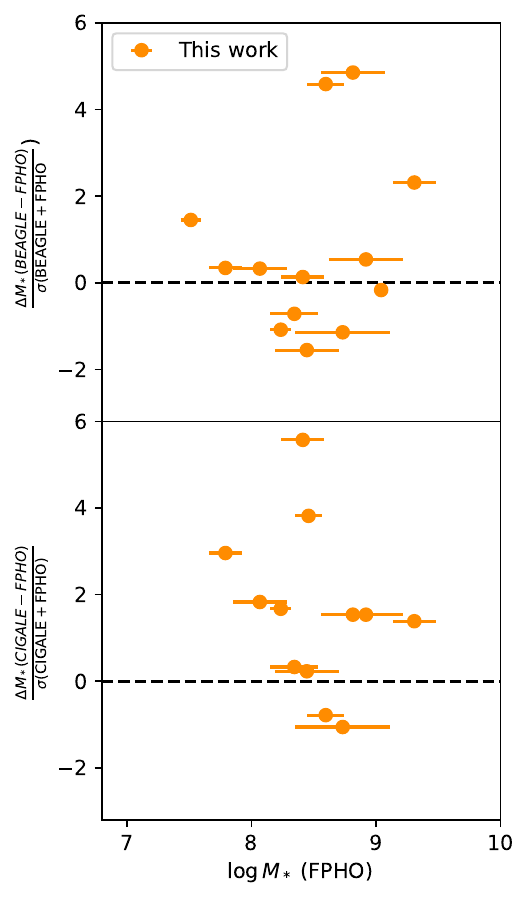}
    \caption{Comparison between stellar mass estimates obtained from \texttt{ForcePho} and those obtained from fitting prism spectra (top) and integrated photometry (bottom). The value on the y axis is the difference in masses obtained by a given estimate (either \cigale or \beagle) and \texttt{ForcePho} photometry, normalized to the total error on both estimates.}
    \label{fig:fpho_comp}
\end{figure}

The fitting of the decomposed point source photometry was carried out using \cigale with the same model components as in \autoref{tab:cigale_params}, except with $f_{AGN}$ being fixed to 0. In addition, we mask all filter bands that include the \OIII$\lambda\lambda$4959,5007 doublet and \Has in their transmission window to avoid residual contamination by AGN photoionization. We utilized the same acceptance criteria as in the previous section, requiring $0.5 < \chi^2_{R} < 3$ for a robust result. This fitting produced stellar masses and sizes for 14 objects out of the whole sample, mostly located at $4 < z < 7$, with a median redshift of 5. We use the stellar masses obtained from \texttt{ForcePho} photometry to assess the potential biases introduced to previous \cigale and \beagle fitting by including the light produced by the central AGN. The comparisons are shown in \autoref{fig:fpho_comp}, where we plot the mass differences ($\Delta M_* = \log{M_{\rm REF}} - \log{M_{\rm FPHO}}$) normalized by the errors on both estimates combined in quadrature as a function of stellar mass estimated by \texttt{ForcePho}. The comparisons indicate that both photometric and spectral fitting, that does not spatially decompose AGN and host galaxy light, tends to systematically overestimate stellar masses by about 2-3$\sigma$, however, spectroscopic fitting by \beagle appears to be slightly more accurate than \cigale. This bias appears uniform and not strongly affected by the host mass. However, \texttt{ForcePho} struggles with fitting bright, morphologically complex sources, thus fits could not be obtained for the brighter, more massive hosts of our sample AGN, which complicates the assessment of stellar mass overestimation in the high mass regime. In principle, $M_*$ estimates for more massive hosts should be less affected by the presence of AGN as most of our sample AGN are relatively faint (\autoref{fig:Edd_rate_mbh}) and stellar light would be expected to dominate over the AGN continuum in cases of massive hosts.

\subsection{Dynamical masses}
\label{sec:Mdyn}

We also utilize morphological information gained from the \texttt{ForcePho} fits to constrain the dynamical masses of our AGN hosts. We utilize the calibrations from \cite{Wel2022} and calculate the $M_{dyn}$ via the following formula:
\begin{equation}
    M_{dyn} = K(n)K(q)\frac{\sigma_*^2R_e}{G},
\end{equation}
where $\sigma_*$ is the stellar velocity dispersion and $R_e$ is the half light radius. $K(n)$ and $K(q)$ are morphological correction factors based on the S\'ersic index ($n$) and axis ratio ($q$) and defined in \cite{Wel2022} with $K(n) = 8.87 - 0.831n + 0.0241n^2$ and $K(q) = \left[0.87 + 0.38e^{-3.71(1-q)} \right]^{2}$.

\section{Host scaling relations and morphologies}
\label{sec:scaling}
In addition to studying the scaling relations of the full sample, in order to assess the potential redshift evolution, we also split it into three redshift bins. The first bin consists of 15 sources at $z > 5$ (including the tentative H$\beta$ objects, Appendix \ref{sec:tent_hbeta}). The second bin contains 13 objects at $3.5 < z < 5$ and the final one - 6 objects at $z < 3.5$. We focus on exploring the BH - stellar mass ($M_{BH}$ - $M_{*}$) and BH - velocity dispersion ($M_{BH}$ - $\sigma_{*}$) scaling relations as those involve the quantities most readily constrained by our data for the majority of our sample. Our \texttt{ForcePho} morphological fits also allow us to constrain dynamical masses for a smaller sub-sample of objects and investigate their positions with respect to the $M_{BH}$ - $M_{dyn}$ scaling relations.

\subsection{The $M_{BH}$ - $M_{*}$ relation}

\begin{figure*}
    \centering
    \includegraphics[width=\textwidth]{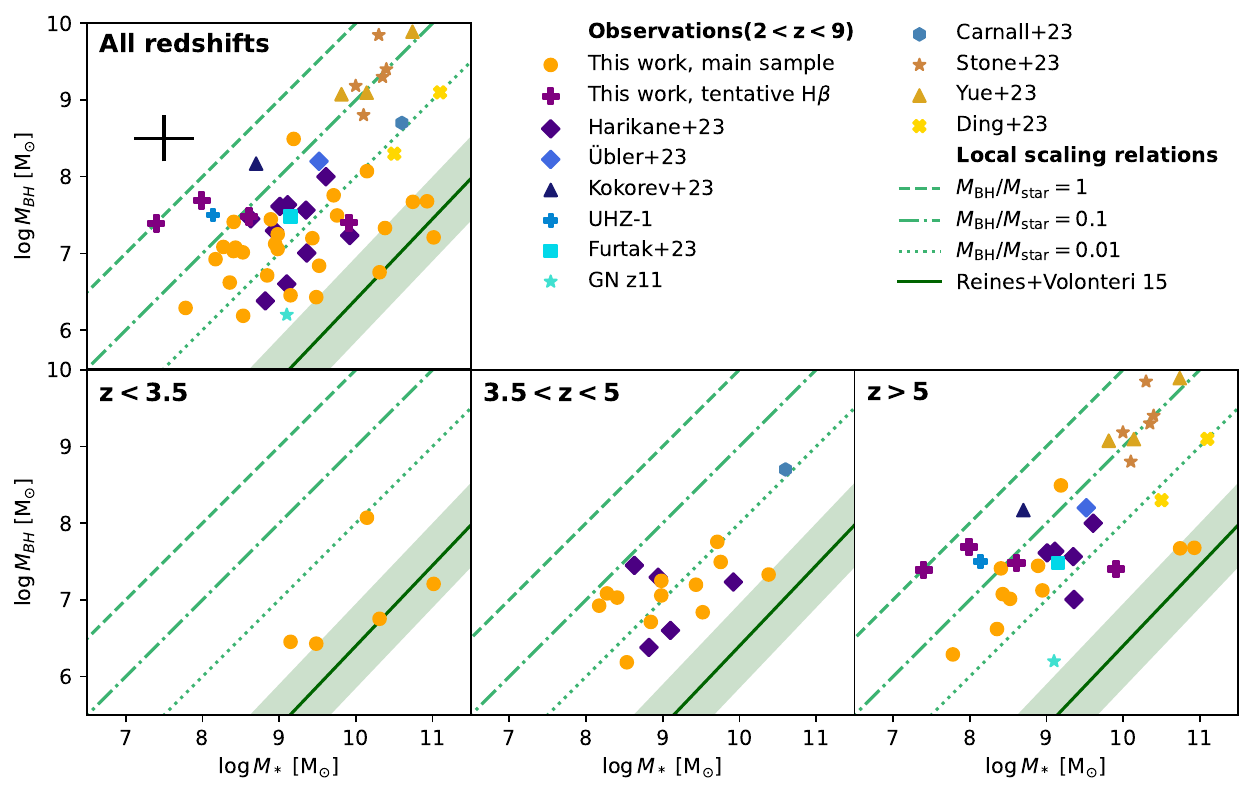}
    
    \caption{$M_{BH}$ - $M_{star}$ relations of our sample sources for the entire sample and the three redshift bins. Main sample sources are shown as orange dots, the tentative \Hbs detections - as purple crosses. Previous JWST observations by \protect\cite{Harikane_AGN, Ubler2023, Goulding2023_AGN} and \protect\cite{Kokorev2023_AGN} are shown in blue colored markers. QSO observations by \protect\cite{Stone2023_QSO, Yue2024_QSO, Ding2023_QSO} are shown as brown stars, triangles and yellow crosses respectively. The solid dark green line shows the local relation from \protect\cite{VolonteriBHmass} with the scatter indicated by green shading. The constant $M_{BH}$/$M_{star}$ ratios are shown by lighter green lines. The average errors on each data point are indicated by the black cross in the upper left panel.}
    \label{fig:Mbh_Mstar}
\end{figure*}

BH mass - stellar mass relations for our entire sample and each redshift bin are presented in \autoref{fig:Mbh_Mstar}. We use the local $M_{BH}$ - $M_{*}$ relation from \cite{VolonteriBHmass} for our comparisons as their BH mass estimation methods are consistent with ours (carried out using \Has measurements together with \autoref{eq:virial_mass}) and their sample focuses on late-type galaxies (i.e. consistent with the prevalent morphologies in our sample and most high-z galaxies), whereas studies like \cite{KormendyHo2013} focus on early type. From the figure, it is apparent that, overall, our sample BHs are overmassive with respect to the local relations, with the mean stellar to BH mass ratio of $\sim 0.01$ placing them $1 - 2$~dex above the local relation from \cite{VolonteriBHmass}, while the tentative \Hbs sources have $M_{BH}/M_*$ approaching unity. When split across the separate redshift bins, the sample exhibits some redshift evolution in the $M_{BH} - M_*$ plane, with sources in the $3.5 < z < 5$ bin being about 0.5~dex closer to the local relation than those in the $z>5$ bin. All objects in the $z < 3.5$ bin appear consistent with the local relation, suggesting that the local relation is being established below this redshift. However, the small number of sources in this bin, due to JADES spectroscopic target selection being biased against low redshift objects, makes it hard to assess significance. It should be noted that the data from \cite{Harikane_AGN} also follow a similar trend between $z>5$ and $3.5<z<5$ redshift bins, however, their sample alone lacks the statistics to establish a robust trend. The higher source count of our sample establishes this evolution more robustly.

The offset from the local relation observed in the higher redshift bins does not appear to depend strongly on the $M_{BH}/M_{*}$ ratio itself, albeit some extra deviation can be seen at the lower end of $M_{BH}$ and $M_{*}$. This is indicative of some observational bias being present, as pointed out by \cite{Li2024}. However, as explored in \cite{Juodzbalis2024}, it is unlikely that all of the observed deviation from the local scaling relations can be attributed to selection effects, especially given that the bias should primarily apply for the most luminous AGN, while our sample includes dormant black holes. Another argument against biases playing a dominant role in the observed offset is that the same orders-of-magnitude offset is not seen in the  $M_{BH}$ - $\sigma_*$ relation, which will be explored in the following section.

We note that the overmassive nature of high-z BHs remains regardless of how the stellar mass is inferred, i.e. regardless of whether it is obtained via spectrosopic or imaging decomposition, and regardless of the stellar fitting code adopted.
Within this context, it should also be pointed out that the stellar mass estimates plotted in \autoref{fig:Mbh_Mstar} may actually be overestimated since, as illustrated by \autoref{fig:fpho_comp}, a fitting that decomposes AGN from host galaxy emission lowers the $M_*$ estimates by about $3\sigma$, potentially making our BHs even more overmassive. Finally, recent results indicating that part, or most, of the Balmer break observed in these AGN is likely non-stellar \citep{Inayoshi2024,Ji2025,DEugenio2025} indicates that the stellar masses are likely overestimated, making the overmassive nature of black holes even more prominent.

\subsection{The $M_{BH}$ - $\sigma_{*}$ and $M_{BH}$ - $M_{dyn}$ relations}
\label{sec:BH-sigma}

\begin{figure*}
    \centering
    \includegraphics[width=\textwidth]{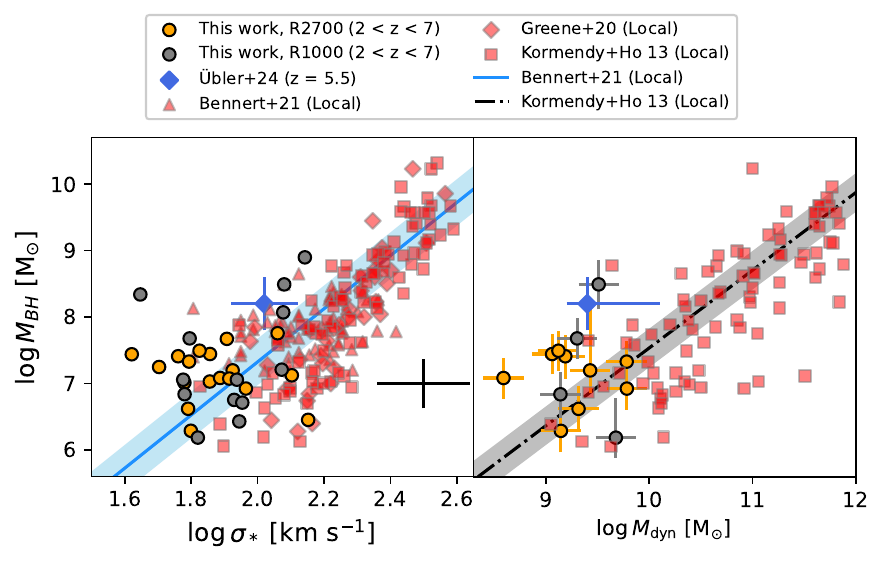}
    \caption{\textbf{Left panel: } $M_{BH}$ - $\sigma_{*}$ relation for our AGN sample sources. The orange points show sources for which $\sigma_*$ was estimated from R2700 data, grey points - those for which we used R1000 measurements. Blue line and the shaded region indicate the local relation from \protect\cite{Bennert2021} and the 0.3~dex scatter on it respectively. The average uncertainties of our data are shown by the black cross. Local measurements by \protect\cite{Bennert2021, KormendyHo2013} and \protect\cite{Greene2020} are shown by the red points. \textbf{Right panel: } $M_{BH}$ - $M_{dyn}$ relation for a subsample of our AGN with robust morphological constraints. The markers have the same meaning as in the left panel. The black dash dotted line is the scaling relation from \protect\cite{KormendyHo2013} with gray shading indicating the scatter.}
    \label{fig:sigma_mbh}
\end{figure*}

We present the $M_{BH}$ - $\sigma_{*}$ for our sources in the left pannel of \autoref{fig:sigma_mbh}. In this case all of our sources are much closer to the local scaling relation. The slight offset present is consistent with the scatter, particularly when taking into account the uncertainties on $\sigma_*$ values. The only significant outlier  (GN-28074) has $\sigma_*$ inferred from R1000 data, which is less reliable due to the lower resolution (\autoref{fig:1000_2700}). This lack of offset in the $M_{BH}$ - $\sigma_{*}$ relation across the whole redshift range, which is not reflected in the $M_{BH}$ - $M_{*}$ relation, seems to imply that the gas required for these sources to reach the local stellar mass scaling relations is already present, but it has not yet been efficiently converted to stars. In this scenario, BH feedback is an attractive solution to explain this inhibiting of star formation either through turbulence or heating. Ejective AGN feedback is instead unlikely as this would remove gas and make the two scaling relations more consistent; lack of significant ejective feedback in these AGN was already pointed out by \cite{Maiolino_xray_weak}, who found a general lack of outflows. Alternatively, or in addition, contribution from dark matter (and with high DM/stellar ratio) may contribute to the $M_{BH}$ - $\sigma_{*}$ relation, while leaving an offset on the $M_{BH}$ - $M_{*}$ relation; this aspect will be explored in a separate paper (McClymont et al., in prep.).

Additionally, this consistency of our sources with the local $M_{BH}$ - $\sigma_{*}$ relation implies that the offset seen in the $M_{BH}$ - $M_{*}$ relation cannot be entirely attributed to selection effects wherein lower mass BHs in more massive (higher $M_*$ and $\sigma_*$) hosts are missed by observations. Indeed, selection effects on the BH masses should be seen as similar, orders of magnitude, offsets on the $M_{BH}$ - $\sigma_{*}$ relation. Selection effect produces a characteristic `banana' shape \citep{Li2024} as BHs in less massive hosts appear to be more overmassive. While both \autoref{fig:Mbh_Mstar} and \autoref{fig:sigma_mbh} exhibit hints of excess deviation in the lower $M_*$/$\sigma_*$ regime, indicative of this selection effect being present, the overall major shift towards local scaling relation when going from $M_*$ to $\sigma_*$ can not be explained this way.

The right panel of \autoref{fig:sigma_mbh} shows the $M_{BH}$-$M_{dyn}$ relation for a subsample of 14 sources for which robust morphological constraints based on \texttt{ForcePho} fitting were possible. We focus on comparing with the $M_{BH}$ - $M_{*}$ relation from \cite{KormendyHo2013} as it is derived for gas-poor early type galaxies for which the stellar mass closely tracks the dynamical mass. As we show in \autoref{fig:sigma_mbh}, our sample AGN are largely consistent with the local relation by \cite{KormendyHo2013} and lie in the same region as most of their data points albeit generally occupying the lower mass range. This consistency with local $M_{BH}$-$M_{dyn}$ and $M_{BH}$-$\sigma_*$ relations together with the offset in the $M_{BH}$-$M_{*}$ one implies either high gas fractions ($f_{gas} > 0.5$) or significant DM contribution present in our sample of AGN hosts. If high gas fraction is primarily responsible for the effect, rather than high DM fraction, this would further indicate that the gas required to bring our sources on to the local $M_{BH}$-$M_{*}$ scaling relation is already present. This implies that the feedback inhibiting star formation in our sample sources is likely of preventative rather than ejective nature.

\begin{figure}
    \centering
    \includegraphics[width=\linewidth]{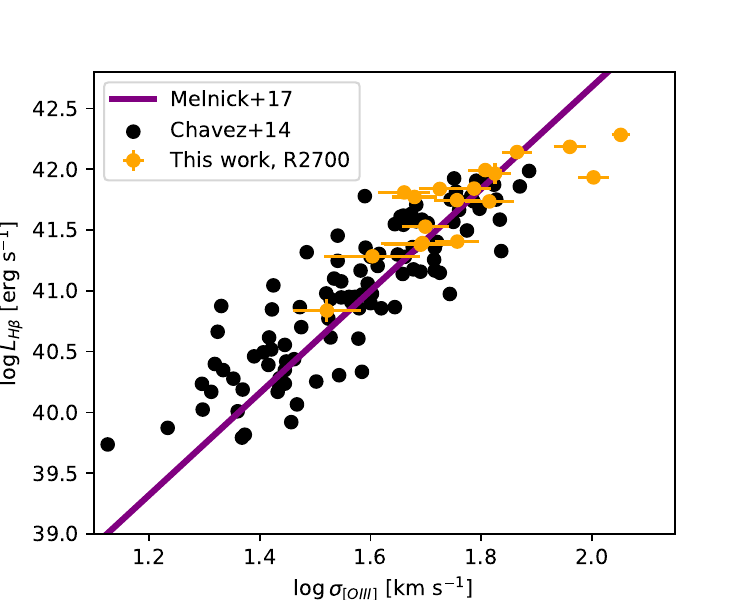}
    \caption{The \Hbs luminosity - \OIIIs velocity dispersion relationship. The purple line shows the empirical relationship of \protect\cite{Melnick2017} while emission line galaxy measurements by \protect\cite{Chavez2014} are shown by black dots. Our Type 1 AGN with R2700 data available are shown with yellow points. The Type 1 AGN are all consistent with the standard relation.}
    \label{fig:lum_sigma}
\end{figure}

As the consistency of our sources with the local $M_{BH}$ - $\sigma_{*}$ and 
$M_{BH}$-$M_{dyn}$ scaling relations might imply the presence of AGN (non-ejective) feedback inhibiting star formation, we endeavor to explore the form this feedback takes by comparing our objects to HII regions and emission line galaxies on the $L_{H\beta}$ - $\sigma$ relation found by  \cite{Melnick2017}. If AGN feedback operates through excess turbulence induced in the NLR, our sources would be expected to deviate from the relation by drifting towards higher $\sigma$ values for the same \Hbs luminosity. We conduct the comparison using the L-$\sigma$ relation derived by \cite{Melnick2017} and individual galaxy measurements by \cite{Chavez2014}. Only sources with available R2700 data are used for this comparison, with the $\sigma_{[OIII]}$ values measured from the \OIII$\lambda\lambda$4959,5007 doublet and corrected for the LSF. We show the comparison in \autoref{fig:lum_sigma}. As it can be seen in the figure, our sample of AGN does not exhibit significant deviations from the scaling relation and falls in line with emission line galaxies from \cite{Chavez2014}. This indicates that the AGN feedback present in our sources is not dominated by turbulence, which would cause them to move towards higher $\sigma_{[OIII]}$ values for a fixed luminosity. Therefore, star formation in our sources is likely inhibited through heating, which is corroborated by the presence of a strong \OIII$\lambda$4363 auroral line emission in a fraction of our sources and its clear detection in the spectral stacks as discussed in Section \ref{sec:stacks}.

\subsection{Morphologies of Type 1 AGN hosts}
\label{sec:morpho}
\begin{figure}
    \centering
    \includegraphics[width=\linewidth]{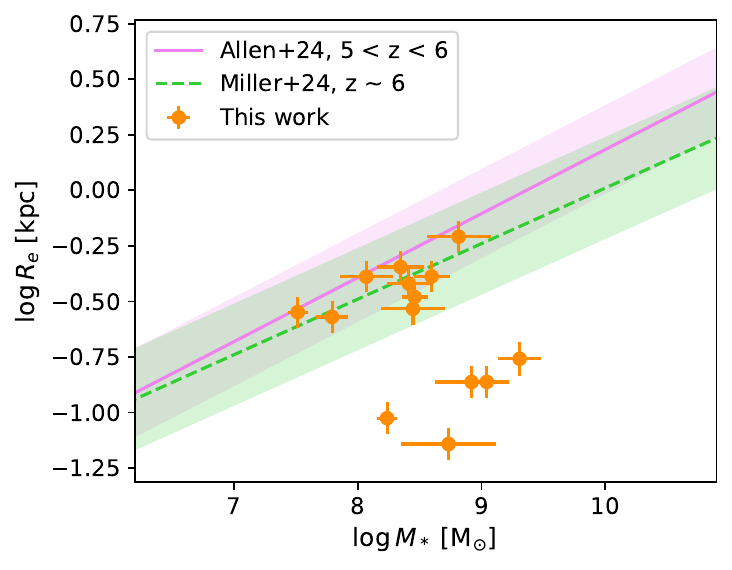}
    \caption{Location of our sample sources (orange points) relative to the mass-radius relations derived for star forming galaxies by \protect\cite{Allen2024} (purple solid line) and \protect\cite{Miller2024} (green dashed line). Colored shading indicates the $\sim 0.2$~dex scatter on both relations.}
    \label{fig:size_mass}
\end{figure}

Morphologies of galaxies hosting AGN are thought to trace the connections between growing BHs and their hosts and have been a subject of extensive research \citep{Kauffmann2003, Povic2012, Woreta2022}. Until the launch of JWST, this has been largely limited to galaxies at $z < 2$ and while a comprehensive study of AGN and star forming galaxy morphology is beyond the scope of this paper, the depth and resolution provided by JWST allows us to make headway.  Our \texttt{ForcePho} fits have produced a subsample of 14 AGN hosts with good constraints on their half-light radii ($R_e$), S\'ersic indices ($n_s$) and stellar masses, and enables comparison with size-mass relations derived for star forming galaxies. The comparison of our objects with the $R_e$ -- $M_*$ relations derived by \cite{Allen2024} and \cite{Miller2024}, around the same redshifts, is shown in \autoref{fig:size_mass}. The bulk of our objects are consistent with the relations; however, there is a significant population of AGN that are considerably more compact than star forming galaxies at similar masses.

\begin{figure}
    \centering
    \includegraphics[width=\linewidth]{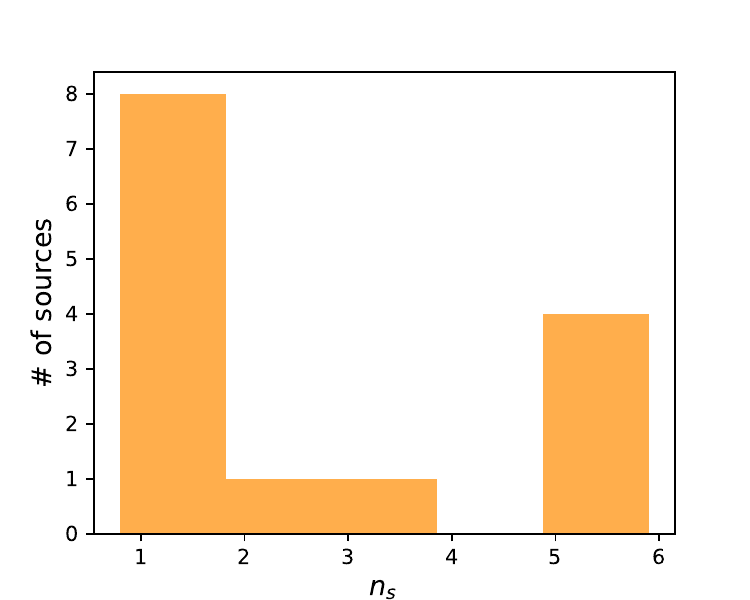}
    \caption{Histogram showcasing the distribution of S\'ersic indices of our sample sources. It can be seen that the majority of our sources have $n_s$ between 1 and 2, with a subset approaching steeper profiles at $n_s = 5$}.
    \label{fig:sersic_dist}
\end{figure}

The apparent population of compact AGN hosts (\autoref{fig:size_mass}) may indicate that these BHs may be fed by the same gas compaction mechanisms that are thought to drive bursty star formation in the early Universe \citep[][and McClymont, in prep.]{Tacchella2016, Emami2021, McClymont2025}.  This compaction is thought to occur due to intense gas inflow events triggered by minor mergers, disk instabilities \citep{Dekel2014,Tacchella2016} or accreting gas with low angular momentum from the halo \citep{Sales2012, Renzini2025}. If such compaction mechanisms also end up feeding the central BH, it would explain the population of compact AGN hosts found in our sample and indicate that high-z AGN grew through different feeding modes than their local counterparts (as already suggested by \citealp{Kocevski2017}), which are primarily fed through major mergers and interactions within galaxy clusters \citep{Bergmann2019} as well as galactic bars \citep{Combes2023, Sliva-Lima2023}.

The S\'ersic index distribution for our sources is shown in \autoref{fig:sersic_dist} and clearly indicates that about $60$\% of our sources are well fit by $n_s \approx 1$, indicating that most of the fitted AGN reside in late type, disky systems. On the other hand, the distribution is clearly bimodal, with about 30\% of AGN hosts fit by more concentrated profiles. While the latter result may indicate that a fraction of AGN are hosted in more early type, spheroidal concentrated systems, we caution that a steeper profile may be the result of residual nuclear AGN emission.

Finally, we caution that the number statistics provided by our sample are small and suffer from biases inherent to the \texttt{ForcePho} fitting procedure (as discussed in Section \ref{sec:fpho}), therefore it is too early to draw sweeping conclusions about the behaviour of AGN hosts at high redshift. Nevertheless, our results indicate the need for dedicated studies exploring AGN host morphologies at high redshifts.

\section{Stacked spectra of Type 1 AGN}
\label{sec:stacks}
In order to further our investigation into the emission line properties of our sample sources, in particular to enhance the weak spectral features, we stack the R1000 spectra in our sample. The stacking was carried out as follows - first, each spectrum was de-redshifted to its rest frame utilizing the narrow line redshifts derived by our pipeline fitting procedure. Afterwards we resample each spectrum onto a common wavelength grid, derived such that its bin widths would be half the size of the narrowest bin of the original spectra. This resampling is carried out using \texttt{spectres} \citep{Carnall2017}. After resampling, we perform the stacking via a weighted sum with weights derived from the inverse rms errors on each bin. This method maximizes the S/N ratio of the stacked spectra, complimenting the median stacking done without weighting, normalized by $F_{[OIII]}$ carried out by \cite{Isobe2025}. We also experiment with normalizing the resampled spectra to $F_{[OIII]}$ and $F_{H\alpha, BLR}$ values, in order to investigate how robust the measured line ratios, on which we base our conclusions, are to different weighting schemes and find no significant differences. The errors on the stacks were estimated as the inverse square root of the sum of weights for each bin.

The above procedure was used to obtain stacks for all sample sources as well as those in each of the previously defined redshift bins. We focus our stacking on the regions surrounding the \Has and \Hbs lines as stacks covering the full wavelength range are prone to producing continuum artifacts due to flux calibration issues between different R1000 gratings.

\begin{figure*}
    \centering
    \includegraphics[width=\textwidth]{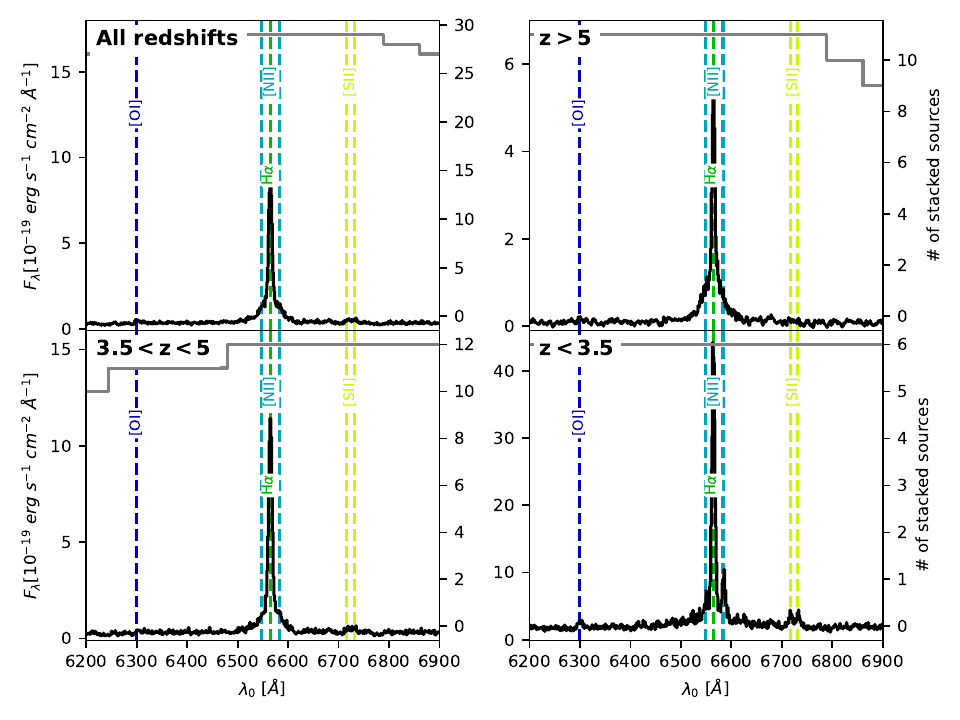}
    \caption{Medium resolution stacks of the \Has region of our sample spectra. The black line shows the flux, while the grey line on top of each plot shows the number of sources that were stacked in each wavelength range. The positions of the \OI$\lambda$6300, \Ha, \NII, and \SIIs lines are indicated by dashed vertical lines.}
    \label{fig:Ha_stacks}
\end{figure*}

The stacked spectra in the region around \Has are shown in \autoref{fig:Ha_stacks} and clearly show a decrease in the strength of \NII$\lambda\lambda$6548,6583 and \SII$\lambda\lambda$6716,6731 doublets relative to \Has as well as an increase in the prominence of the broad component relative to the narrow \Ha, when going from lower to higher redshifts.

\begin{figure*}
    \centering
  \includegraphics[width=\textwidth]{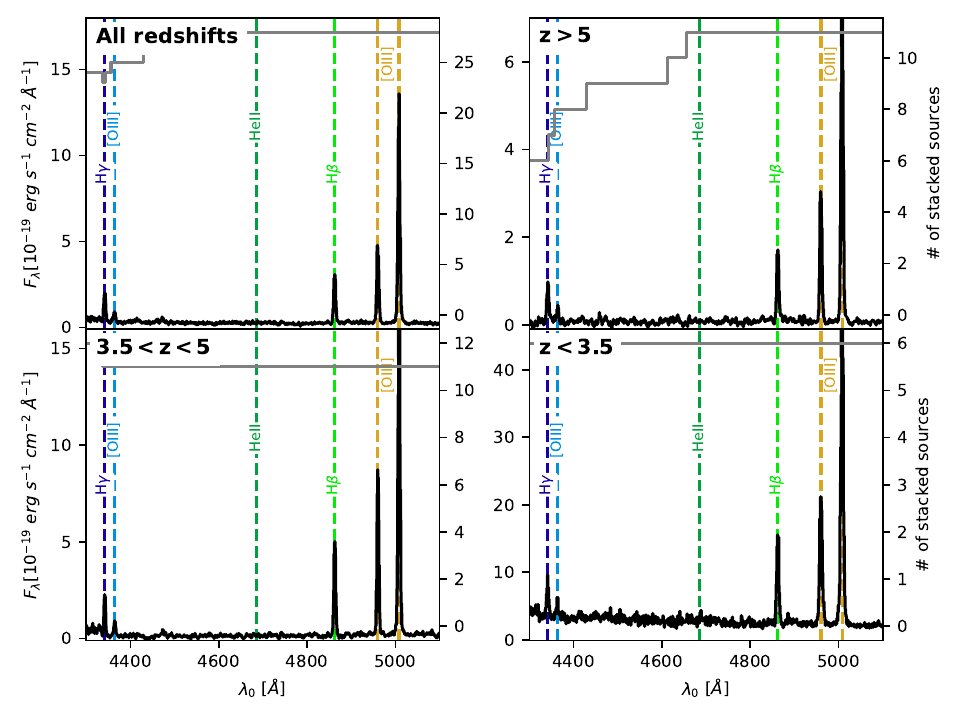}
    \caption{Medium resolution stacks of the \Hbs region of our sample spectra. The lines hold the same meaning as in \autoref{fig:Ha_stacks}. Dashed vertical lines indicate the locations of the \Hg, \Hb, \OIII, and \HeII\ emission lines.}
    \label{fig:Hb_stacks}
\end{figure*}

The stacked spectra in the region around \Hbs are shown in \autoref{fig:Hb_stacks} and illustrate the notable presence of the auroral \OIII$\lambda$4363 line across all redshifts. However, the \HeII$\lambda$4686 emission is entirely absent from the stacked spectra; the implications of this  will be discussed in Section \ref{sub:line_ratios}. In addition, the \OIII$\lambda$5007 emission profiles are narrow - suggesting a systematic lack of prominent large scale outflows in our sample of AGN, a feature also found in other Type 1 AGN JWST-selected samples \citep{Maiolino_xray_weak}.

We perform Gaussian fits to the lines found in our stacked spectra in order to obtain line fluxes from our stacked spectra. The fitting procedure followed that of Section \ref{sec:fitting}, with the exception of the redshift of the narrow lines being set to zero and the \OII$\lambda\lambda$3726,3728 doublet being fitted with a single Gaussian profile centered on 3727~\AA\ as our spectral resolution is insufficient to disentangle its components. The results from this fitting are summarized in \autoref{tab:stack_meas}. It should be noted that the fluxes for the \OII$\lambda\lambda$3726,3728 doublet were obtained from additional stacks centered on the line as, due to \OII$\lambda\lambda$3726,3728 ending up on different R1000 gratings to \Hb, stacks centered on \Hbs contained too few sources in the \OII$\lambda\lambda$3726,3728 region to be representative of the whole sample. We also list the measured \SIIs doublet ratios which change little across the different redshift bins, corresponding to $n_e$ values between $\sim 500$ and $\sim 800$~cm$^{-3}$ when estimated using the calibrations from \cite{Kaasinen2017}.

\subsection{Balmer decrement and dust extinction}

\begin{figure}
    \centering
    \includegraphics[width=\linewidth]{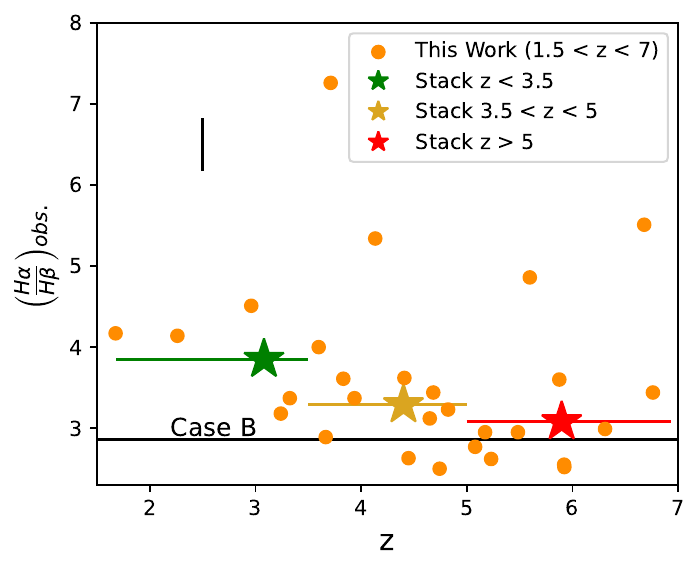}
    \caption{Showcase of the \Has to \Hbs ratio evolution for our sample AGN. The orange points show the individual sources with the black bar displaying the median error. Sample stacks are shown as colored stars placed at the median redshift values of each stack. The y errors on the stacks are smaller than the markers, while the x errors show the width of each stacked bin.  The Case B value is shown by the solid black line.}
    \label{fig:balmer_decrement_evo}
\end{figure}

The (narrow) \Has to \Hbs ratios measured from our stacked spectra show clear redshift evolution and range from 
$3.85_{-0.20}^{+0.08}$
to $3.29_{-0.03}^{+0.03}$ and 
$3.08_{-0.05}^{+0.05}$ 
for the $z < 3.5$,
$3.5<z<5$ and
$z > 5$    bins, respectively (\autoref{fig:balmer_decrement_evo}). This is consistent with redshift evolution of dust obscuration, corresponding to $A_V$ decreasing from $\sim 1$ to $\sim 0.2$, if the SMC extinction curve is assumed. These values are broadly consistent with those of individual sources in \autoref{tab:all_objecs}, for which the measured Balmer decrements generaly indicate minor to moderate dust obscuration. Although red outliers appear present across the redshift range. Either way, our sample sources do not exhibit Balmer decrements below the Case B value of 2.86 as found in a Type 1 AGN sample from \cite{Brooks2024} and emission line galaxies by \cite{Scarlata2024}. The interpretation of \cite{Brooks2024} for obtaining $F_{H\alpha}/F_{H\beta} < 2.86$ was increased temperature in the ISM. Following this interpretation, our ratios of $ > 3$ may be explained by lower ISM temperatures or higher dust obscuration in the NLR masking the lower ratio. However, our stacks display clear \OIII$\lambda$4363 emission, which indicates $T_e \approx 2\times 10^4$~K across all bins, when using \texttt{pyNeb} \citep{PyNeb} and assuming $n_e = 600$~cm$^{-3}$. This indicates that the ISM in our sources is warm and may suggest that the $A_V$ values in \autoref{tab:all_objecs} may be underestimated as they were calculated assuming a standard Case B ratio of $2.86$. However, as pointed out by \cite{Osterbrock2006, Smith2022} and \cite{Sandles2024}, the intrinsic \Has and \Hbs flux ratio varies by less than 10\% in the range $\rm 5000\ K\leq T_e \leq 20000\ K$ and $\rm 500\ cm^{-3}\leq n_e \leq 10000\ cm^{-3}$. Thus, a significant underestimation of $A_V$ is unlikely.

It should be noted that the \Has and \Hbs ratio for the stack of all sample sources is $F_{H\alpha}/F_{H\beta} \approx 4.2$, this is considerably higher than the stacks in the individual redshift bins and is inconsistent with the $F_{H\gamma}/F_{H\beta} \approx 0.53$, above the Case B value, for the same stack. Since this inconsistency is not present in the stacks of individual redshift bins, a likely reason for this discrepancy is flux calibration issues between the different grating/filter combinations of R1000. In order to check for any issues with our stacking procedure itself, we generate mock spectra with a fixed Case B $F_{H\gamma}/F_{H\beta}$ of 0.47 and find that the stacked spectra recover the set value. Thus the deviations in the stack of the entire sample are likely due to systematic errors in the flux calibrations (\cite{Bunker2024} points our $\sim 10$\% flux differences between prism and gratings). These systematics do not affect the individual redshift bins as much since the stacked lines are more likely to end up in the same grating/filter. Comparing the $F_{H\alpha}/F_{H\beta}$ and $F_{H\gamma}/F_{H\beta}$ values of the entire sample stack to a weighted mean of the results in the individual redshift bins we find that the magnitude of this systematic uncertainty is in the 10-20\% range. We thus add 20\% to the fit uncertainties presented in the `All redshifts' column of \autoref{tab:stack_meas} and caution that any stacking of involving the full range of grating/filter combinations of R1000 must carefully account for the flux calibration differences.

\subsection{Narrow line ratio diagnostics}
\label{sub:line_ratios}
Both recent and pre-JWST era observations have shown that traditional narrow line diagnostic diagrams, such as the BPT \citep{bpt} and VO87 \citep{vo} diagrams, struggle to robustly identify Type 2 AGN at higher redshifts due to the sensitivity of their lines to metallicity and more intense starburst activity \citep{Masters2014, Coil2015, Ubler2023, Scholtz2023}. As a significant fraction of AGN are expected to be in the Type 2 regime \citep{Vijarnwannaluk2022, Scholtz2023}, exploring alternative narrow line diagnostics is crucial in working towards a complete census of high redshift AGN in both obscured and unobscured regimes. The statistics and redshift coverage of our sample allow for thorough testing of classical and newly established Type 2 AGN diagnostics and assessing their redshift evolution. 

\begin{figure*}
    \centering
    \includegraphics[width=\textwidth]{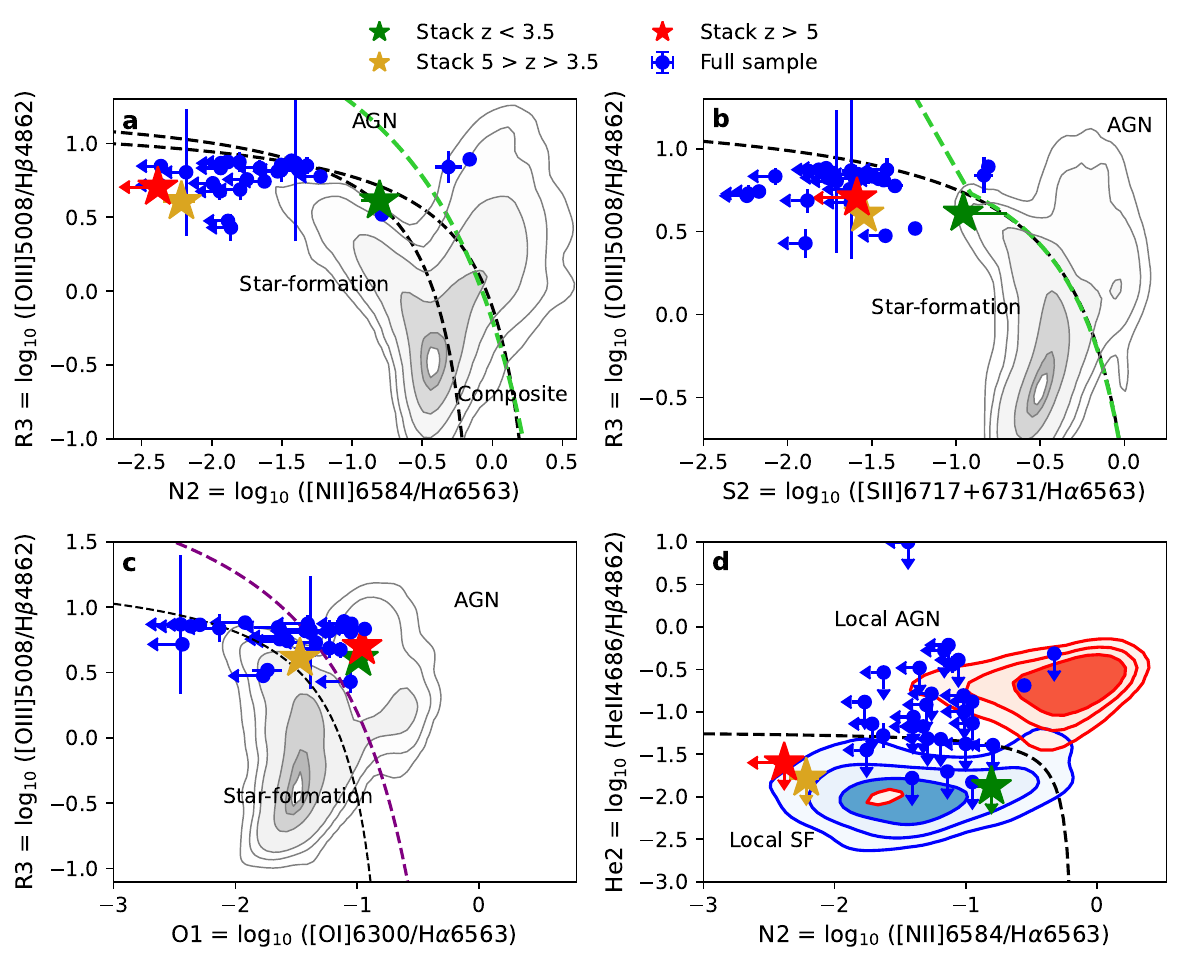}
    \caption{\textbf{a.} Showcase of our sample sources (blue circles) and the stacks across the three redshift bins (colored stars) on the classic BPT diagnostic diagram. The black dashed lines represent the AGN - star forming demarcation lines from \citet{Kauffmann2003,Kewley2001}, while the green dashed line is the one used in \protect\cite{Scholtz2023} for high-$z$ galaxies. The gray contours are SDSS galaxies. It can be seen that most of our sources lie in the star-forming region due to weak \NIIs emission, while the stacked spectra show a strong redshift evolution of the N2 ratio. \textbf{b.} Position of our sample sources and stacks on the S2 diagram. The dashed black line is the demarcation from \protect\cite{Kewley2001}, while other symbols have the same meanings as in panel \textbf{a}. As in the previous panel, most of our sources lie in the star-forming region of the plot and a significant redshift evolution in the stacks is observed. \textbf{c.} Same as panel \textbf{b}, but for the O1 diagram with the purple line being the boundary from \protect\cite{Mazzolari2024CEERS}. This diagnostic does not exhibit consistent redshift evolution, however, most of our sources straddle the boundary. \textbf{d.} Location of the stacked and individual sample spectra on the \HeII\ diagram. SDSS AGN and star-forming galaxies selected through BPT are shown in red and blue contours, respectively. The dashed demarcation line is from \protect\cite{Shirazi2012}}
    \label{fig:bpt}
\end{figure*}

We therefore utilize our broad-line AGN sample to check their location on the new and old narrow line ratio diagnostic diagrams. We first focus on the traditional $\rm N2 = \log{[NII]/H\alpha}$, $\rm R3 = \log{[OIII]/H\beta}$ BPT diagram \citep{bpt} as well as similar diagrams incorporating \OI$\lambda$6300, \SIIs and \HeII$\lambda$4686 lines \citep{vo, Kewley2001, Shirazi2012}. Panel \textbf{a} of \autoref{fig:bpt} shows the distribution of our sources in the traditional N2-R3 BPT diagram \citep{bpt}. Most of them lie firmly in the star-forming regions of the plot as found in other studies of high redshift Type 1 and Type 2 AGN \citep{Ubler2023, Maiolino_AGN, Scholtz2023, Harikane_AGN, Mazzolari2024, Mazzolari2024CEERS, Backhaus2025}. We also utilize the stacked R1000 spectra to investigate the presence of faint UV/optical lines. The N2 and R3 ratios from the stacks are also shown in \autoref{fig:bpt} and showcase that, even if detected, the \NIIs emission is too weak to classify our sources as AGN. In addition, the stacks in \autoref{fig:bpt} also show strong evolution of the N2 ratio with redshift as the $z < 3.5$ stack is considerably closer to the boundary than the higher redshift ones, albeit still not in the AGN region of the diagram. This evolution is likely caused by higher redshift sources generally being more metal poor and the \NIIs doublet's sensitivity to metallicity, although the magnitude of the decrease shown by the stacks may be overestimated due to blending of weak \NIIs with broad \Ha.
More generally, photoionization models of the NLR of AGN have shown that the decreasing metallicity of the high-z AGN host is indeed expected to result into the steady decrease of both [NII]/H$\alpha$ and [OIII]/$\beta$ \citep{Nakajima2022,Maiolino_AGN,Ubler2023}.

A similar, although lesser, effect is seen in the $\rm S2 = \log{[SII]/H\alpha}$ {\it vs.\ }R3 diagram \citep{vo} shown in panel \textbf{b} of \autoref{fig:bpt}, where the \SIIs doublet gets progressively weaker with redshift \citep[see also][]{Ubler2023, Scholtz2023}. This offset, while still present, is considerably less extreme in $\rm O1 = \log{[OI]/H\alpha}$ {\it vs.\ }R3 \citep{vo}, shown in panel \textbf{c} of \autoref{fig:bpt}, and does not show the same consistent evolution with redshift across the stacks. This would indicate that lowering metallicity is a strong contributor to the decrease in effectiveness of the standard emission line diagnostic diagrams, as both nitrogen and sulphur have delayed enrichment with respect to that of oxygen \citep{Kobayashi_2020}.

We also investigate the positions of our sample sources on the He2-N2 diagram from \cite{Shirazi2012} and \cite{Tozzi2023}, which replaces the \OIIIs in the BPT with \HeII$\lambda$4686, as used in recent JWST studies. However, the \HeII\ line is robustly detected in only two sources across the entire sample and is otherwise absent even when the spectra are stacked across the full redshift range and individual bins. These stacks yield an upper limit on the \HeII$\lambda$4686 to \Hbs ratio of $\sim0.02$ and place our sources in the star-forming regions of the He2-N2 diagram (panel \textbf{d} of \autoref{fig:bpt}). While the offset on the N2, S2 and O1 BPT diagrams may be attributed to low metallicity \citep{Nakajima2022}, the absence of metallicity-insensitive high ionization lines (such as \HeII$\lambda$4686) may be indicative of our sample AGN being inefficient at producing high energy ionizing photons. Whether this inefficiency is intrinsic to the structures of the accretion disks of these AGN and/or is caused by dense gas blanketing the BLR and NLR remains to be investigated \citep[see e.g.][]{Lambrides2024, Madau2024, Inayoshi2024b, Maiolino_xray_weak}. It is important to notice that there are also noticeable exceptions; for instance the prominent type 1.8 AGN at z=5.5 studied by \cite{Ubler2023} and \cite{Ji2024GS3073} does show clear HeII emission, both broad and narrow, despited being totally undetected in deep X-ray data. Additionally, as seen in panel \textbf{d} of \autoref{fig:bpt}, some SDSS AGN are similarly weak in \HeII$\lambda$4686, therefore, local analogues of this higher redshift population likely exist. It is also important to note that,  the scenario of nuclear blanketing of the ionizing radiation would provide an alternative or additional explanation for the offset of our AGN on the BPT diagrams, whereby the AGN ionizing source does not reach the circumnuclear region to form a NLR, hence the observed narrow lines would be dominated by star formation in the host galaxy.

\begin{figure}
    \centering
    \includegraphics[width=\columnwidth]{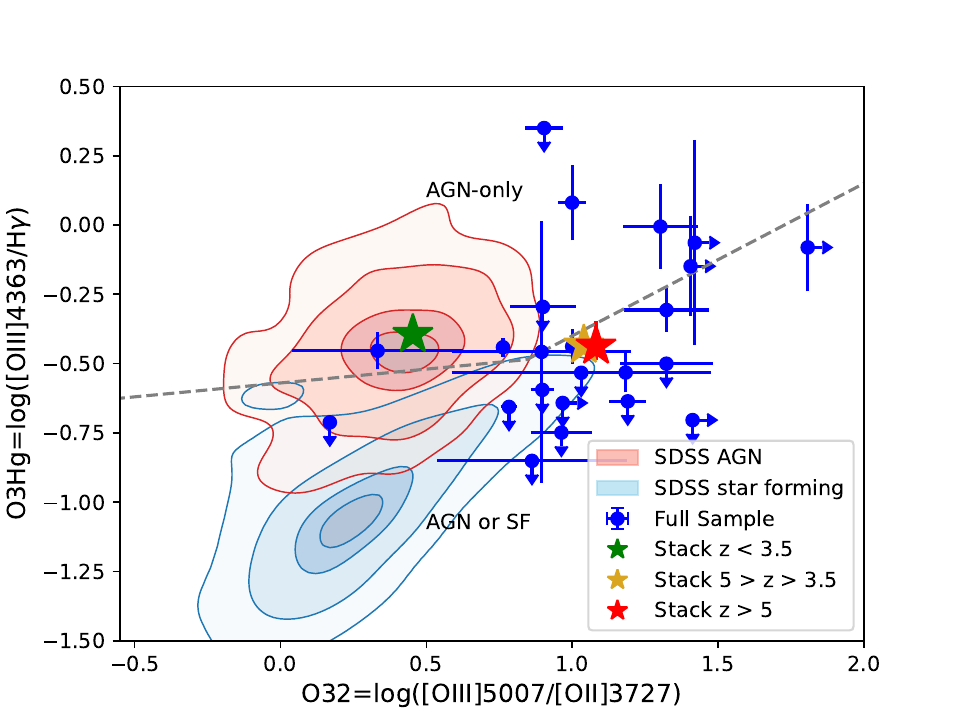}
    \caption{Diagnostic diagram of the \OIII$\lambda$4363/\Hg\ and \OIII$\lambda$5007/\OII$\lambda$3727 from \protect\cite{Mazzolari2024}. The AGN and star forming galaxies from SDSS are shown in red and blue contours respectively. Our sample sources and stacks are shown with blue circles and colored stars as in \autoref{fig:bpt}. The grey dashed line indicates the boundary between the region populated by AGN only (above it) and the one with a mixture of AGN and star forming galaxies (below it).}
    \label{fig:Mazzolari_AUR}
\end{figure}

In addition to the traditional strong emission line diagnostics, we also utilize our sample to test some newly proposed AGN diagnostics diagrams, mainly those based on the \OIII$\lambda$4363 \citep{Mazzolari2024} line. Specifically, these authors propose a log[OIII]4363/H$\gamma$ (O3Hg) vs log[OIII]5007/[OII]3727 (O3O2) diagram, where they identify a region of high O3Hg populated only by AGN, while other parts of the diagram is populated by both AGN and SF galaxies hence inconclusive. The O32-O3Hg diagram from \cite{Mazzolari2024} with our sample is shown in \autoref{fig:Mazzolari_AUR}. As can be seen there, most of our individual sources do not have strong enough detections of the \OIII$\lambda$4363 or \OII$\lambda$3727 lines to be reliably classified using this diagnostic. However, all required lines are clearly detected in the stacked spectra. The lowest ($z < 3.5$) stack is largely coincident with the position of the SDSS AGN, while the two $z > 3.5$ stacks showcase strong evolution in \OIII/\OII\ ratio going from $z\sim2$ to $z\sim4$, which moves both of them slightly beyond the decision boundary and into the region populated by both SFGs and AGN. This weakening of the \OIIs relative to the \OIIIs is likely indicative of more extreme ionization conditions in higher redshift galaxies or lower metallicities \citep{Cameron2024}. These results indicate that, while the O32-O3Hg diagnostic appears to be more efficient in in selecting AGN than the BPT, it still struggles at higher redshifts. It should however be noted that \cite{Mazzolari2024} are quite conservative in defining the boundary identifying the AGN-only region of the diagram, and that the high-z stack would  be inconsistent with the envelope of their star-forming photoionization models. 

Overall, our investigation of the narrow emission line diagnostics shows that there is a considerable population of AGN that are unidentifiable by narrow emission lines only. This results in the AGN fraction in galaxy evolution estimated in previous studies \citep[e.g. ][]{Scholtz2023, Mazzolari2024CEERS} being a lower limit on the overall number of AGN, as there is a considerable number of AGN that can only be identified through broad component in their permitted lines.

\section{Shapes of the broad emission lines}
\label{sec:BLR_shape}
Recent works \citep{DEugenio2025,Maiolino_AGN,Rusakov2025} have found that many AGN found by JWST exhibit extended wings in their broad line profiles that cannot be accommodated by a single Gaussian. \cite{Santos2025GravitySinfoni, Rusakov2025} suggested that these profiles are exponential in shape and result from Compton scattering in a dense medium. These results would indicate significant overestimation of BH masses and underestimation of accretion rates. However, `wingy' profiles, not well reproduced by a single Gaussian, have been seen in more generally in most luminous quasars, where they are well reproduced by either broken power laws, double Gaussians  \citep{Kollatschny2013,Nagao2006, Santos2025GravitySinfoni} and are largely attributed to the complex internal dynamics of the BLR. If this latter scenario is true, then detecting the more extended wings even in less luminous AGN may be a matter of signal-to-noise. We thus attempt to test these scenarios by exploring the shapes of our Type 1 candidates by investigating the performance of single Gaussian, double Gaussian, and exponential profiles.

We construct the double Gaussian model as a simple superposition of two Gaussians centred on the same wavelength. The widths of the two components were constrained to the same priors as in Section \ref{sec:fitting}. It should be noted that we do not interpret the second Gaussian profile as tracing the presence of a second BH \citep[although this could be the case in some objects, see][]{Ubler24} - this component is introduced to empirically model the extended wings not fit by a single Gaussian. As for the exponential profile, this was constructed following the prescriptions of \cite{Rusakov2025} and consisted of an intrinsically Gaussian component, modelling the unscattered BLR emission, convolved with a broken symmetric exponential of the form:
\begin{equation}
\label{eq:expo}
    N(\lambda_0, B, W)  = B\exp(-|\lambda - \lambda_0|/W),
\end{equation}
where $\lambda_0$ is the central wavelength, $B$ - the amplitude and $W$ - the e-folding length. The full BLR profile of \cite{Rusakov2025} would then have the form:
\begin{equation}
    f_sN(\lambda)*G(\lambda)+(1-f_s)G(\lambda),
\end{equation}
where $f_s$ is the fraction of light scattered, $N(\lambda)$ - the scattering exponential from \autoref{eq:expo} and $G(\lambda)$ - the intrinsic Gaussian. The prior on the FWHM of the exponential profile (${\rm FWHM} \equiv 2W\ln{2}$) is the same as that set on the broad Gaussians in the Section \ref{sec:fitting} fits.

\begin{figure}
    \centering
    \includegraphics[width=\linewidth]{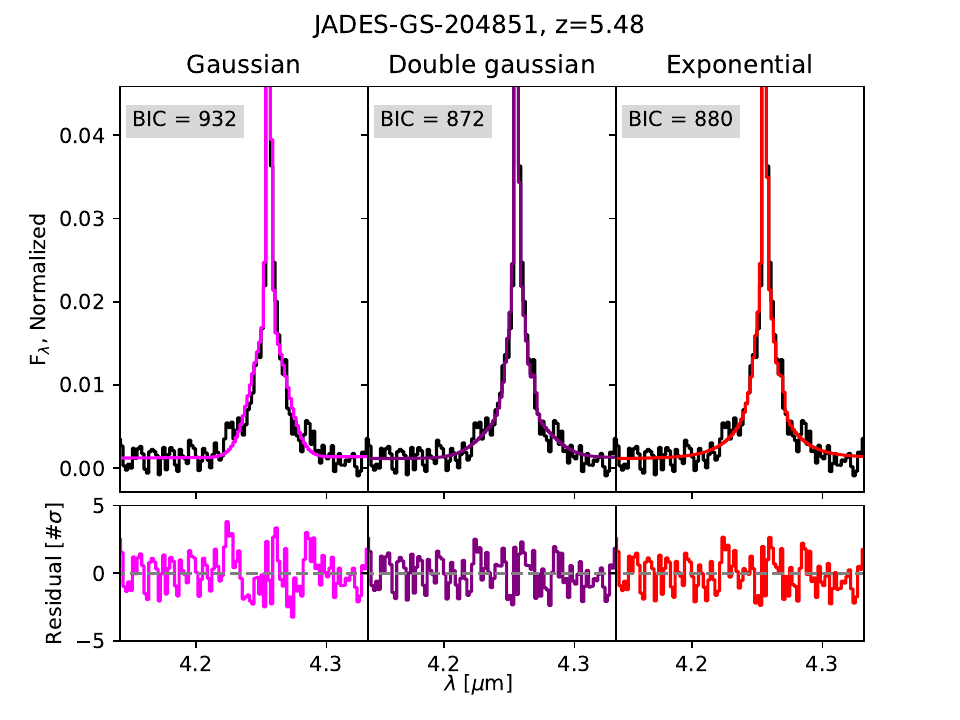}
    \caption{An example comparison of the single Gaussian, double Gaussian and exponential fits to one of our objects exhibiting extended BLR wings. It can be seen that the single Gaussian model leaves symmetric systematic residuals that are considerably reduced when fitting a double Gaussian or Exponential profiles.}
    \label{fig:wingy_example}
\end{figure}

We then refit all of our main sample sources (excluding the 6 tentative objects) for a total of 28 fitted, with the two models described above and quantify the relative performance using the BIC (\autoref{eq:bic}). We find that for most of our sample sources (22) a single-Gaussian fit is preferred. However, 6 of our sample sources (GS-49729, GN-73488, GN-28074, GS-204851, GS-38562 and GS-159717) exhibit broad wings that cannot be accommodated by a single Gaussian fit (see \autoref{fig:wingy_example} for an illustration). For these sources, the double Gaussian and exponential profiles generally fit equally well, with the double Gaussian outperforming the exponential in terms of BIC for 5 of them and performing equally well for the remaining sixth (see \autoref{tab:gauss_expo} for a summary). Treating the entire profile of the sources with extended BLR wings as a result of virial broadening does change the BH mass estimates, however, the resulting 0.3 - 0.4~dex differences are well within the scatter on the virial estimators. In addition, the second Gaussian component in GN-28074 was found by \cite{Juodzbalis2024b} to kinematically match an outflow in \OIII, thus interpreting it as originating in the BLR is likely inappropriate.

\begin{figure}
    \centering
    \subfloat{\includegraphics[width=\linewidth]{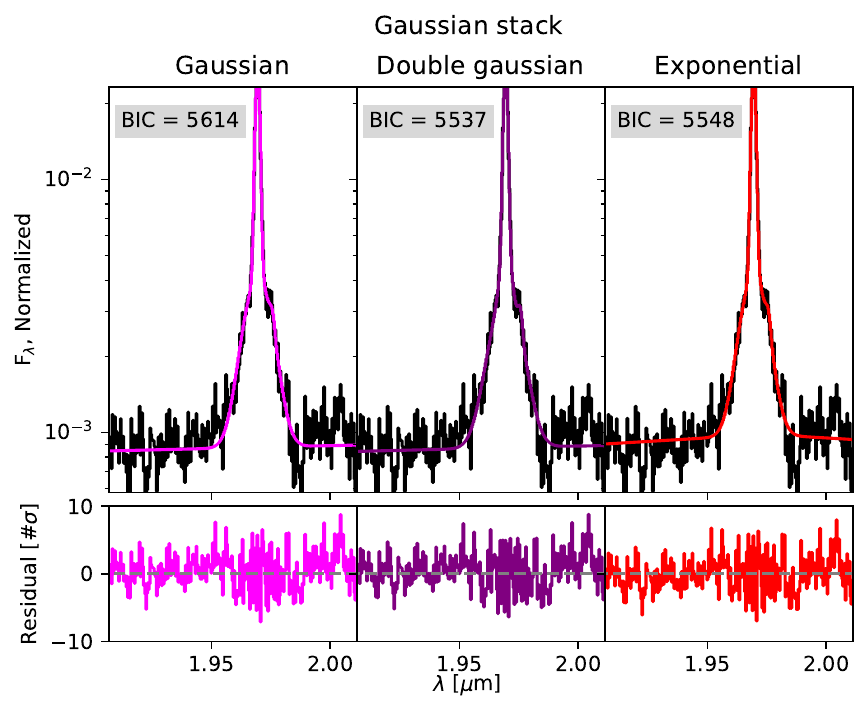}}
    \hfill
    \subfloat{\includegraphics[width=\linewidth]{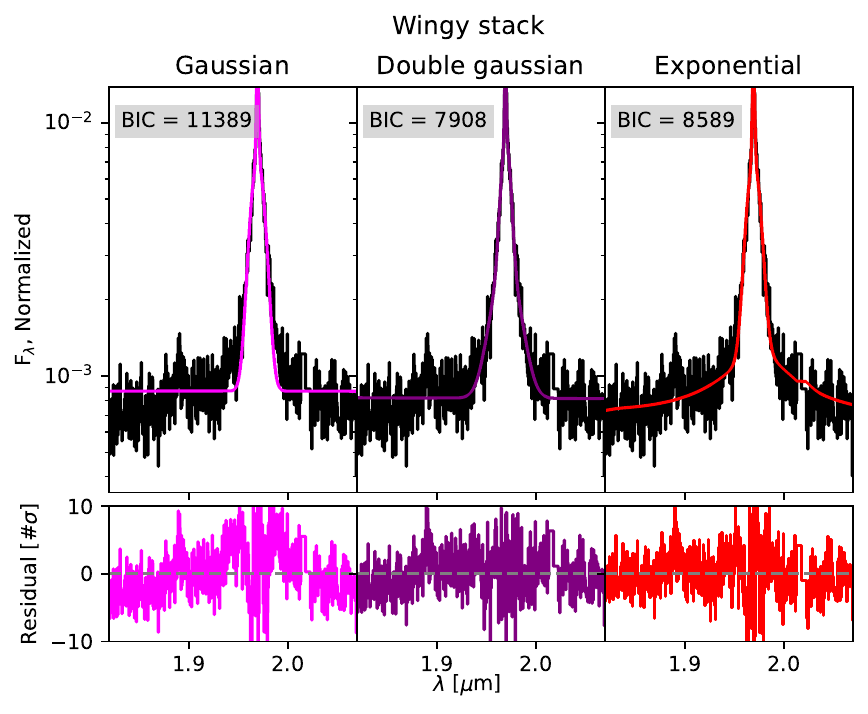}}
    \caption{Showcase of the performance of the three models on the stacks of the objects with simple Gaussian profiles (top) and those with extended BLR wings (bottom). The stacks are all redshifted to $z=2$ and the narrow lines not part of the \Ha-\NIIs complex are excluded from fitting and BIC calculation. It can be seen that the double Gaussian model is preferred for both stacks.}
    \label{fig:stack_shape}
\end{figure}

We note that the 6 AGN in \autoref{tab:gauss_expo} are among the most luminous in our sample, suggesting that the reason for the apparent simplicity of the remaining broad profiles may be low signal-to-noise. To test this, we stack the objects best-fitted by a single Gaussian profile. The stacking was performed with the same methods as in Section \ref{sec:stacks}. We also stack the objects in \autoref{tab:gauss_expo} to obtain further constraints on the shapes of their profiles. The results for each stack are shown in \autoref{fig:stack_shape}. As shown in the figure, the double Gaussian is strongly preferred over the simple Gaussian or the exponential models for both stacks. Given the diversity in luminosities, BH masses, and line widths of the AGN that ended up in the Gaussian profile stack, there is a possibility of the extended wings being spuriously introduced by stacking faint broad components and brighter, narrower ones. To mitigate this, we experiment with weighting the stacks by the inverse FWHM of the \Has line as well as removing BHs with $M > 10^8$~M$_{\odot}$ from the stack and find that the feature persists. This result suggests that non-Gaussianity may be intrinsic to the vast majority of AGN BLR and only apparently identified in the brightest quasars due to S/N constraints. In \autoref{fig:stack_shape} (bottom) we also stack those objects for which the single Gaussian fit was not satisfactory, not surprislingly showing broad wings. Interestingly, as shown in the same figure, a fit with a double-Gaussian is preferred relative to the exponential fit in both stacks. 

The origin of the broader component is currently unclear - it may be due to the inner part of the BLR \citep{Sulentic_2002},
turbulence in the BLR \citep{Kollatschny2013}, a combination of rotation and turbulence and/or probing different regions of the BLR \citep{Santos2025GravitySinfoni}, or an outflowing component \citep[e.g.][]{Matthews2023}.
Another possibility is that such wings are due to electron scattering, as suggested by \cite{Laor2006} and \cite{Rusakov2025}. The latter actually propose that most of the broad line profile of JWST-selected AGN is produced by electron scattering.
However, we discuss in the following section that this scenario is highly unlikely for our sample.

\subsection{A scattering origin of the broad lines?}
\label{sec:scattering_scen}

Most models assume that the clouds of the BLR are in virial equilibrium around the black hole, with possible contributions from an outflow or inflow component. This assumption is at the basis of the virial relations between black hole mass, width (or velocity dispersion) of the permitted broad line and continuum luminosity, derived from reverberation mapping \citep[e.g.][]{Du2015, Du2018, Li2021, DallaBonta2020}. This assumption has been verified by comparing these AGN with the black hole masses inferred from the $M_{BH}-\sigma _*$ relation for black holes with direct dynamical measurements \citep[e.g.][]{Park2012,Woo2015}.

However, recently \cite{Rusakov2025} and \cite{Naidu2025} have suggested that the observed widths of the broad lines seen in the high-z AGN and LRDs discovered by JWST are not tracing the motions of the BLR clouds, but scattering from a medium located outside the BLR. According to their models, the intrinsic  width of the broad line would be 5 - 10 times narrower compared to the observed width, and claim that the observed width is mostly the result of scattering by the outer medium.
Given that in the virial relations the mass of the black hole scales quadraticaly with the line width, this would imply that the black hole masses might be overestimated by about two orders of magnitude. While a more extensive discussion will be presented in a separate paper, in this section, we briefly illustrate that these scenarios are unlikely to apply to the majority of our sample sources.

\cite{Rusakov2025} fit the H$\alpha$ profiles in the high S/N spectra of 13 JWST-discovered AGN at z$\sim$3--7. They find that, with the exception of one object, an exponential profile of the broad line is preferred to a single Gaussian (although a double Gaussian fit is not attempted by them). They suggest that this is evidence for an intrinsically very narrow ``broad'' H$\alpha$ associated with the nuclear BLR, which is embedded in an outer ionized medium whose free electrons scatter the line into a much broader exponential profile. In this scenario, the outer ionizing medium embedding the BLR must have a column density of ionized gas on the order of $10^{24}~{\rm cm}^{-2}$ to explain the observed broad \Ha\ wings in their sample. This scenario has difficulties in explaining our sample, as discussed in the following paragraphs.

The previous section has shown that an exponential profile is not necessarily the best fit -- we have seen in our sample that a double-Gaussian profile often fits better the individual profiles and stacks. In addition, as discussed in the previous section, profiles with extended wings that can be approximated with power laws or exponential profiles are commonly seen in many AGN \citep{Kollatschny2013,Nagao2006,Collin2006,Santos2025}, especially those approaching quasar-like luminosities, which include nearby AGN with reverberation studies, which were used to calibrate the virial relations. Therefore, the proposed electron scattering scenario should apply to the a significant fraction of AGN/quasars, including those used for calibrating the reverberation mapping \citep{Collin2006,DallaBonta2020}. This would seem to imply that the black hole masses are overestimated by orders of magnitude for most black holes in AGN/quasars. However, the consistency with the $M_{BH}-\sigma_*$ relation \citep[][]{Park2012,Woo2015}, defined by cases for which the black hole mass is measured directly, together with the consistency with direct BH mass measurements in AGN \citep{Winkel2025}, excludes that black holes are generally overestimated by orders of magnitude via the virial relations.
Moreover, low-z quasars with ``wingy'' profiles are generally not Compton thick to X-ray radiation 
\citep[which is another prediction of the scattering scenario proposed by ][]{Rusakov2025} and exhibit the same X-ray properties as other type 1 AGN \citep{Lusso2020,Maiolino_xray_weak}.

\begin{figure}
    \centering
    \subfloat{\includegraphics[width=\linewidth]{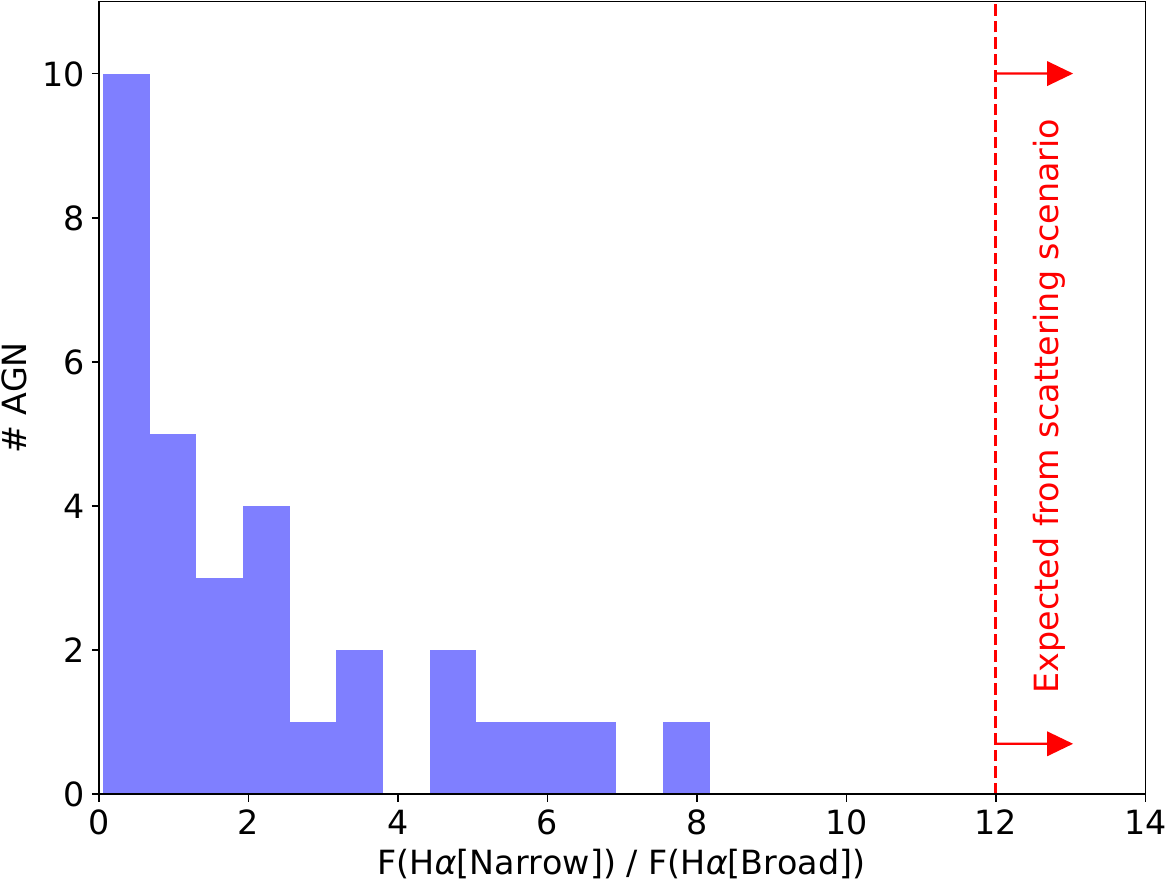}}
    \caption{Distribution of the ratio between narrow and broad component of H$\alpha$ in our sample. The vertical dashed line indicates the lower limit of the ratio expected in the scenario proposed by \citet{Rusakov2025} whereby the broad component results from electron scattering of a proposed much narrower BLR component from a hypothetical ionionized medium surrounding the BLR}
    \label{fig:Han_Hab_ratio}
\end{figure}

Another  physical problem that remains to be addressed for the electron scattering scenario is the recombination within the scattering medium. The hypothetical ionized scattering medium located outside the BLR proposed by \cite{Rusakov2025} should have a large amount of $\rm H^+$ ($\sim 10^{24}~{\rm cm^{-2}}$ based on the large $\rm N_{\rm e}$ from $\tau _{\rm scatter}\sim 1$), therefore producing
its own H$\alpha$ emission, independent of the \Ha\ emission from the BLR that has been scattered. The highly ionized medium located outside the BLR must be photoionized by the same central accretion disc and, in order to explain the ubiquity of the scattering scenario proposed by \cite{Rusakov2025}, must be nearly entirely covering the BLR (and, therefore, the accretion disc). 
However, to allow a significant fraction of ionizing photons to reach such an outer ionized (electron scattering) medium, the covering factor of the inner BLR should be much smaller, or otherwise it would prevent ionizing photons from reaching the outer ionized medium. 
Both the BLR and the hypothesized outer ionized medium emit H$\alpha$ via recombination. The width of the H$\alpha$ emitted by the outer medium is presumably narrower than the H$\alpha$ intrinsically coming from the BLR, hence it can only be part of the narrow component of H$\alpha$ (given that in the best-fit models of \citealp{Rusakov2025} the widths of the BLR lines are typically as narrow as $\rm FWHM_{intrinsic}\lesssim600~{ km~s^{-1}}$). 
Meanwhile, in the scenario proposed by \cite{Rusakov2025}, the H$\alpha$ coming from BLR is usually not seen directly (only a small fraction of this is transmitted and unscattered), while the bulk of the BLR emission is scattered in the putative exponential wings (i.e. what we see as very broad lines in the JWST spectra). 
To the first order, the ratio of the H$\alpha$ fluxes of the outer scattering medium (produced by recombination) and the inner BLR is roughly proportional to their covering factor relative to the accretion disc that is responsible for the photoionzation. 
The covering fraction ratio should be reflected by the occurrence rate of exponential-profile BLR among all BLRs, barring selection effects.
In the high S/N sample of \cite{Rusakov2025} only one source out of 13 has a broad line profile dominated by a single Gaussian, while all others are dominated by the exponential wings, hence seen through the putative scattering medium. Therefore, according to the scenario proposed by \cite{Rusakov2025}, the ratio between the narrow component of H$\alpha$ (coming from recombination in the proposed scattering medium) and the broad component of H$\alpha$ (coming from the electron scattering of the BLR) should be at least 12:1. This is completely inconsistent with our observations. \autoref{fig:Han_Hab_ratio} shows the distribution  of the narrow-to-broad H$\alpha$ flux ratio in our sample. All of our AGN have this ratio smaller than eight, and smaller than four for 80\% of the sampole. The high S/N sample in \cite{Rusakov2025} has a distribution even more skewed towards low values. The vertical dashed line shows the expected value in the scenario proposed by \cite{Rusakov2025}. This ratio of 12:1 is a lower limit, since the narrow component must also include the standard narrow component from the NLR, \Has emission by star formation in the host, and, in some cases, by the residual emission from the BLR that is transmitted and not scattered.

In addition to the caveats discussed above, the electron scattering scenario appears to require very high ionization to produce $N_{\rm e}\sim10^{24}~{\rm cm}^{-2}$. For reference, \citet{Laor2006} inferred an ionization parameter of $\log(U)\sim -0.5$ for the scattering medium of NGC 4395, which has only optically thin scattering with $\tau_{\rm scatter}\approx 0.34$, lower than the typical $\tau_{\rm scatter} \sim 1$ inferred for \citet{Rusakov2025}'s sample. Thus, the proposed electron scatter dominated scenario would appear to require $\log(U)\sim 0$, larger than typically inferred for the BLR, despite being at larger distances. Additionally, such a large ionization parameter has implications for the radiation pressure and implications for the stability of such ionized medium, which according to \cite{Rusakov2025} should be common to all newly discovered AGN (which are a significant fraction of the whole galaxy population, as discused in the next section). Detailed exploration of this aspect is, however, beyond the scope of this work and will be presented in a separate paper.

\cite{Naidu2025} proposed a different scattering scenario. In contrast to \cite{Rusakov2025}, in the \cite{Naidu2025} scenario the scattering medium is largely neutral, but warm, with $N_{\rm H}\sim10^{26}$~cm$^{-2}$ and $n_{\rm H}\sim10^{11}$~cm$^{-3}$. Such a
neutral, warm medium would be responsible for producing a strong Balmer absorption, similarly to the scenario proposed by \cite{Inayoshi2024} and \cite{Ji2025} for LRDs. However, \cite{Naidu2025} further suggest that the same medium introduces Balmer scattering of H$\beta$ and \Ha,
analogous to the Ly$\alpha$ scattering in observations, but from hydrogen atoms whose $n=2$ state is populated by collisional excitation and Ly$\alpha$ trapping. According to the scenario proposed by \cite{Naidu2025}, the intrinsic width of the Balmer lines in the BLR could be much narrower than observed and that the double peaked Balmer line profiles produced by this scattering would resemble Balmer absorption, and give apparent widths much broader than the intrinsic BLR. They propose that this mechanism applies to all LRDs, hence resulting into an overestimation of the black hole masses by orders of magnitude. The scenario proposed by \cite{Naidu2025} can be tested via lower redshift LRDs in our sample, as discussed in the following.

\begin{figure}
    \centering
    \includegraphics[width=\linewidth]{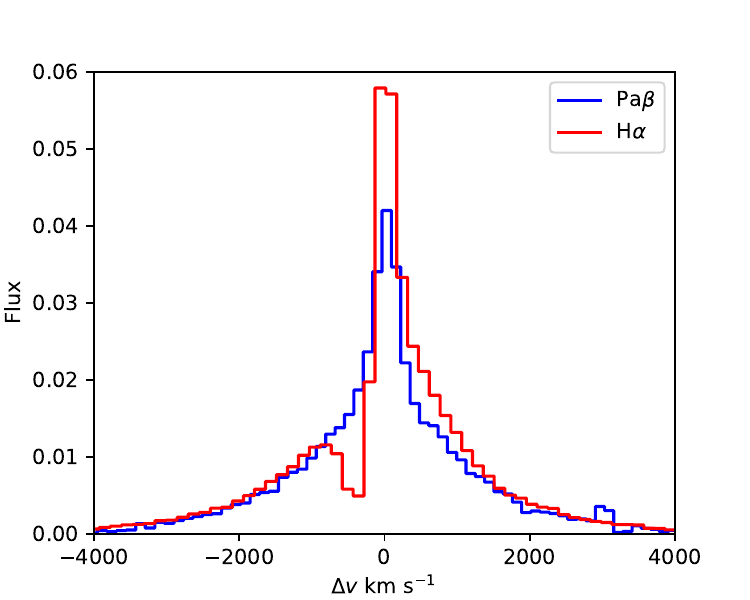}
    \caption{Comparison of the \Has and Pa$\beta$ line profiles of GN-28074 \citep{Juodzbalis2024b}, to evaluate the resonant-scattering scenario. The two lines are shown in velocity space, and have been normalized to the same flux in the wings. In the
    resonant-scattering scenario of \citet{Naidu2025}, Pa$\beta$ should have a considerably narrower line profile than \Ha, contrary to
    observations.
    }
    \label{fig:Ha_PaB}
\end{figure}

While a more extensive analysis will be presented in a separate paper
(Brazzini et al. in prep.), here we focus specifically on the case of GN-28074 at z=2.3. This object was already presented in \cite{Juodzbalis2024b} as a more luminous ``Rosetta Stone'' for understanding higher-z LRDs. This object presents broad permitted lines, a Balmer break, and absorption of H$\alpha$ and H$\beta$ (and HeI$\lambda$10830), which \cite{Juodzbalis2024b} demonstrate requires absorption by gas with densities $n>10^9~cm^{-3}$. The same mechanism likely occurs in most LRDs \citep{Inayoshi2024, Ji2025}, including the source presented in \cite{Naidu2025}. However, the detection of bright Paschen lines in GN-28074 offers the possibility of testing the Balmer scattering scenario proposed by \cite{Naidu2025}. If the broad wings observed in H$\alpha$ and H$\beta$ were due to Balmer scattering of a much narrower BLR line, then the same wings should not be seen in Pa$\beta$, which would mostly be tracing the intrinsic BLR emission. In the \cite{Naidu2025} scenario the population of n=3 of hydrogen is much lower than n=2, even in the case of high temperatures and densities (making the population of levels thermalized in a Maxwellian distribution), hence making the column density of hydrogen atoms in the n=3 state much smaller than n=2. This should make the wings seen in the Balmer lines not observable in Pa$\beta$ and make it much narrower than \Has or \Hb, due to the much smaller cross-section of Paschen scattering. However, the Pa$\beta$ line of GN-28074 shows prominent broad wings and an overall similar shape to H$\alpha$. This is illustrated in \autoref{fig:Ha_PaB} where the H$\alpha$ and Pa$\beta$ profiles of GN-28074 are overlaid, after normalizing them to the same flux of the wings. Pa$\beta$ does not show evidence of absorption -- this is expected in the simpler scenario proposed by \cite{Juodzbalis2024b}, \cite{Ji2025} and \cite{Inayoshi2024}, in which the high density in the absorbing medium populates n=2, but not n=3 levels. However, the most important feature is the presence of broad wings of Pa$\beta$ -- as shown by the figure, Pa$\beta$ has proportionally more flux in the wings than H$\alpha$, in contradiction to what is expected by the Balmer scattering scenario. 

Finally, if the black hole masses were overestimated by two orders of magnitude, this would certainly help to bring them down to the local $M_{BH}-M_{star}$ relation. However, since they are located close to the local $M_{BH}-\sigma$ relation (Section\ref{sec:BH-sigma}), reducing their black hole mass by two orders of magnitude would make them severely undermassive on the  latter relation, which is considered more fundamental and universal \citep{KormendyHo2013,Newman2025}.

In summary, both the electron scattering and Balmer scattering scenarios for explaining  the broad wings of the Balmer lines appear untenable for the vast majority of our objects, although they may apply to a few rare cases. Therefore, there is no evidence that the black hole masses have been systematically overestimated by orders of magnitude as proposed by \cite{Rusakov2025} and \cite{Naidu2025}.

\section{UV luminosity function of AGN and their hosts}
\label{sec:uvlf}
\begin{figure*}
    \includegraphics[width=\textwidth]{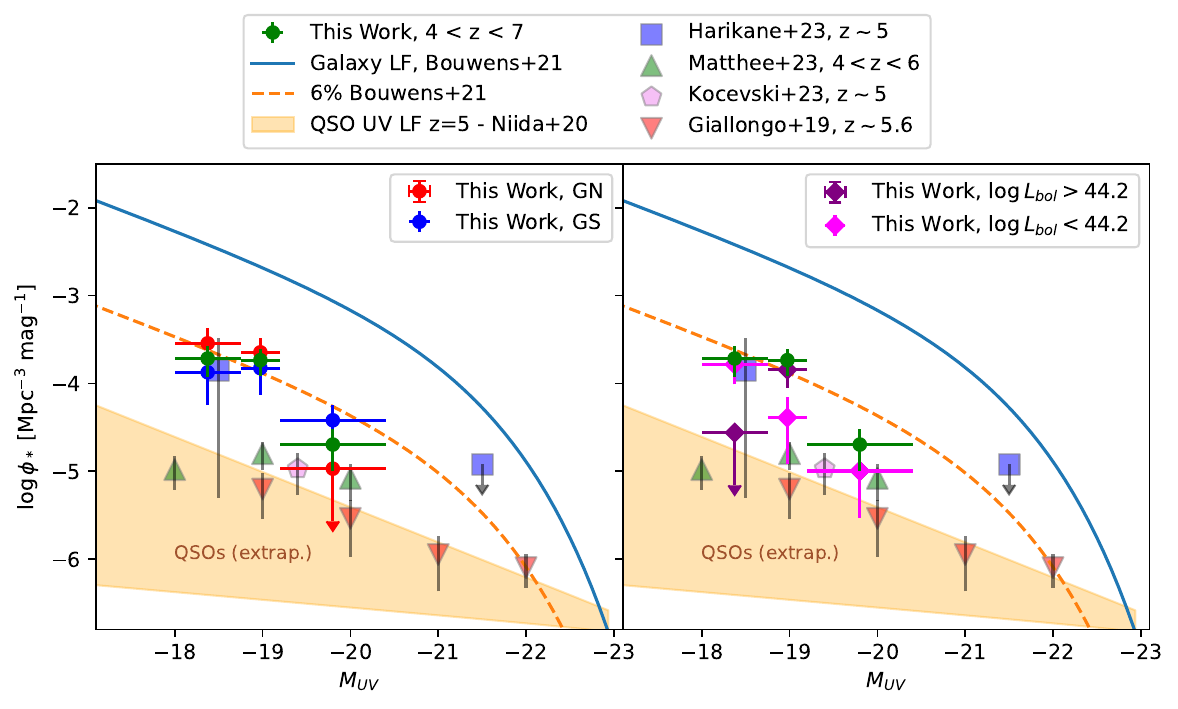}
    \caption{UVLF of our Type 1 AGN hosts between z = 4 and z = 7. Density estimates for the whole sample are shown with green points. The left panel shows the variance in the LF when the sample is split across the two fields - GN and GS with red and blue points respectively. The right panel shows results from a split in $L_{bol}$ with high and low luminosity bins shown in purple and magenta diamonds, respectively. The highest MUV bin contains the same amount of sources across both $L_{bol}$ ranges thus the points overlap. The solid blue line shows the galaxy LF at $z \sim 5$ from \protect\cite{Bouwens2021}, while the dashed orange line shows the same LF scaled down to 6\%. The orange shaded region shows the range of extrapolated QSO LFs from \protect\cite{Niida2020}. Results from X-ray AGN studies by \protect\cite{Giallongo2019} are shown in orange triangles. The remaining points showcase results from other JWST surveys as indicated in the legend.}
    \label{fig:UVLF}
\end{figure*}

We utilize the increased statistical power afforded by the expanded sample to estimate the contribution of Type 1 AGN and their hosts to the UV luminosity function (LF) at $4 < z < 7$ \citep[note that this redshift bin is slightly wider and extended to higher redshift than the $4 < z <6$ one used in the previous work by][]{Maiolino_AGN}. The total number densities of AGN hosts were estimated by rescaling the galaxy UVLF at z = 6 from \cite{Bouwens2021} by the ratios of number of AGN hosts to that of star forming galaxies in JADES in each magnitude bin. The UV magnitude bins were chosen to include roughly the same amount of AGN hosts and were centered on $M_{UV} = -18.375$, $M_{UV} = -18.975$ and $M_{UV} = -19.8$ with the brightest bin introduced to include a small subset of UV bright AGN. We find that AGN make up 5\%, 9\% and 2\% of all galaxies in corresponding magnitude bins although these values are likely lower limits due to our conservative sample selection. The source counts are 7, 9 and 4 for the faintest, medium and brightest bins respectively. The overall UVLF estimate is plotted as green points in \autoref{fig:UVLF}. While our sample statistics are not constraining enough to obtain a reasonable fit to a functional form, rescaling the \cite{Bouwens2021} results down to 6\% can reasonably reproduce the densities observed in the two fainter bins. The brightest bin, however, sits significantly below this rescaled curve and may indicate a steepening of the UVLF of faint AGN hosts relative to the bright population, however, statistics there are small. 

Overall, our sample contains 20 AGN at $4 < z < 7$, almost double the amount published in \cite{Maiolino_AGN}. This allows us to explore variations in number density estimates due to cosmic variance. We investigate cosmic variance first, by splitting our sample across GOODS-S and GOODS-N fields. Both fields end up containing 10 AGN at the redshift range considered thus any variation in bin densities is due to variance in galaxy counts per field and the AGN UV luminosity. As shown in the left panel of \autoref{fig:UVLF}, GOODS-N field contains an excess of faint AGN when compared to GOODS-S. On the contrary, GOODS-S has  more sources at higher $M_{UV}$ relative to GOODS-N. Clearly, these differences highlight that cosmic variance plays an important role in the  number density and luminosity distribution of AGN in small fields and can affect the statistics by at least a factor of a few. This also highlights that the evolutionary patterns of AGN (as imprinted in the luminosity function) are likely different in different environments and different primeval conditions.

\autoref{fig:UVLF} also showcases a comparison between our results and those from QSO \citep{Niida2020} and X-ray AGN \citep{Giallongo2019} surveys together with results from the first years of JWST observations \citep{Harikane_AGN, Matthee2024, Kocevski2024}. We find that density estimates lie significantly above the extrapolated QSO luminosity functions as well as those from X-ray based AGN surveys (\autoref{fig:UVLF}) as found by \cite{Maiolino_AGN} for a subsample of our sources. Our estimated AGN fractions, however, are somewhat lower than theirs, with AGN hosts contributing about 6\% of galaxy density, as opposed to $\sim$10\% found by \cite{Maiolino_AGN}. This could likely be attributed to an increase in the number of galaxy spectra taken by the JADES survey along with an expanded (and higher) redshift range considered here. However, confirming that the UV luminosity function of the newly discovered AGN is one or two orders of magnitude higher than the extrapolation of quasars, indicated that JWST is likely uncovering a new population of AGN possibly formed through a different route. Additionally, confirming that the new population of AGN  is much more numerous than X-ray selected AGN is in line with the finding that they are X-ray weak \citep{Maiolino_xray_weak} and that X-ray surveys have likely found only the tip of the iceberg of the population, either less (X-ray) absorbed or intrinsically more X-ray loud.

In order to investigate the variance of the UVLF with varying AGN bolometric luminosity, we split our sample into high ($L_{bol} > 10^{44.2}$~erg~s$^{-1}$) and low ($L_{bol} < 10^{44.2}$~erg~s$^{-1}$) luminosity halves, containing 10 sources each. The resulting density estimates are shown in the right panel of \autoref{fig:UVLF}. As it is shown there, the differences in the luminous end are not drastic, while the high luminosity sample contains far fewer sources in the faintest $M_{UV}$ bin. This may indicate some level of contribution of AGN emission to the UV magnitude for more luminous AGN or a correlation between $L{bol}$ and SFR of the host, which brightens the UV luminosity of the hosts of luminous AGN and ``depletes" the lowest UV luminosity bin. However the discrepancy between the luminosity bins is not large. This indicates that the AGN component is either sub-dominant due to low accretion rates, as is clearly exemplified by GN-38509 \citep{Juodzbalis2024} or due to their UV emission being weak. Dust obscuration provides a tempting explanation for this weakness, however, most of our sources, with the exception of GN-209777, show low to moderate attenuation ($0 < A_V < 2.0$) of the NLR and host galaxy. Thus either the dust is concentrated towards the BLR or the AGN have intrinsically red SEDs if the lack of AGN contribution to the UV emission is to be explained by a physical deficiency in UV photons.

We conclude by mentioning that our updated UV luminosity function confirms that AGN and their hosts can potentially contribute to the reionization of the universe, although a more quantitative assessment requires a proper disentangling between AGN and host galaxy contribution to the observed UV radiation, and should also take into account of the cosmic variance revealed by our study \citep{Madau2024reionization,Grazian2024reioniz,Dayal2024,Asthana2024reioniz}; such an analysis is beyond the scope of this paper.

\section{Conclusions}
\label{sec:conclusion}
In this paper we have presented a robust sample of Type 1 AGN spanning redshifts from $z = 1.7$ to $z = 7$ with the tentative broad \Hbs emitters reaching $z = 9$, enabling us to probe AGN from Cosmic Noon to the Epoch of Reionization. The luminosity regime investigated is $>2$~dex fainter than that of all-sky quasar surveys and about $\sim 0.5$~dex below that of other JWST AGN surveys. Thus our sample presents a deep view into the low-luminosity regime of AGN activity. It should be noted that only $\sim$30\% of our sample sources exhibit LRD like spectra \citep{Hainline2025} thus, while LRDs make up a significant fraction of the Type 1 AGN population, most Type 1 AGN are not LRDs.

The low luminosities of most of our AGN sample allow for reasonable estimates of the stellar masses of their hosts. Most BHs in our sample are significantly overmassive with respect to the local $M_{BH}-M{star}$ relations, with the highest excess occurring at $z>5$, while all AGN at $z < 3.5$ are consistent with the local $M_{BH}$-$M_{*}$ scaling relation (although our statistics in the lower redshift regime are less constraining). This is nevertheless indicative of the local $M_{BH}$-$M_{*}$ relation being established at $z < 4$, consistent with the findings of some BH growth models \citep{Trinca2024} and observational work by \cite{Sun2025}.

When comparing our BH masses to the inferred stellar velocity dispersions of each of our sources, we find that all our objects, irrespective of redshift, are consistent with local scaling relations. For a subsample we could also infer the dynamical mass of the host galaxy, and finding also in this case that our sources are consistent with the local $M_{BH}-M_{dyn}$ relation. This suggests that most of the gas required to bring our sources to agreement with the local scaling relations is likely already present, but star formation is being inhibited by AGN feedback. Alternatively, the dynamical mass of these early systems could be Dark Matter dominated. However, both of these scenarios require additional testing in simulations. In addition, the lack of a significant,  offset relative to local $M_{BH}$-$\sigma_{*}$ and $M_{BH}$-$M_{dyn}$ relations (\autoref{fig:sigma_mbh}) suggests that the overmassive nature of BHs at high redshifts can not be ascribed entirely to selection effects although these are certainly present \citep{Li2024, Juodzbalis2024}.

Comparison of the NLR of our AGN to the $L_{H\beta} - \sigma$ relation from \cite{Melnick2017} shows no significant deviation between them and star forming galaxies. This is suggestive of a lack of turbulent or ejective feedback present in our sample AGN and implies that star formation in our sources is likely inhibited through heating of the gas.

Our evaluation of different narrow line diagnostics reveals that no single narrow line diagnostic method is capable of selecting a complete and pure sample of Type II AGN at high redshifts. The \OIII$\lambda$4363 diagnostics from \cite{Mazzolari2024} are promising in terms of purity, however, their completeness suffers at higher redshift due to weakening of the \OII$\lambda$3727 with respect to \OIII$\lambda$5007. 

The lack of \HeII$\lambda$4686 or any other high ionization line emission in our sources, even in the stacked spectra, is curious and indicates a deficiency in high energy photons reaching the NLR. This may be connected to the relative prevalence of Balmer absorption features of JWST-discovered broad line AGN \citep{Wang2024, Matthee2024, Juodzbalis2024b}, which traces high density gas surrounding their nuclei. This gas, in addition to producing Balmer absorption lines, could likely attenuate UV emission by bound - free absorption from n=1 and n=2 states of hydrogen, reducing the ionizing photon budget and potentially leading to non-stellar Balmer breaks as theoretically shown by \cite{Inayoshi2024} and demonstrated observationally by \cite{Ji2025} and \cite{Naidu2025}. However, more in-depth studies investigating the prevalence of Balmer absorption lines in low luminosity Type 1 AGN as well as follow-up observations exploring their nature are needed to robustly establish this scenario.

Our luminosity function confirms that the new population of AGN revealed by JWST is one or two orders of magnitude more numerous than expected from the extrapolation of the quasars' luminosity function, suggesting that JWST is uncovering a different population, possibly formed through different processes. They are also about one or two order of magnitude more numerous than X-ray selected AGN, which is in line with the finding that they are X-ray weak \citep{Maiolino_xray_weak}.

We also investigated the recent claims that the broad lines observed in the JWST-discovered AGN are driven by electron scattering by an ionized medium \citep{Rusakov2025}, or Balmer scattering by a neutral medium \citep{Naidu2025} around the BLR. In these scenarios the intrinsic width of H$\alpha$ produced by the BLR would be much narrower than observed. As a consequence, the black hole masses estimated via the virial relations applied to the observed width would have been systematically overestimated by orders of magnitude. We have shown that both of these scattering effects do not contribute significantly to the BLR line widths for the majority of our sample sources. A more complete analysis of the physical implications of these scenarios is deferred to a future paper.

Lastly, we have assessed the UV luminosity function of the AGN in our sample and their host galaxies. We confirm that these AGN are at least 1-2 orders of magnitude more abundant than the extrapolation of luminous quasars and X-ray selected AGN. Moreover, our abundance estimates are likely lower limits due to conservative sample selection. We found substantial cosmic variance, by a factor of a few, between GOODS-S and GOODS-N, which should be taken into account when assessing the fraction of AGN in the early Universe. We also find a dependence of the UV luminosity function on the AGN bolometric luminosity, which should also be taken into account when assessing their space density.

\clearpage
\begin{table}
    \centering
    \begin{tabular}{cc}
    \hline
        Disperser + Filter &  Nominal wavelength range [$\mu$m] \\
    \hline
        G140M/F070LP & 0.70 -- 1.27\textdagger \\
        G235M/F170LP & 1.66 -- 3.07\\
        G395M/F290LP & 2.87 -- 5.10 \\
        G395H/F290LP & 2.87 -- 5.14 \\
    \hline
    \end{tabular}
    \caption{A summary of grating/filter configurations for the R1000 and R2700 dispersers of NIRSpec. \textdagger The data processing pipeline employed by JADES recovers G140M/F070LP spectra beyond the nominal wavelength range, see text for details.}
    \label{tab:R1000}
\end{table}

\begin{table*}
    \centering
    \tabcolsep 3pt 
    \begin{tabular}{cccccccccc}
    \hline
        Object ID & R.A. & Dec & z & $\log{M_{BH}/M_{\odot}}$ & $\log{L_{bol}}$/(erg~s$^{-1}$) & $\lambda_{Edd}$ & F$_{H\alpha;\ nr}$/F$_{H\beta;\ nr}$ &   A$_V$ & Ref. \\
    \hline
    \multicolumn{10}{c}{Robust} \\
    \hline
    GS-30148179 & 53.14208 & -27.77985 & 5.922 & $7.12_{-0.35}^{+0.34}$ & $44.25_{-0.08}^{+0.06}$ & $0.11_{-0.03}^{+0.05}$ & $2.52_{-0.19}^{+0.18}$ &  0 & --\\
    GS-10013704 & 53.12654 & -27.81809 & 5.919 & $7.44_{-0.31}^{+0.31}$ & $44.29_{-0.02}^{+0.02}$ & $0.055_{-0.006}^{+0.007}$ &  $2.55_{-0.16}^{+0.20}$ & 0 & 1.\\
    GS-210600 & 53.16611 & -27.77985 & 6.306 & $7.41_{-0.34}^{+0.33}$ & $44.29_{-0.11}^{+0.09}$ & $0.057_{-0.013}^{+0.016}$ & $2.99_{-0.25}^{+0.22}$ &  $0.12_{-0.23}^{+0.19}$ & -- \\
    GS-209777 & 53.15847 & -27.77405 & 3.709 & $8.90_{-0.30}^{+0.30}$ & $45.42_{-0.01}^{+0.01}$ & $0.027_{-0.001}^{+0.001}$ & $7.26_{-1.63}^{+2.52}$* &  $2.59_{-0.70}^{+0.83}$* & -- \\
    GS-204851 & 53.13859 & -27.79025 & 5.480 & $7.68_{-0.31}^{+0.32}$ & $44.97_{-0.12}^{+0.17}$ & $0.16_{-0.02}^{+0.03}$ & $2.95_{-0.28}^{+0.44}$ &  $0.08_{-0.28}^{+0.38}$ & 2.  \\
    GS-179198 & 53.08898 & -27.86069 & 3.830 & $7.23_{-0.33}^{+1.42}$ & $43.92_{-0.08}^{+4.46}$ & $0.04_{-0.01}^{+9.44}$ & $3.61_{-0.16}^{+136.0}$ &  $0.62_{-0.12}^{+9.8}$ & -- \\
    GS-172975 & 53.08773 & -27.87124 & 4.741 & $7.25_{-0.32}^{+0.32}$ & $44.07_{-0.04}^{+0.04}$ & $0.05_{-0.01}^{+0.01}$ & $2.50_{-0.40}^{+0.65}$ &  0 & -- \\
    GS-159717 & 53.09753 & -27.90126 & 5.077 & $7.44_{-0.30}^{+0.30}$ & $45.13_{-0.008}^{+0.008}$ & $0.38_{-0.04}^{+0.04}$ & $2.77_{-0.38}^{+0.61}$ &  0. & -- \\
    GS-159438 & 53.05447 & -27.90246 & 3.239 & $6.47_{-0.31}^{+0.32}$ & $44.11_{-0.04}^{+0.03}$ & $0.35_{-0.08}^{+0.05}$ & $3.18_{-0.07}^{+0.08}$ &  $0.28_{-0.06}^{+0.07}$ & -- \\
    GN-77652 & 189.29323 & 62.19900 & 5.229 & $6.62_{-0.32}^{+0.34}$ & $44.11_{-0.05}^{+0.04}$ & $0.24_{-0.08}^{+0.09}$ & $2.62_{-0.53}^{+1.45}$ &  0 & 1. \\
    GN-73488 & 189.19740 & 62.17723 & 4.133 & $7.83_{-0.30}^{+0.30}$ & $45.22_{-0.05}^{+0.07}$ & $0.20_{-0.01}^{+0.01}$ & $5.34_{-0.24}^{+0.33}$ &  $1.68_{-0.12}^{+0.16}$ & 1.  \\
    GN-62309 & 189.24898 & 62.21835 & 5.172 & $6.29_{-0.33}^{+0.36}$ & $43.56_{-0.12}^{+0.13}$ & $0.15_{-0.05}^{+0.05}$ & $2.95_{-0.26}^{+0.33}$ &  $0.08_{-0.24}^{+0.28}$ & 1. \\
    GN-61888 & 189.16802 & 62.21701 & 5.874 & $7.04_{-0.32}^{+0.33}$ & $44.38_{-0.07}^{+0.09}$ & $0.18_{-0.03}^{+0.05}$ & $3.60_{-0.31}^{+0.43}$ &  $0.62_{-0.24}^{+0.30}$ & 1.  \\
    GN-53757 & 189.26978 & 62.19421 & 4.447 & $7.33_{-0.31}^{+0.31}$ & $44.29_{-0.02}^{+0.02}$ & $0.07_{-0.009}^{+0.01}$ & $2.63_{-0.34}^{+0.42}$ &  0 & 1.  \\
    GS-49729 & 53.17850 & -27.78411 & 3.189 & $7.67_{-0.30}^{+0.30}$ & $44.83_{-0.02}^{+0.02}$ & $0.115_{-0.004}^{+0.005}$ & -- &  -- & -- \\
    GS-38562 & 53.13586 & -27.87165 & 4.822 & $7.51_{-0.31}^{+0.30}$ & $44.70_{-0.06}^{+0.06}$ & $0.12_{-0.01}^{+0.01}$ & $3.23_{-0.21}^{+0.22}$ &  $0.32_{-0.19}^{+0.18}$ & -- \\
    GN-38509 & 189.09144 & 62.22811 & 6.678 & $8.57_{-0.38}^{+0.37}$ & $44.84_{-0.14}^{+0.14}$ & $0.015_{-0.005}^{+0.008}$ & $5.51_{-0.69}^{+0.86}$ &  $1.74_{-0.37}^{+0.38}$ & 3.  \\
    GN-29648 & 189.20920 & 62.26427 & 2.960 & $6.81_{-0.35}^{+0.39}$ & $43.90_{-0.06}^{+0.08}$ & $0.11_{-0.03}^{+0.12}$ & $4.51_{-0.16}^{+0.26}$ &  $1.22_{-0.10}^{+0.15}$ & -- \\
    GN-28074 & 189.06457 & 62.27382 & 2.259 & $8.55_{-0.30}^{+0.30}$ & $45.76_{-0.02}^{+0.02}$ & $0.129_{-0.003}^{+0.004}$ & $4.14_{-0.14}^{+0.16}$ &  $1.78_{-0.06}^{+0.07}$ & 4.\\
    GN-20621 & 189.12252 & 62.29285 & 4.682 & $7.05_{-0.34}^{+0.35}$ & $44.17_{-0.13}^{+0.16}$ & $0.11_{-0.03}^{+0.05}$ & $3.44_{-0.50}^{+0.72}$ &   $0.50_{-0.42}^{+0.50}$ & 1.  \\
    GS-17341 & 53.08727 & -27.72962 & 3.598 & $6.76_{-0.37}^{+0.38}$ & $44.01_{-0.12}^{+0.16}$ & $0.15_{-0.05}^{+0.09}$ & $4.00_{-0.61}^{+0.93}$ &  $0.90_{-0.45}^{+0.56}$ & -- \\
    GS-13329 & 53.13904 & -27.78443 & 3.936 & $6.86_{-0.37}^{+0.34}$ & $44.11_{-0.10}^{+0.09}$ & $0.14_{-0.03}^{+0.06}$ & $3.37_{-0.40}^{+0.42}$ &  $0.44_{-0.34}^{+0.31}$ & -- \\
    GN-11836 & 189.22059 & 62.26368 & 4.409 & $6.06_{-0.32}^{+0.32}$ & $44.11_{-0.05}^{+0.05}$ & $0.11_{-0.02}^{+0.03}$ & $3.62_{-0.11}^{+0.12}$ &  $0.63_{-0.08}^{+0.09}$ & 1. \\
    GS-9598 & 53.16181 & -27.77072 & 3.324 & $6.48_{-0.35}^{+0.45}$ & $43.85_{-0.11}^{+0.09}$ & $0.18_{-0.08}^{+0.09}$ & $3.37_{-0.40}^{+0.42}$ &  $1.01_{-0.16}^{+0.16}$ & -- \\
    GS-8083 & 53.13284 & -27.80186 & 4.753 & $7.10_{-0.31}^{+0.31}$ & $44.03_{-0.04}^{+0.04}$ & $0.07_{-0.009}^{+0.01}$ & $3.12_{-0.23}^{+0.25}$ &  $0.23_{-0.11}^{+0.11}$ & 1.  \\
    GN-2916 & 189.10774 & 62.26952 & 3.664 & $7.05_{-0.43}^{+0.44}$ & $43.91_{-0.14}^{+0.13}$ & $0.06_{-0.03}^{+0.07}$ & $2.89_{-0.31}^{+0.33}$ &  $0.03_{-0.31}^{+0.29}$ & -- \\
    GN-1093 & 189.17974 & 62.22463& 5.594 & $7.14_{-0.33}^{+0.34}$ & $44.32_{-0.11}^{+0.13}$ & $0.12_{-0.03}^{+0.04}$ & $4.86_{-0.59}^{+0.80}$ &  $1.43_{-0.35}^{+0.41}$ & 1.  \\
    GN-954 & 189.15197 & 62.25964& 6.759 & $7.70_{-0.32}^{+0.31}$ & $45.17_{-0.17}^{+0.15}$ & $0.23_{-0.04}^{+0.03}$ & $3.44_{-0.46}^{+0.44}$ &  $0.49_{-0.39}^{+0.32}$ & 1.  \\
    \hline
    \multicolumn{10}{c}{Tentative} \\
    \hline
    GS-200679 & 53.11392 & -27.80620 & 4.547 & $6.19_{-0.30}^{+0.60}$ & $43.63_{-0.39}^{+0.09}$ & $0.23_{-0.19}^{+0.02}$ & -- &  -- & -- \\
    GN-23924 & 189.03205 & 62.25089 & 1.676 & $7.22_{-0.33}^{+0.32}$ & $43.61_{-0.11}^{+0.10}$ & $0.020_{-0.007}^{+0.010}$ & $4.17_{-0.23}^{+0.25}$ &  $0.25_{-0.13}^{+0.13}$ & -- \\
    \hline
    \multicolumn{10}{c}{Tentative \Hbs} \\
    \hline
    GS-20057765 & 53.04080 & -27.85901 & 8.913 & $7.33_{-0.70}^{+0.62}$ & $44.16_{-0.30}^{+0.19}$ & $0.051_{-0.036}^{+0.090}$ & -- &  -- & -- \\
    GS-20030333 & 53.05373 & -27.87789 & 7.891 & $7.42_{-0.48}^{+0.65}$ &$44.44_{-0.14}^{+0.11}$& $0.08_{-0.056}^{+0.078}$ & -- &  -- & -- \\
    GS-164055 & 53.08168 &  -27.88857 & 7.397 & $7.63_{-0.66}^{+0.59}$ & $44.21_{-0.21}^{+0.16}$ & $0.03_{-0.020}^{+0.048}$ & -- &  -- & -- \\
    GN-4685 & 189.09629 &  62.23914 & 7.415 & $7.36_{-0.42}^{+0.45}$ & $44.13_{-0.12}^{+0.10}$ & $0.045_{-0.020}^{+0.020}$ & -- &  -- & -- \\
    \hline
    
    \end{tabular}
    \caption{Summary of the key derived AGN properties for all objects in our catalog. The first column provides the catalog ID, the second and third - their coordinates. Column four gives the best-fit redshift estimates, with uncertainties of order 0.001. The following columns list the main properties of the BLR - the BH mass in Solar masses (with the uncertainties including the 0.3~dex scatter on \autoref{eq:virial_mass}), bolometric luminosity in erg~s$^{-1}$ and the Eddington ratio. Columns eight and nine give the observed ratio of narrow \Has and \Hbs lines and the derived A$_V$ respectively. An A$_V$ of 0 is listed for sources with negative A$_V$ values that were consistent with 0. A dash in a column indicates that the value could not be measured due to the relevant lines falling in a detector gap or outside the NIRSpec coverage.  The final column contains the literature reference for previously published sources (1. \protect\cite{Maiolino_AGN}, 2. \protect\cite{Matthee2024}, 3. \protect\cite{Juodzbalis2024}, 4. \protect\cite{Juodzbalis2024b}). The observed Balmer decrement for the narrow lines of GS-209777 (marked with '*') was steeper than that of the broad lines, we thus use the uncorrected estimates for this source.}
    \label{tab:all_objecs}
\end{table*}

\begin{table*}
    \centering
    \begin{tabular}{c c c c}
    \hline
       Component & Parameter & Description & Prior\\
       \hline
        
        SFH & $\mathrm{log}(M_{tot}/\mathrm{M_{\odot}}$) & Integrated SFH mass & Uniform $\in [5,12]$\\
            & $\mathrm{log}(\tau_{SFR}/\mathrm{yr})$ & Exponential decline in delayed SFH & Uniform $\in [6,12]$\\
            &                                        & ($SFR\propto t\,\mathrm{exp}[-t/\tau_{SFR}]$ for $t>10$Myr) \\
            & $\mathrm{log}(t/\mathrm{yr})$ & Maximum age of stars & Uniform $\in [7,10.8]$ \\
            & $\mathrm{log}(SFR/\mathrm{M}_{\odot}\,\mathrm{yr}^{-1})$ & Constant SFR for $t<10$Myr & Uniform $\in [-4,4]$ \\
            & $\mathrm{log}(Z_*/\mathrm{Z}_{\odot})$ & Metallicity of the stars & Uniform $\in [-2.2, 0.24]$\\
            & $m_{\textrm{up}}$/$\mathrm{M_{\odot}}$ & IMF upper mass cutoff  & Fixed to 300\\
       \hline
        Nebular & $\mathrm{log}(Z_{gas}/\mathrm{Z}_{\odot})$ & Nebular gas metallicity & Fixed to $Z_*$\\
                & log$U$ & Ionisation parameter & Uniform  $\in [-4,-1]$\\
                & $\xi_d$ & Dust-to-metal mass ratio & Fixed to 0.3\\
       \hline
        Attenuation & $\hat{\tau}_V$ & Effective V-band optical depth to stars & Exponential $\in [0,6]$ \\
                    & $\mu$ & Fraction of $\hat{\tau}_V$ from dust in the diffuse inter-stellar medium & Fixed to 0.4\\
                    & $\hat{\tau}_{V,AGN}$ & Effective V-band optical depth of the AGN power-law component &  Exponential $\in [0,6]$ \\
       \hline
        AGN emission &  $\mathrm{log}(f_{AGN})$ & fractional contribution of AGN to 1500\AA & Uniform $\in [-3,3]$\\
                     &  ${\rm slope}_{\rm PL}$ & AGN continuum power law slope & Fixed -1.54 or -2.33 \\
    \hline
    \end{tabular}
    \caption{Parameters and  priors used in \beagle\ fitting}
    \label{tab:beagleparams}
\end{table*}

\begin{table}
    \centering
    \begin{tabular}{ccc}
    \hline
        Component & Model & Parameters \\
    \hline
        SFH & Delayed SFH & $T_{main} = 100, 1000, 5000, 10000$~Myr \\
        & & $T_{burst} = 5, 20$~Myr\\
        & &    $f_{burst} = 0, 0.2$\\
    \hline
        SSP & BC03 & $Z_* = 0.02, 0.0001$\\
    \hline
        Nebular &  & $\log{U} = -1.0, -2.0, -3.8$\\
        & & $Z_{gas} = 0.02, 0.0004$ \\
        & & $n_e = 100, 1000$~cm$^{-3}$ \\
        & & $FWHM = 300, 600, 2000$~km~s$^{-1}$\\
    \hline
        Attenuation & Calzetti00 & $E(B-V) = 0, 0.3, 1.0, 1.5, 3.0$ \\
    \hline
        Dust emission & Dale2014 & $\alpha = 1.3125, 2.0, 4.0$\\
    \hline
        AGN emission &  Skirtor & $i = 30, 70$ \\
        & & $f_{AGN} = 0, 0.1, 0.5, 0.6, 0.8, 0.9$ \\
        & & $E(B-V) = 0.03, 0.5, 1.5, 3.0$ \\
    \hline
    \end{tabular}
    \caption{A summary of \cigale fitting parameters used. First column indicates the emission component, second column lists the exact model while the third column lists parameters used to compute model grid. Parameters not listed were kept to \cigale defaults.}
    \label{tab:cigale_params}
\end{table}

\begin{table*}
    \centering
    \setlength{\tabcolsep}{4pt}
    \begin{tabular}{cccccccccc}
    \hline
        Object ID &  $\log{M_{*}/M_{\odot}}$ \beagle & $\chi^2_R$ \beagle & $\log{M_{*}/M_{\odot}}$ \cigale & $\chi^2_R$ \cigale &   $\log{\sigma_*}$ &$M_{UV}$& $R_e$ [pc] & $\log{M_{*}/M_{\odot}}$ FPHO & $n_s$\\
    \hline
    \multicolumn{10}{c}{Robust} \\
    \hline
    GS-30148179 & $8.95_{-0.58}^{+0.58}$ & $1.46$ & $8.78_{-0.04}^{+0.04}$ & $0.36$ & $2.10_{-0.13}^{+0.13}$ & -20.36 & -- & -- & --\\
    GS-10013704 & $8.90_{-0.90}^{+0.90}$ & $0.85$ & $8.28_{-0.08}^{+0.08}$ & $0.62$ & $1.86_{-0.14}^{+0.14}$ & -18.95 & $137_{-23}^{+23}$ & $9.04_{-0.02}^{+0.02}$ & $0.89_{-0.14}^{+0.14}$\\
    GS-210600 & $8.40_{-0.61}^{+0.61}$ & $1.49$ & $8.26_{-0.02}^{+0.02}$ & $1.59$ & $1.76_{-0.14}^{+0.14}$ & -18.61 & $283_{-45}^{+45}$ & $7.51_{-0.08}^{+0.08}$& $0.81_{-0.13}^{+0.13}$\\
    GS-209777 & -- & -- & -- & -- & $2.14_{-0.21}^{+0.21}$* & -18.17 & -- & -- & --\\
    GS-204851 & $10.74_{-0.09}^{+0.09}$ & $0.93$ & $8.94_{-0.04}^{+0.04}$ & $1.75$ & $1.91_{-0.13}^{+0.13}$  & -18.85 & -- & --& --\\
    GS-179198\textsuperscript{\textdagger} & $8.44_{-0.05}^{+0.05}$ & $0.79$ & $9.43_{-0.06}^{+0.06}$ & $0.53$ & $1.92_{-0.13}^{+0.13}$ & -19.10 & $380_{-61}^{+61}$ & $8.41_{-0.17}^{+0.17}$& $5.88_{-0.94}^{+0.94}$\\
    GS-172975 & $8.98_{-0.14}^{+0.14}$ & $0.81$ & $8.30_{-0.70}^{+0.70}$ & $1.21$ & $1.70_{-0.16}^{+0.16}$ & -17.74 & -- & --& --\\
    GS-159717 & -- & -- & -- & -- & $1.62_{-0.14}^{+0.14}$ & -19.09 & -- & --& --\\
    GS-159438\textsuperscript{\textdagger} & $8.35_{-0.13}^{+0.13}$ & $2.64$ & $9.14_{-0.18}^{+0.18}$ & $1.97$ & $2.15_{-0.13}^{+0.13}$ & -20.98 & -- & --& --\\
    GN-77652 & $8.36_{-1.64}^{+1.64}$ & $1.82$ & $8.20_{-0.04}^{+0.04}$ & $1.42$ & $1.79_{-0.31}^{+0.31}$ & -18.27 & $269_{-44}^{+44}$ & $7.79_{-0.13}^{+0.13}$& $0.85_{-0.14}^{+0.14}$\\
    GN-73488 & $9.71_{-0.33}^{+0.33}$ & $1.83$ & $8.50_{-0.53}^{+0.53}$ & $2.26$ & $2.06_{-0.15}^{+0.15}$ & -18.87 & -- & --& --\\
    GN-62309 & $7.78_{-0.34}^{+0.34}$ & $0.94$ & $8.52_{-0.18}^{+0.18}$ & $0.94$ & $1.80_{-0.13}^{+0.13}$ & -18.58 & $293_{-48}^{+48}$ & $8.44_{-0.26}^{+0.26}$& $3.11_{-0.50}^{+0.50}$\\
    GN-61888 & $8.53_{-1.73}^{+1.73}$ & $0.84$ & $8.25_{-0.10}^{+0.10}$ & $0.56$ & $1.78_{-0.14}^{+0.14}$ & -18.78 & -- & --& --\\
    GN-53757 & $10.38_{-0.19}^{+0.19}$ & $0.63$ & $9.44_{-0.31}^{+0.31}$ & $1.19$ & $1.79_{-0.15}^{+0.15}$ & -18.86 & $618_{-99}^{+99}$ & $8.81_{-0.26}^{+0.26}$& $0.80_{-0.13}^{+0.13}$\\
    GS-49729 & -- & -- & $10.14_{-0.25}^{+0.25}$ & $3.14$ & $2.08_{-0.21}^{+0.21}$* & -21.38& -- & --& --\\
    GS-38562 & $9.76_{-0.09}^{+0.09}$ & $1.09$ & $9.57_{-0.08}^{+0.08}$ & $1.19$ & $1.83_{-0.14}^{+0.14}$ & -18.98 & $174_{-30}^{+30}$ & $9.31_{-0.17}^{+0.17}$& $1.02_{-0.16}^{+0.16}$\\
    GN-38509 & $9.19_{-0.40}^{+0.40}$ & $1.05$ & $9.43_{-0.31}^{+0.31}$ & $0.87$ & $2.08_{-0.23}^{+0.23}$* & -19.15 & $137_{-23}^{+23}$ & $8.92_{-0.30}^{+0.30}$& $0.94_{-0.15}^{+0.15}$\\
    GN-29648\textsuperscript{\textdagger} & $9.71_{-0.01}^{+0.01}$ & $1.38$ & $10.31_{-0.06}^{+0.06}$ & $1.40$ & $1.93_{-0.21}^{+0.21}$* & -19.00 & -- & --& --\\
    GN-28074 & -- & -- & -- & --  & $1.65_{-0.24}^{+0.24}$* & -20.15 & -- & --& --\\
    GN-20621 & $8.41_{-1.58}^{+1.58}$ & $1.38$ & $8.08_{-0.10}^{+0.10}$ & $1.11$ & $1.86_{-0.13}^{+0.13}$ & -18.61 & -- & --& --\\
    GS-17341\textsuperscript{\textdagger} & $8.54_{-0.32}^{+0.32}$ & $1.09$ & $8.84_{-0.03}^{+0.032}$ & $1.14$ & $1.95_{-0.21}^{+0.21}$* & -19.54 & -- & --& --\\
    GS-13329 & $9.52_{-0.14}^{+0.14}$ & $0.75$ & $8.26_{-0.41}^{+0.41}$ & $1.16$ & $1.78_{-0.14}^{+0.14}$ & -18.59 & $409_{-66}^{+66}$ & $8.60_{-0.15}^{+0.15}$& $5.86_{-0.94}^{+0.94}$\\
    GN-11836 & $8.17_{-0.15}^{+0.15}$ & $1.94$ & $8.52_{-0.50}^{+0.50}$ & $2.98$ & $1.96_{-0.13}^{+0.13}$ & -18.89 & $451_{-74}^{+74}$ & $8.35_{-0.19}^{+0.19}$& $0.86_{-0.14}^{+0.14}$\\
    GS-9598\textsuperscript{\textdagger} & $9.08_{-0.38}^{+0.38}$ & $0.96$ & $9.48_{-0.02}^{+0.02}$ & $3.11$ & $1.93_{-0.21}^{+0.21}$* & -18.70 & -- & --& --\\
    GS-8083 & $8.27_{-0.13}^{+0.14}$ & $1.08$ & $8.30_{-0.15}^{+0.15}$ & $2.11$ & $1.89_{-0.13}^{+0.13}$ & -18.67 & $72_{-11}^{+11}$ & $8.73_{-0.38}^{+0.38}$& $5.90_{-0.94}^{+0.94}$\\
    GN-2916 & $8.98_{-0.97}^{+0.97}$ & $0.95$ &  $9.57_{-0.05}^{+0.05}$ & $1.30$ & $1.78_{-0.14}^{+0.14}$ & -19.42 & -- & --& --\\
    GN-1093 &  $8.43_{-0.63}^{+0.63}$ & $1.55$ &  $8.77_{-0.39}^{+0.39}$ & $1.29$ & $1.91_{-0.14}^{+0.14}$ & -18.03 & -- & --& --\\
    GN-954 &  $10.93_{-0.11}^{+0.11}$& $1.60$ &  $9.68_{-0.30}^{+0.30}$ & $0.67$ & $1.79_{-0.21}^{+0.21}$* & -19.94 & $331_{-53}^{+53}$ & $8.46_{-0.11}^{+0.11}$& $0.80_{-0.13}^{+0.13}$\\
    \hline
    \multicolumn{10}{c}{Tentative} \\
    \hline
    GS-200679\textsuperscript{\textdagger} & $8.16_{-0.20}^{+0.20}$ & $1.67$ & $8.53_{-0.13}^{+0.13}$ & $0.98$ & $1.82_{-0.22}^{+0.22}$* & -20.26 & $408_{-65}^{+65}$ &$8.07_{-0.21}^{+0.21}$& $1.91_{-0.31}^{+0.31}$\\
    GN-23924\textsuperscript{\textdagger} & -- & -- & $11.01_{-0.03}^{+0.03}$ & $1.81$ & $2.07_{-0.13}^{+0.13}$* & -20.05 & -- & --&\\
    \hline
    \multicolumn{10}{c}{Tentative \Hbs} \\
    \hline
    GS-20057765 & $7.40_{-0.97}^{+0.78}$ & $0.53$ & $8.08_{-0.30}^{+0.30}$ & $1.45$ & -- & -19.21 & -- & --&--\\
    GS-20030333 & -- & -- & $8.61_{-0.20}^{+0.20}$ & $1.76$ & -- & -19.29 & $94_{-15}^{+15}$ & $8.24_{-0.08}^{+0.08}$&$5.49_{-0.88}^{+0.88}$\\
    GS-164055\textsuperscript{\textdagger} & $7.27_{-0.87}^{+0.68}$ &  $0.55$ & $7.99_{-0.23}^{+0.23}$ & $1.60$ & -- & -18.75 & -- & --&--\\
    GN-4685 & $9.91_{-0.21}^{+0.22}$ &  $0.80$ & $9.92_{-1.24}^{+1.24}$ & $0.55$ & -- & -18.15 & -- & --&--\\
    \hline

    \end{tabular}
    \caption{Summary of host properties derived for our sample AGN. The first column lists the IDs of each object in the same order as \autoref{tab:all_objecs}. Second and third columns contain $M_*$ estimates and $\chi^2_R$ values given by \beagle fitting. Columns 4 and 5 contain the same values, but obtained from \cigale fits. Column 6 gives the estimates of stellar velocity dispersions for each host. The remaining three columns show the half-light radii obtained from \texttt{ForcePho} decompositions, the $M_*$ values obtained from decomposed photometry and S\'ersic indices. Sources marked with `\textdagger' are significantly extended, thus \cigale fitting results were used as best estimates for their $M_*$. Values of $\log{\sigma_*}$ marked with `*' were inferred from fitting R1000 rather than R2700 data. The uncertainties on them also reflect the $\sim$0.15~dex intrinsic scatter on the calibrations, which ends up dominating the combined errors in most cases.}
    \label{tab:host_prop}
\end{table*}

\begin{table}
    \centering
    \begin{tabular}{ccccc}
\hline
Value                 & All redshifts      & $z < 3.5$          & $3.5 < z < 5$      & $z > 5$          \\
\hline
$F_{H\alpha, nr}$       & $676_{-135}^{+135}$    & $3130_{-150}^{+19}$    & $832_{-3}^{+3}$        & $295_{-3}^{+3}$      \\
$F_{H\alpha, br}$       & $413_{-83}^{+83}$      & $2236_{-439}^{+46}$    & $307_{-4}^{+4}$        & $518_{-5}^{+5}$      \\
$F_{H\beta, nr}$        & $160_{-32}^{+32}$      & $815_{-6}^{+7}$        & $253_{-2}^{+2}$        & $95.9_{-1.3}^{+1.2}$ \\
$F_{H\gamma, nr}$       & $85.1_{-17.1}^{+17.1}$ & $288_{-8}^{+6}$        & $99.3_{-1.8}^{+1.9}$   & $45.5_{-4.6}^{+2.2}$ \\
$F_{[OII]\lambda3727}$  & $97.8_{-19.6}^{+19.6}$ & $1172_{-8}^{+8}$       & $93.2_{-2.6}^{+2.8}$   & $40.1_{-2.0}^{+2.0}$ \\
$F_{[OIII]\lambda4363}$ & $30.9_{-6.3}^{+6.3}$   & $116_{-7}^{+5}$        & $36.7_{-1.6}^{+1.7}$   & $16.7_{-1.8}^{+3.7}$ \\
$F_{[OIII]\lambda5007}$ & $695_{-139}^{+139}$    & $3333_{-10}^{+8}$      & $1022^{+4}_{-4}$       & $485^{+2}_{-2}$      \\
$F_{HeII\lambda4686}$   & $<3.2$             & $<10.8$            & $<4.2$             & $<2.4$           \\
$F_{[SII]\lambda6716}$  & $14.3_{-2.9}^{+2.9}$   & $168_{-15}^{+8}$       & $11.1_{-1.0}^{+0.9}$   & $< 3.0$          \\
$R_{[SII]}$             & $0.90_{-0.07}^{+0.07}$ & $0.96_{-0.58}^{+0.06}$ & $0.88_{-0.11}^{+0.13}$ & –                \\
$F_{[NII]\lambda6583}$  & $5.70_{-1.46}^{+1.46}$ & $492_{-147}^{+10}$     & $5.1_{-1.3}^{+1.3}$    & $<1.5$           \\
$F_{[OI]6300}$          & $60.1_{-12.5}^{+12.5}$ & $322_{-13}^{+13}$& $28.1_{-4.6}^{+5.3}$   & $32.3_{-5.1}^{+5.9}$ \\
\hline
\end{tabular}
    \caption{Table summarizing the measurements obtained from the stacked spectra and their variation across the redshift bins. The first column lists the values being measured, second column - measurements for the stack of all sample sources. The subsequent columns give the measurements for $z > 5$, $3.5 < z < 5$ and $z < 3.5$ bins respectively. The line fluxes are all given in $10^{-20}$~erg~s$^{-1}$~cm$^{-2}$. In case of a non-detection, a 3$\sigma$ upper limit is given. The line flux uncertainties on the values in column two reflect the 20\% systematic error arising from flux calibration differences between different R1000 grating/filter combinations.}
    \label{tab:stack_meas}
\end{table}

\begin{table*}
    \centering
    \begin{tabular}{cccccc}
    \hline
        Object ID & BIC$_{\rm Gauss}$ & BIC$_{\rm 2Gauss}$ & BIC$_{\rm Exp}$& $\log{M_{BH({\rm Gauss})}}$ & $\log{M_{BH({\rm 2Gauss})}}$\\
    \hline
         GS-159717 & 3766 & 3559 & 3857& $7.85_{-0.30}^{+0.30}$ & $7.44_{-0.30}^{+0.30}$ \\
         GN-73488& 901 & 744 & 882 & $7.95_{-0.30}^{+0.30}$& $7.55_{-0.30}^{+0.30}$\\
         GN-28074& 6819 & 4321 & 6868 &$8.55_{-0.30}^{+0.30}$&$8.09_{-0.30}^{+0.30}$\\
         GS-204851& 932 & 872& 880 &$7.68_{-0.32}^{+0.31}$& $7.23_{-0.30}^{+0.31}$\\
         GS-38562& 839 & 804 & 799 & $7.53_{-0.30}^{+0.31}$& $7.23_{-0.30}^{+0.31}$\\
         GS-49729& 2682 & 1305 & 1659 &$8.07_{-0.30}^{+0.30}$& $7.67_{-0.30}^{+0.30}$ \\
        \hline
    \end{tabular}
    \caption{Comparison of the BIC values between the three tested models for the sources exhibiting a wingy BLR as well as the BH masses obtained from single and double Gaussian fits. The first column gives the source ID, columns one to three - the BIC values for the Gaussian, Double-Gaussian and Exponential models respectively. The final two columns list the BH masses obtained from single Gaussian and double Gaussian profiles respectively.}
    \label{tab:gauss_expo}
\end{table*}

\clearpage
\section*{Acknowledgements}

This work is based on observations made with the NASA/ESA/CSA James Webb Space Telescope. The data were obtained from the Mikulski Archive for Space Telescopes at the Space Telescope Science Institute, which is operated by the Association of Universities for Research in Astronomy, Inc., under NASA contract NAS 5-03127 for JWST. These observations are associated with program 1180, 1181, 1210, 1286, 3215.This work was performed using resources provided by the Cambridge Service for Data Driven Discovery (CSD3) operated by the University of Cambridge Research Computing Service (www.csd3.cam.ac.uk), provided by Dell EMC and Intel using Tier-2 funding from the Engineering and Physical Sciences Research Council (capital grant EP/T022159/1), and DiRAC funding from the Science and Technology Facilities Council (www.dirac.ac.uk). IJ acknowledges support by the Huo Family Foundation through a P.C. Ho PhD Studentship. RM, IJ, JS, FDE and G.C.J. acknowledge support by the Science and Technology Facilities Council (STFC), by the ERC through Advanced Grant 695671 “QUENCH”, and by the UKRI Frontier Research grant RISEandFALL. RM also acknowledges funding from a research professorship from the Royal Society. WMB gratefully acknowledges support from DARK via the DARK fellowship. This work was supported by a research grant (VIL54489) from VILLUM FONDEN. ECL acknowledges support of an STFC Webb Fellowship (ST/W001438/1). YI is supported by JSPS KAKENHI Grant No. 24KJ0202. ST acknowledges support by the Royal Society Research Grant G125142. AJB acknowledges funding from the "FirstGalaxies" Advanced Grant from the European Research Council (ERC) under the European Union’s Horizon 2020 research and innovation programme (Grant agreement No. 789056). SC, GV, and BT acknowledge support by European Union’s HE ERC Starting Grant No. 101040227 - WINGS. M.P. acknowledges support from the research project PID2021-127718NB-I00 of the Spanish Ministry of Science and Innovation/State Agency of Research (MICIN/AEI/ 10.13039/501100011033), and the Programa Atracci\'on de Talento de la Comunidad de Madrid via grant 2018-T2/TIC-11715. BER acknowledges support from the NIRCam Science Team contract to the University of Arizona, NAS5-02015, and JWST Program 3215. H\"U acknowledges funding by the European Union (ERC APEX, 101164796). Views and opinions expressed are however those of the authors only and do not necessarily reflect those of the European Union or the European Research Council Executive Agency. Neither the European Union nor the granting authority can be held responsible for them.

\section*{Data Availability}
The Data used in this study has been made public on the JWST DAWN archive as part of JADES DR3. The catalogue table can be provided by the corresponding author upon reasonable request.



\bibliographystyle{mnras}
\bibliography{example} 




\appendix

\section{Fits and spectra of the sample AGN}
\label{sec:all_fits}
Here we present the \Has fits to all sources in our sample not presented in the main text.

\begin{figure*}
    \centering
    \subfloat{\includegraphics[width=0.5\textwidth]{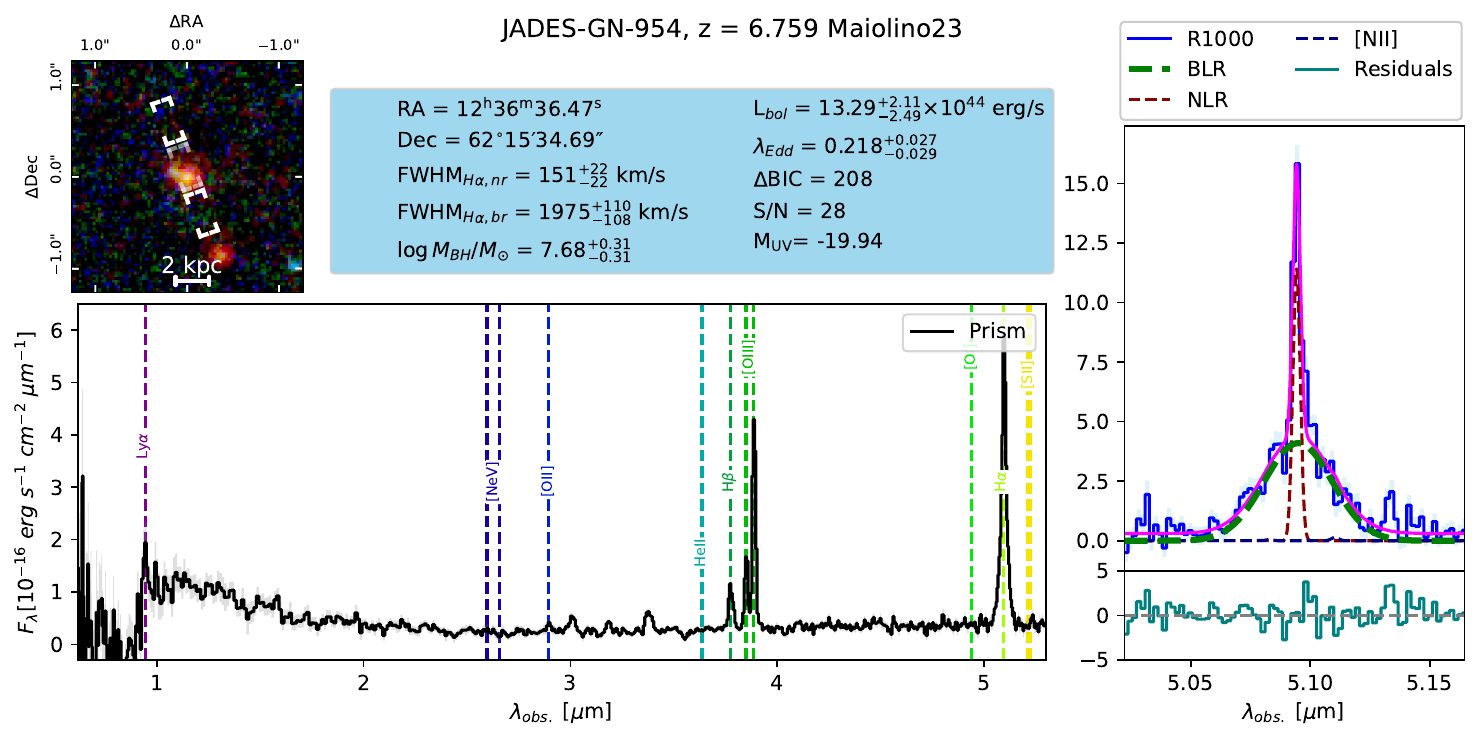}}
    \hfill
    \subfloat{\includegraphics[width=0.5\textwidth]{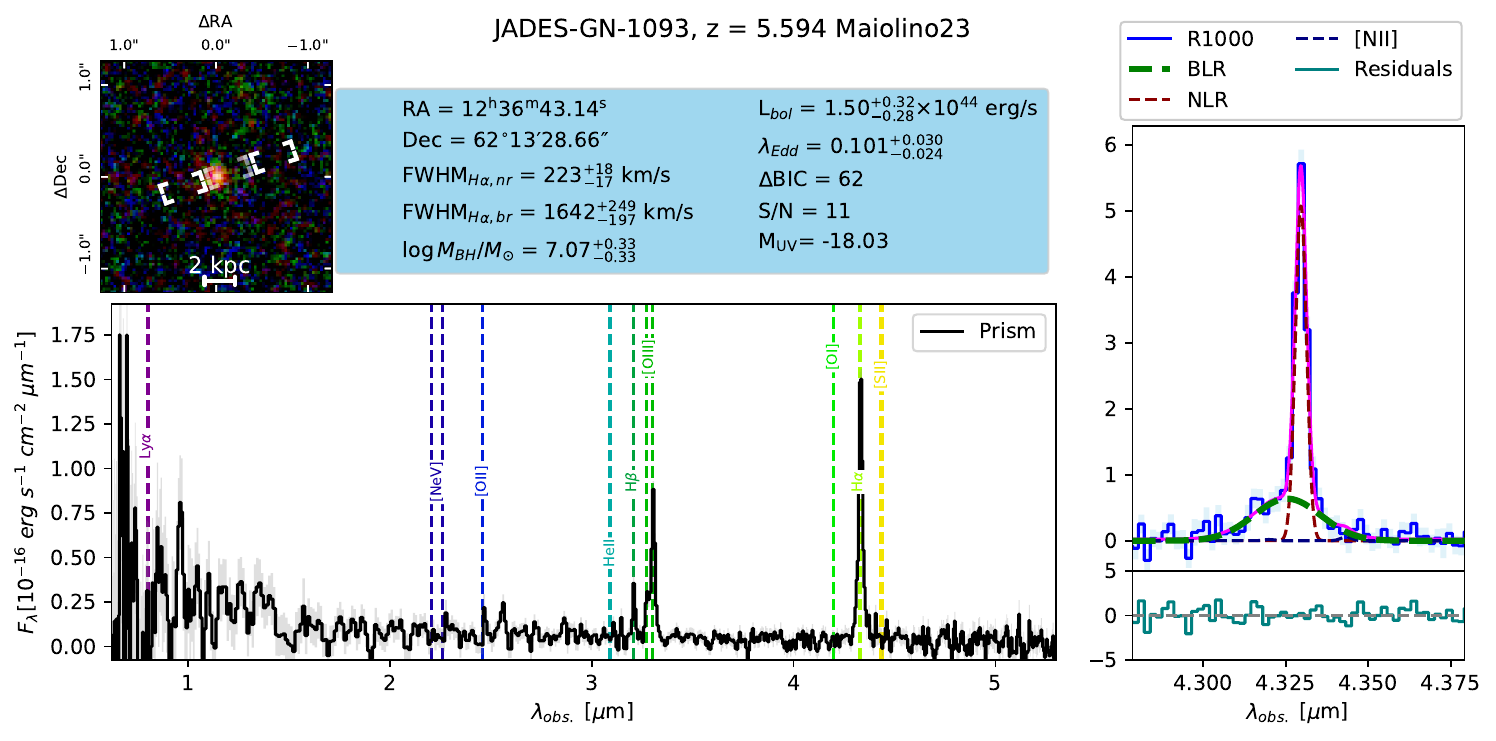}}
    \hfill
    \subfloat{\includegraphics[width=0.5\textwidth]{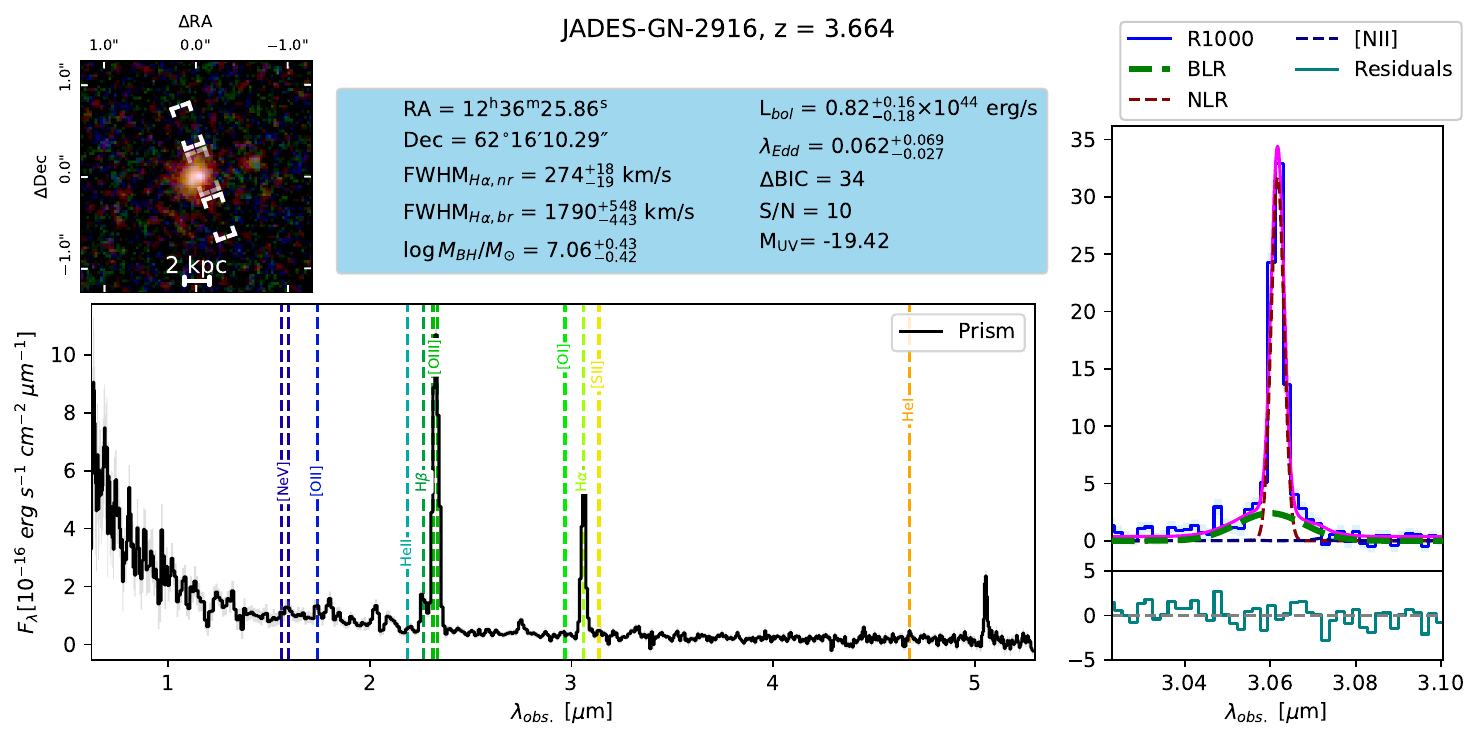}}
    \hfill
    \subfloat{\includegraphics[width=0.5\textwidth]{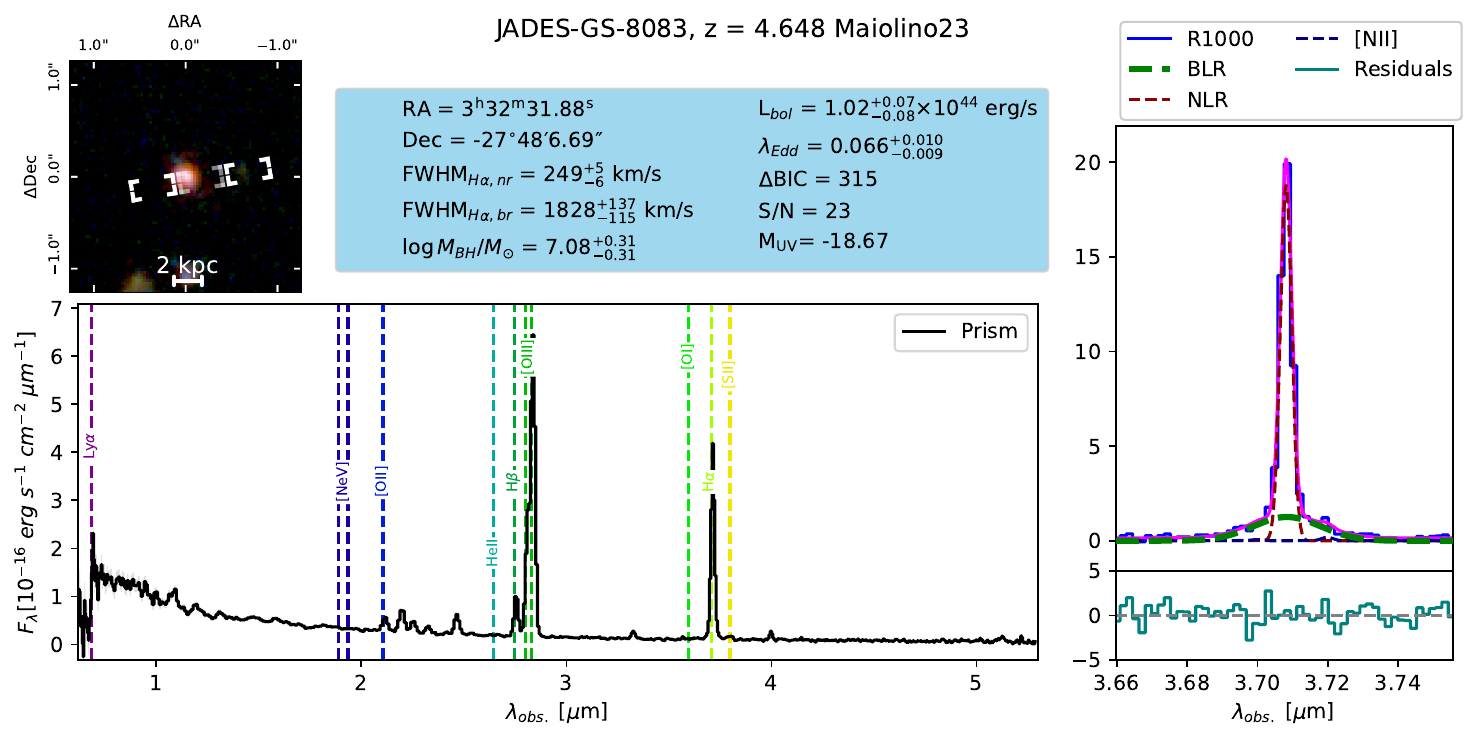}}
    \hfill
    \subfloat{\includegraphics[width=0.5\textwidth]{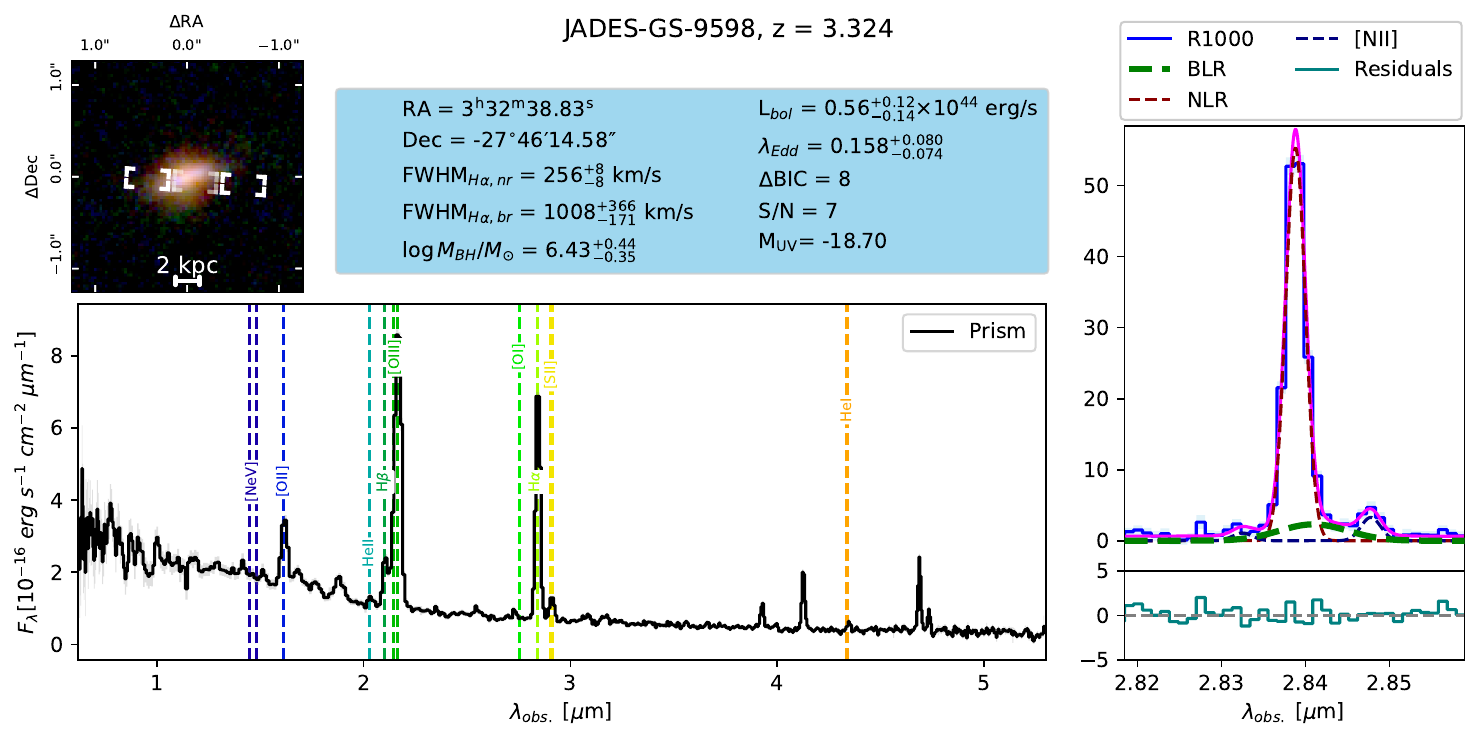}}
    \hfill
    \subfloat{\includegraphics[width=0.5\textwidth]{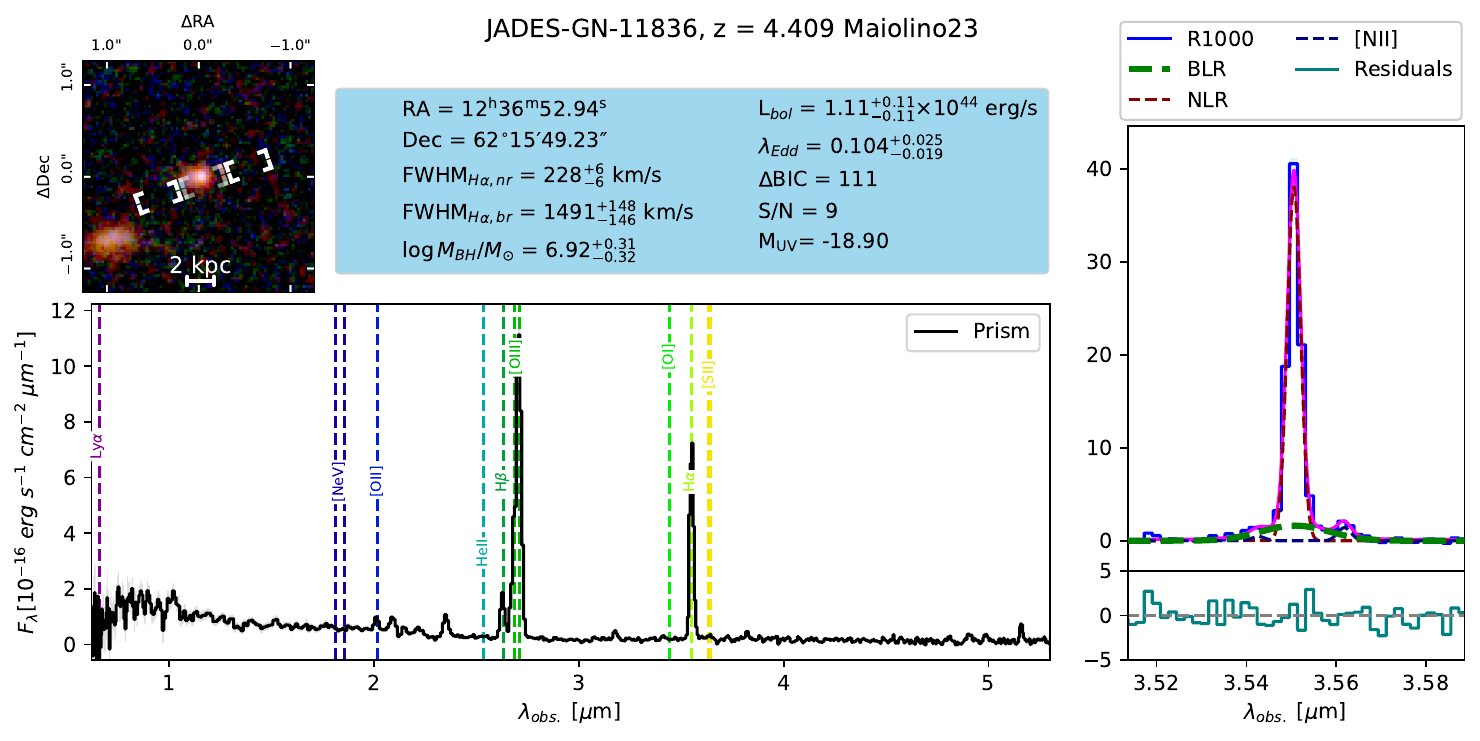}}
    \hfill
    \subfloat{\includegraphics[width=0.5\textwidth]{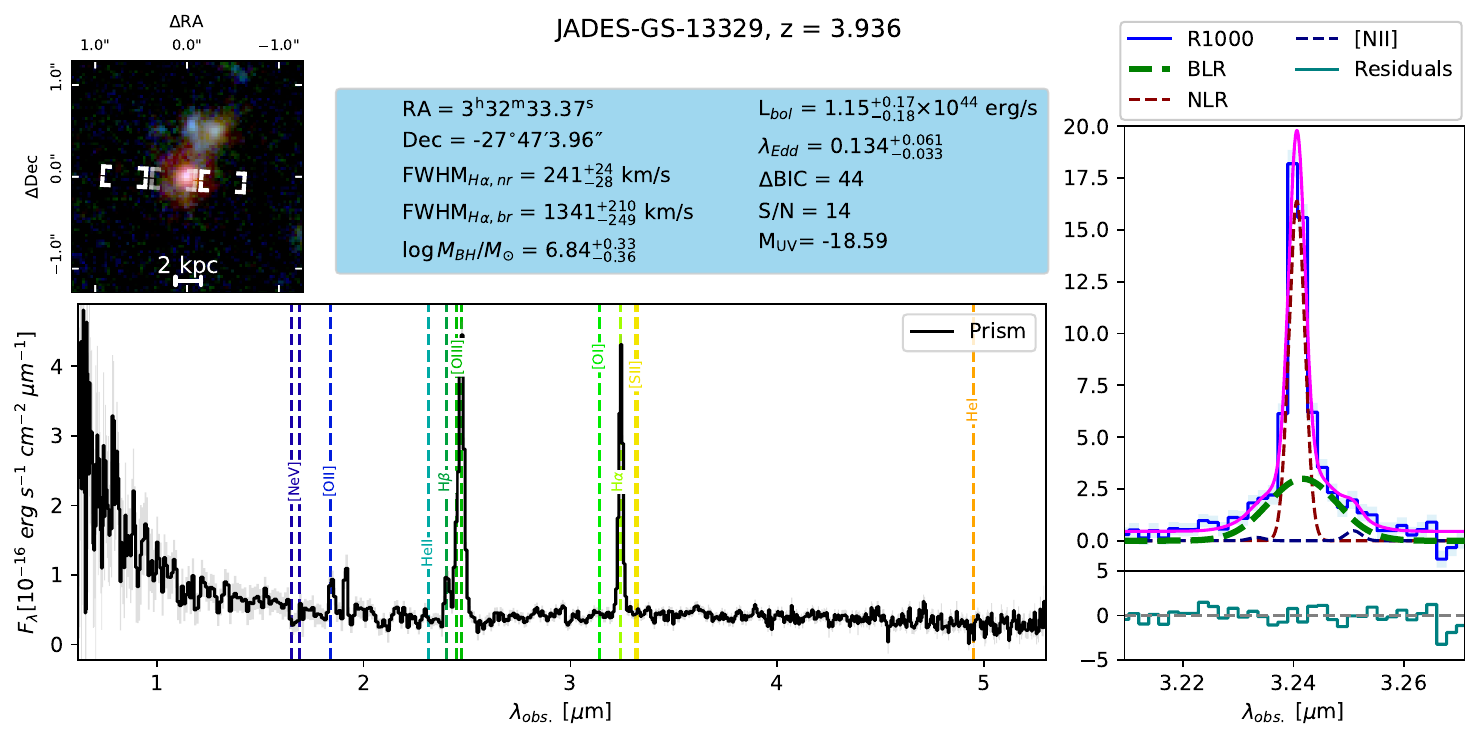}}
    \hfill
    \subfloat{\includegraphics[width=0.5\textwidth]{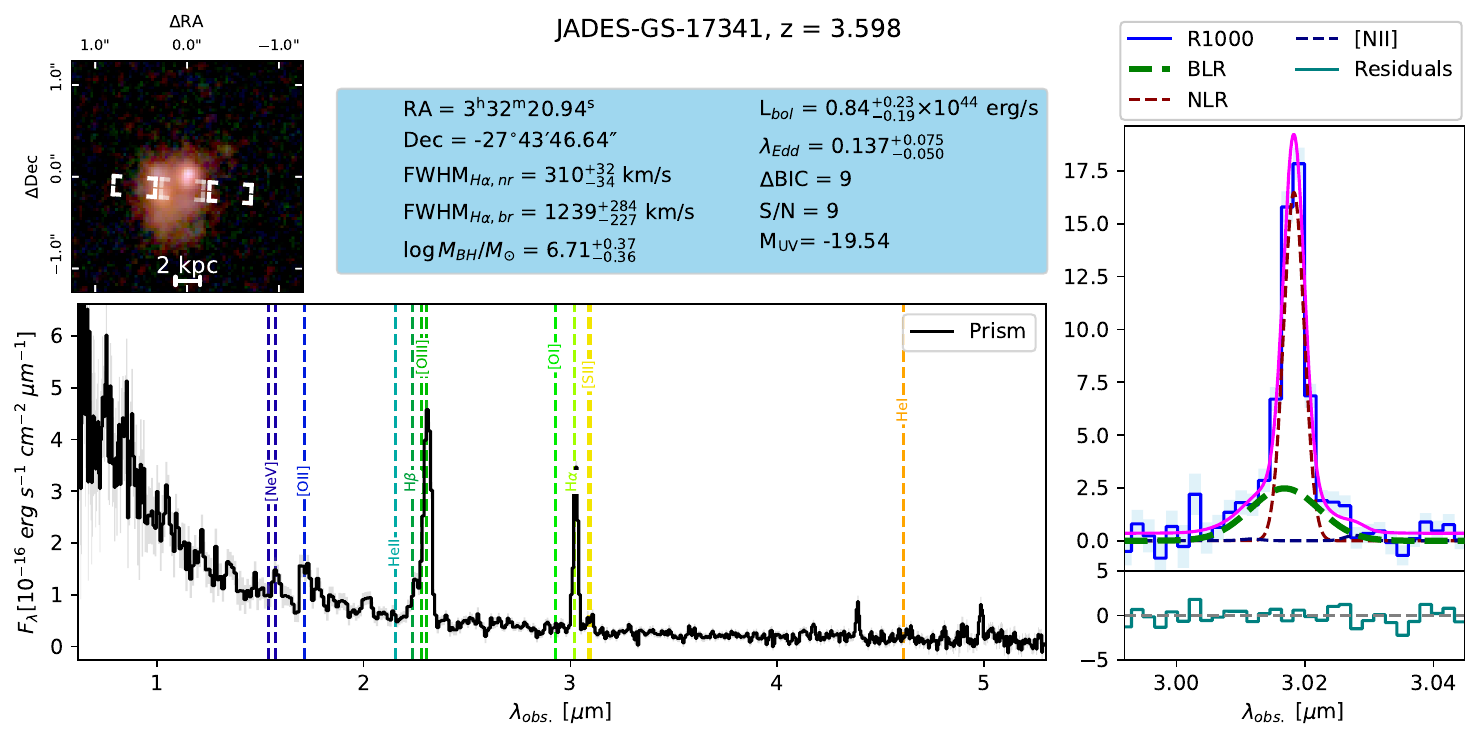}}
    \hfill
    \caption{Same structure as \autoref{fig:representative_fits} in the main text.}
    \label{fig:all_fits1}
\end{figure*}

\begin{figure*}
    \centering
    \subfloat{\includegraphics[width=0.5\textwidth]{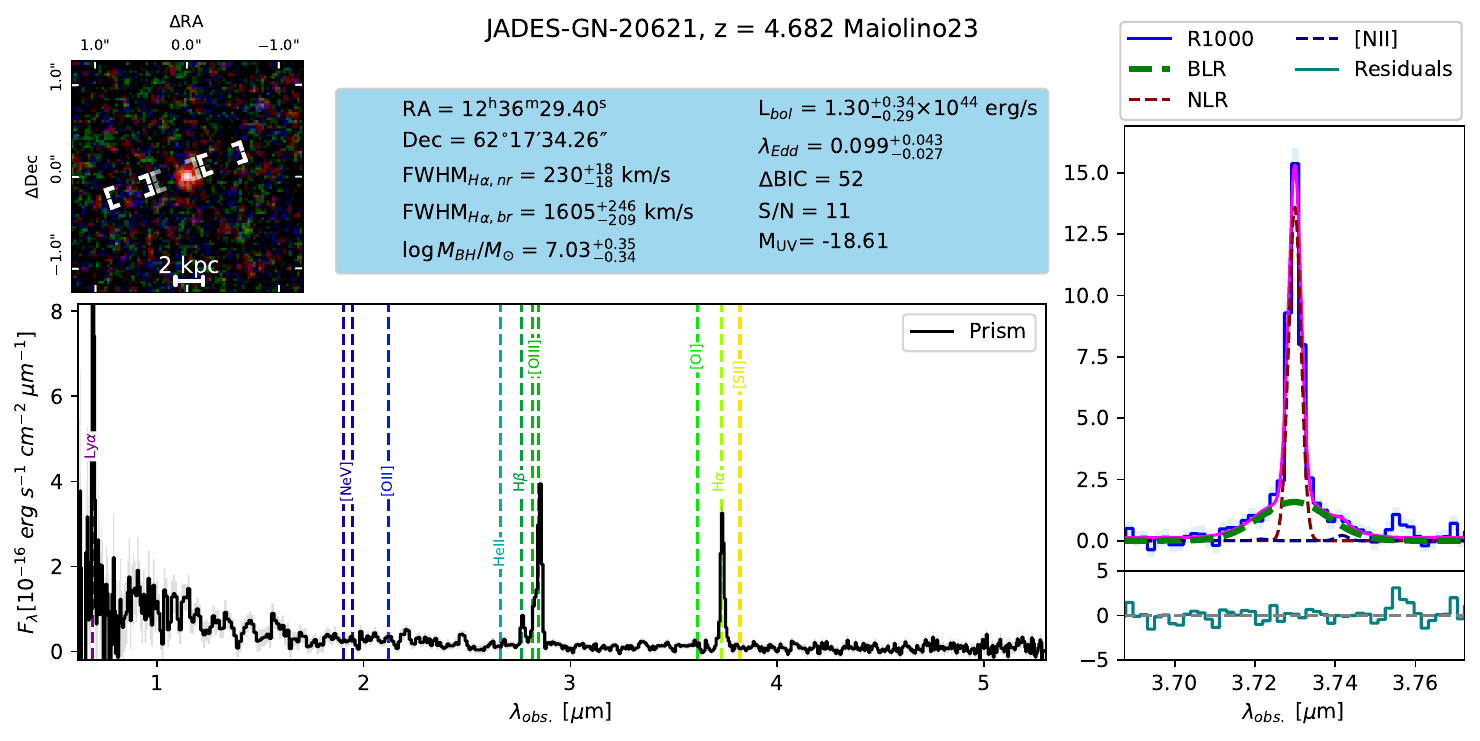}}
    \hfill
    \subfloat{\includegraphics[width=0.5\textwidth]{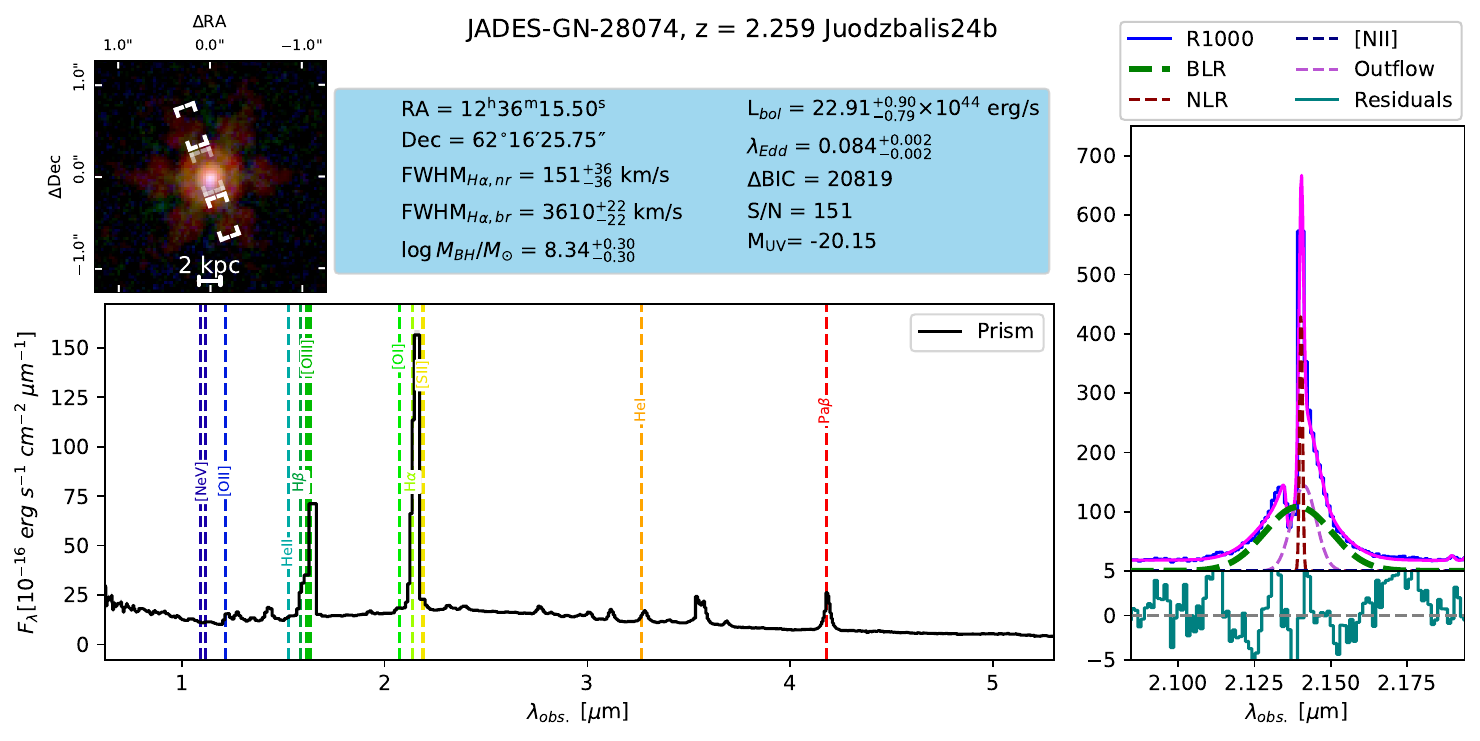}}
    \hfill
    \subfloat{\includegraphics[width=0.5\textwidth]{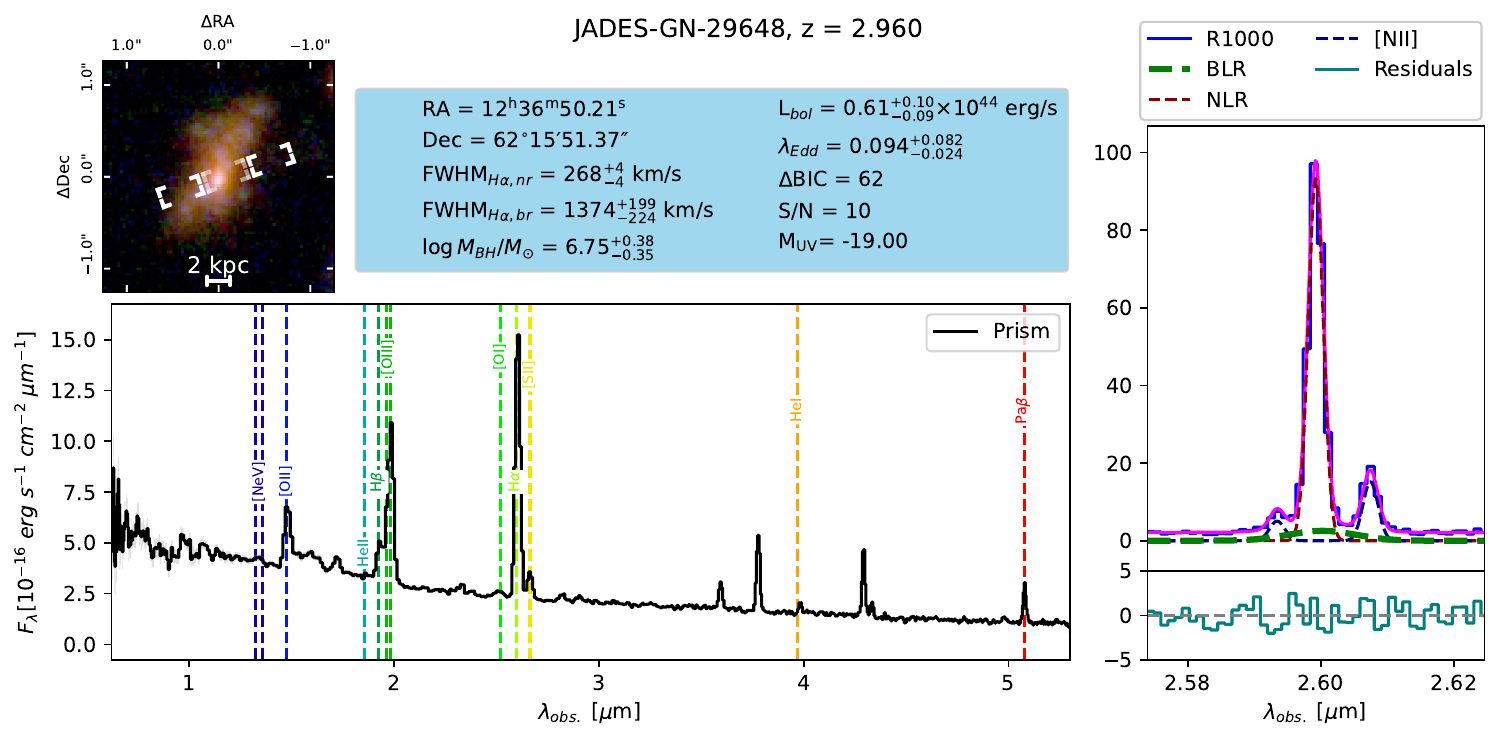}}
    \hfill
    \subfloat{\includegraphics[width=0.5\textwidth]{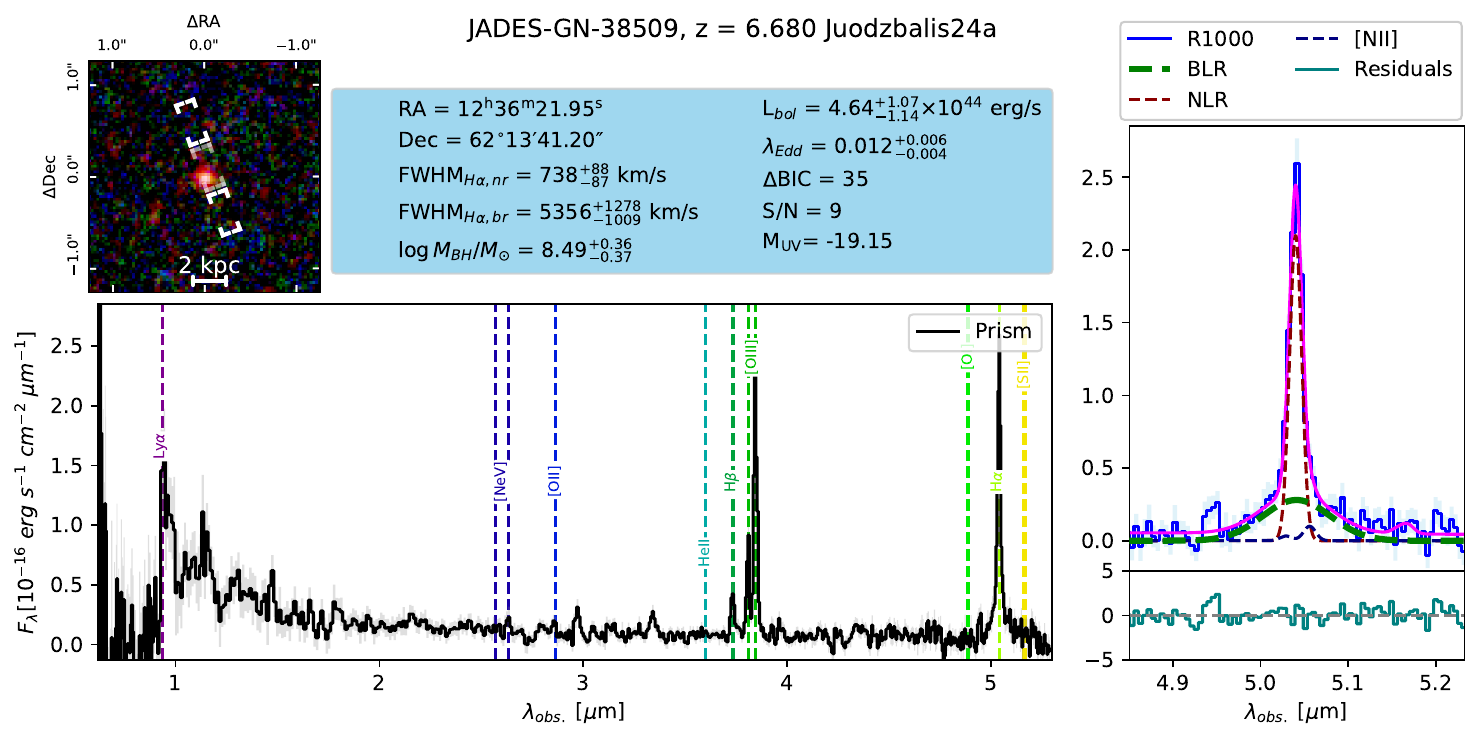}}
    \hfill
    \subfloat{\includegraphics[width=0.5\textwidth]{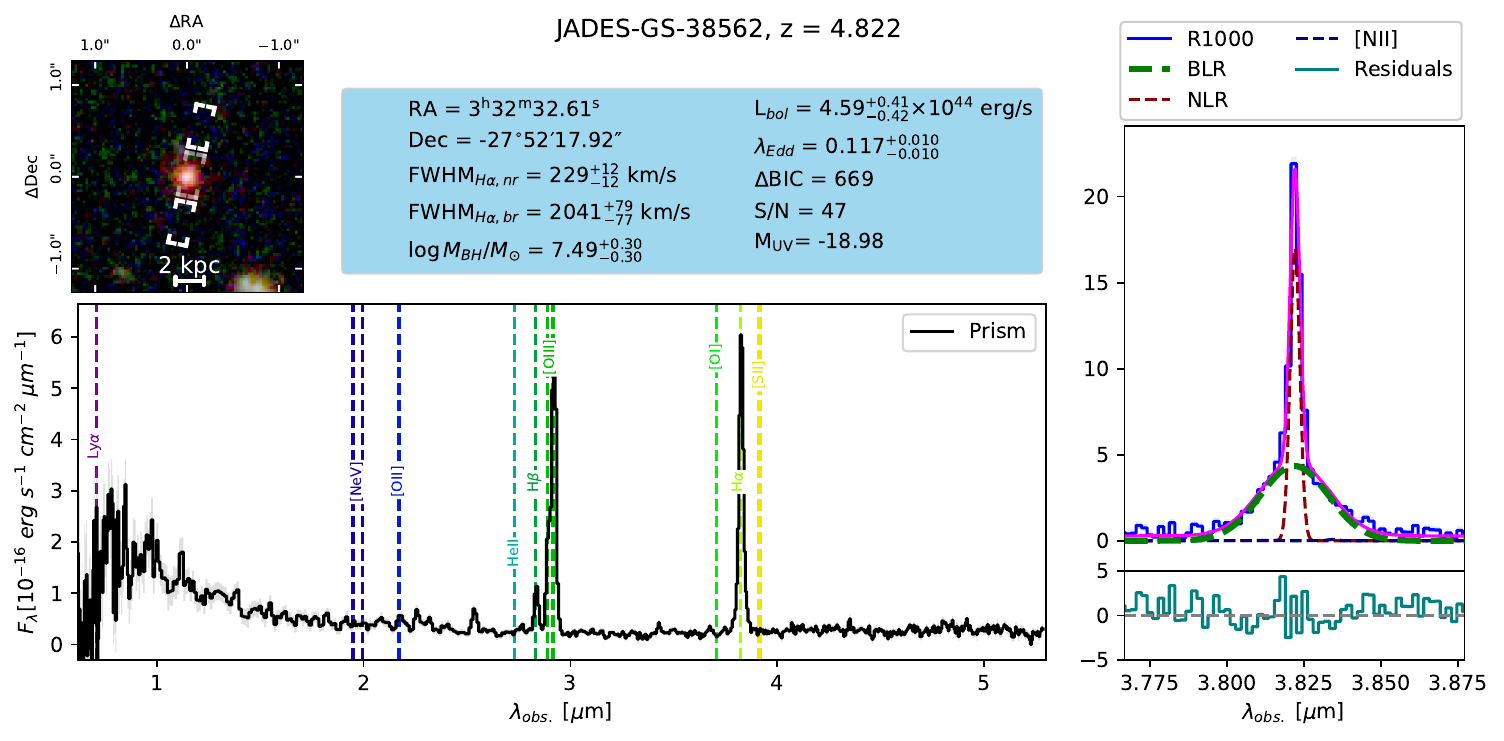}}
    \hfill
    \subfloat{\includegraphics[width=0.5\textwidth]{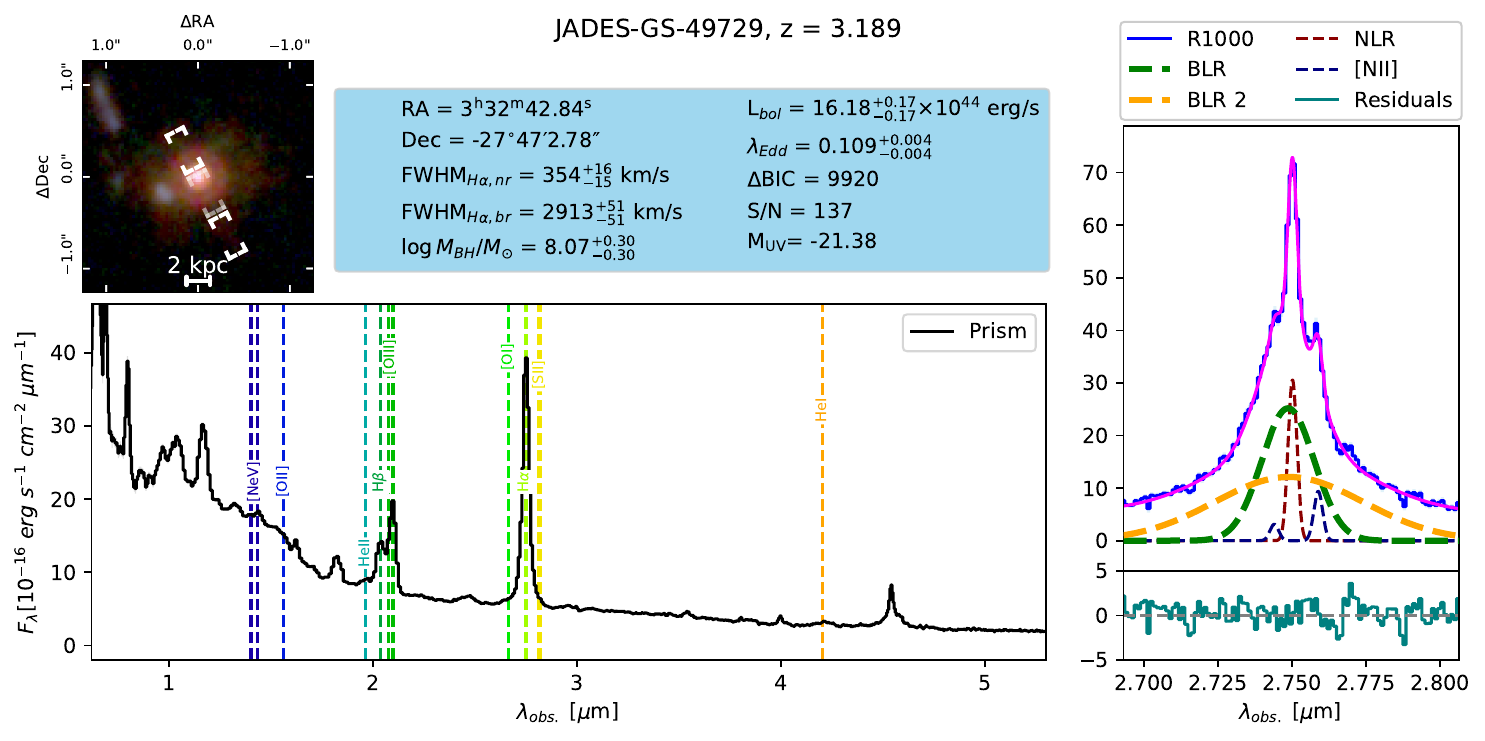}}
    \hfill
    \subfloat{\includegraphics[width=0.5\textwidth]{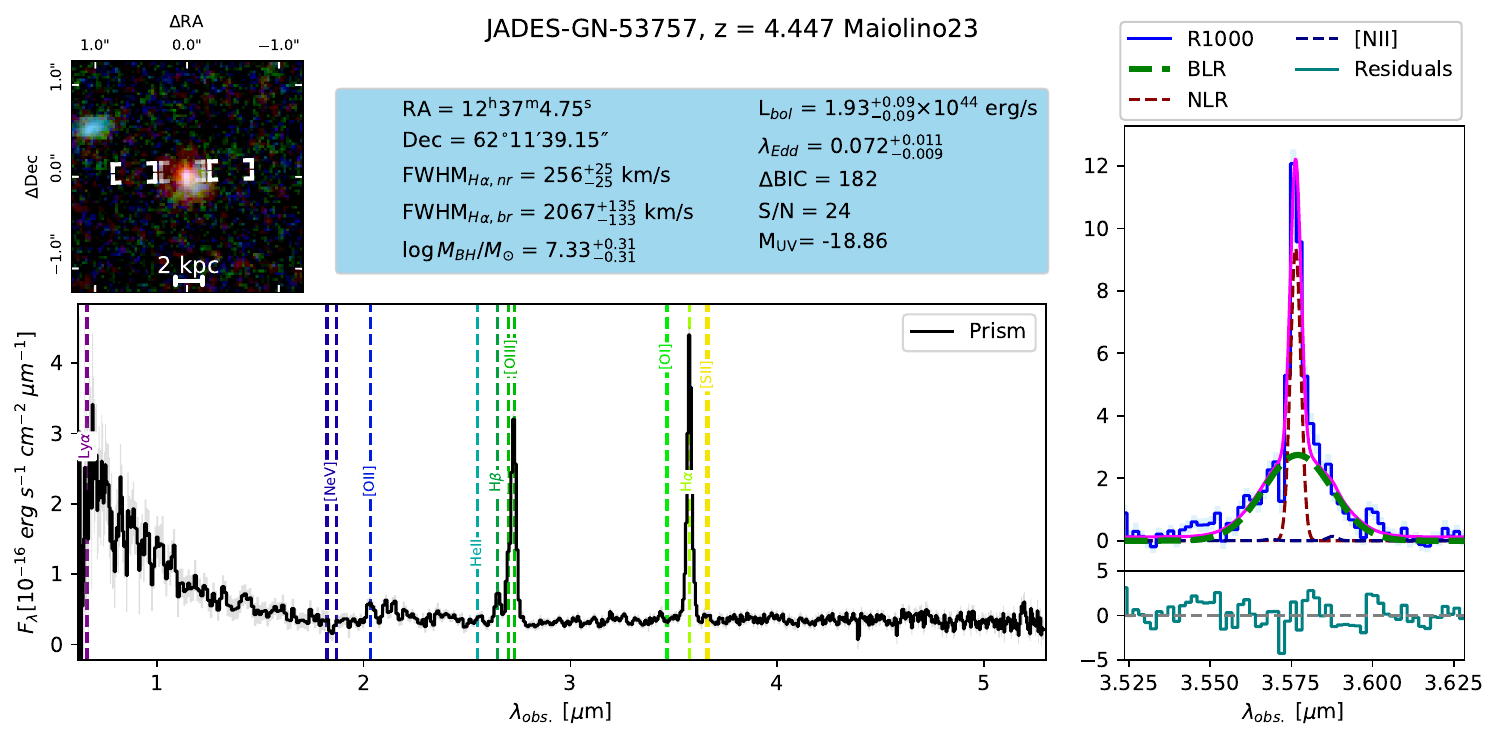}}
    \hfill
    \subfloat{\includegraphics[width=0.5\textwidth]{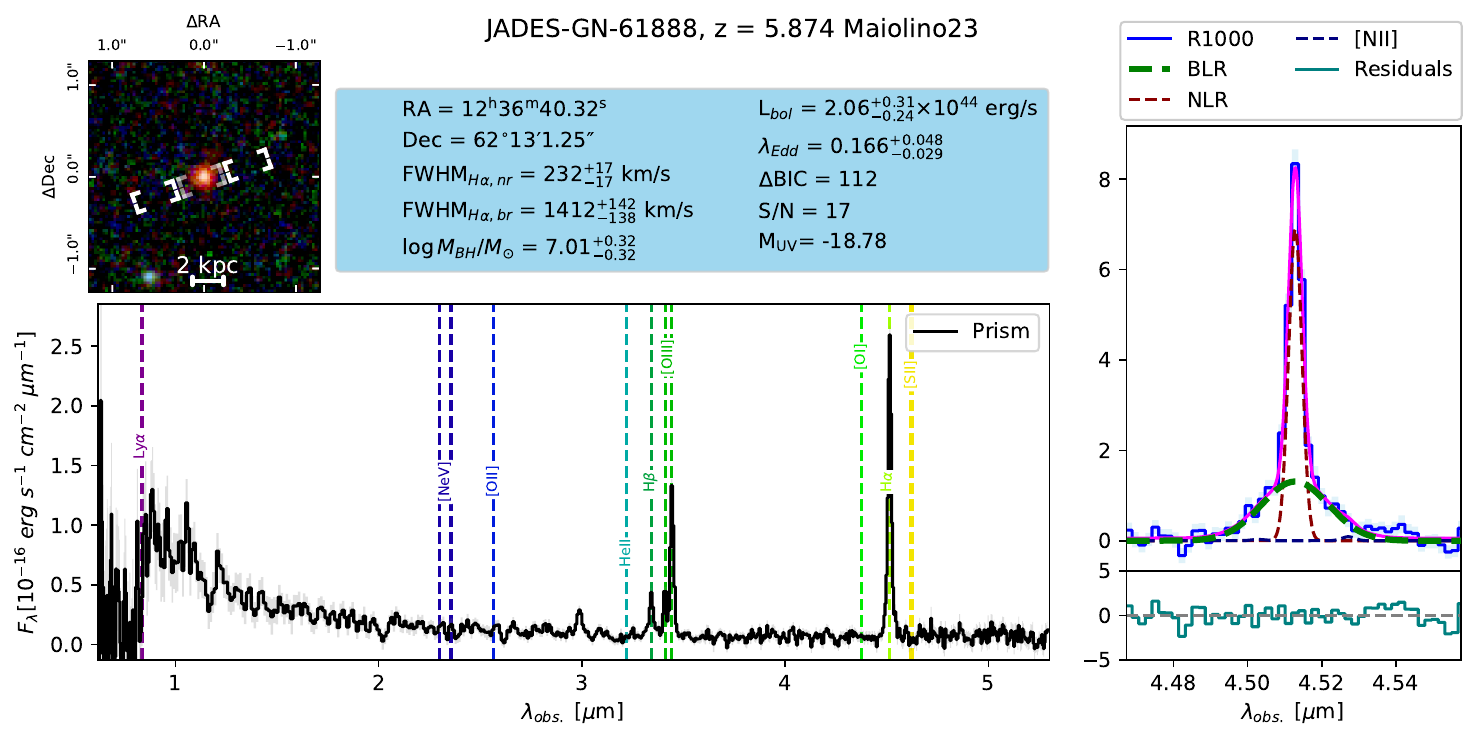}}
    \hfill
    \caption{Same structure as \autoref{fig:representative_fits} in the main text.}
    \label{fig:all_fits2}
\end{figure*}

\begin{figure*}
    \centering
    \subfloat{\includegraphics[width=0.5\textwidth]{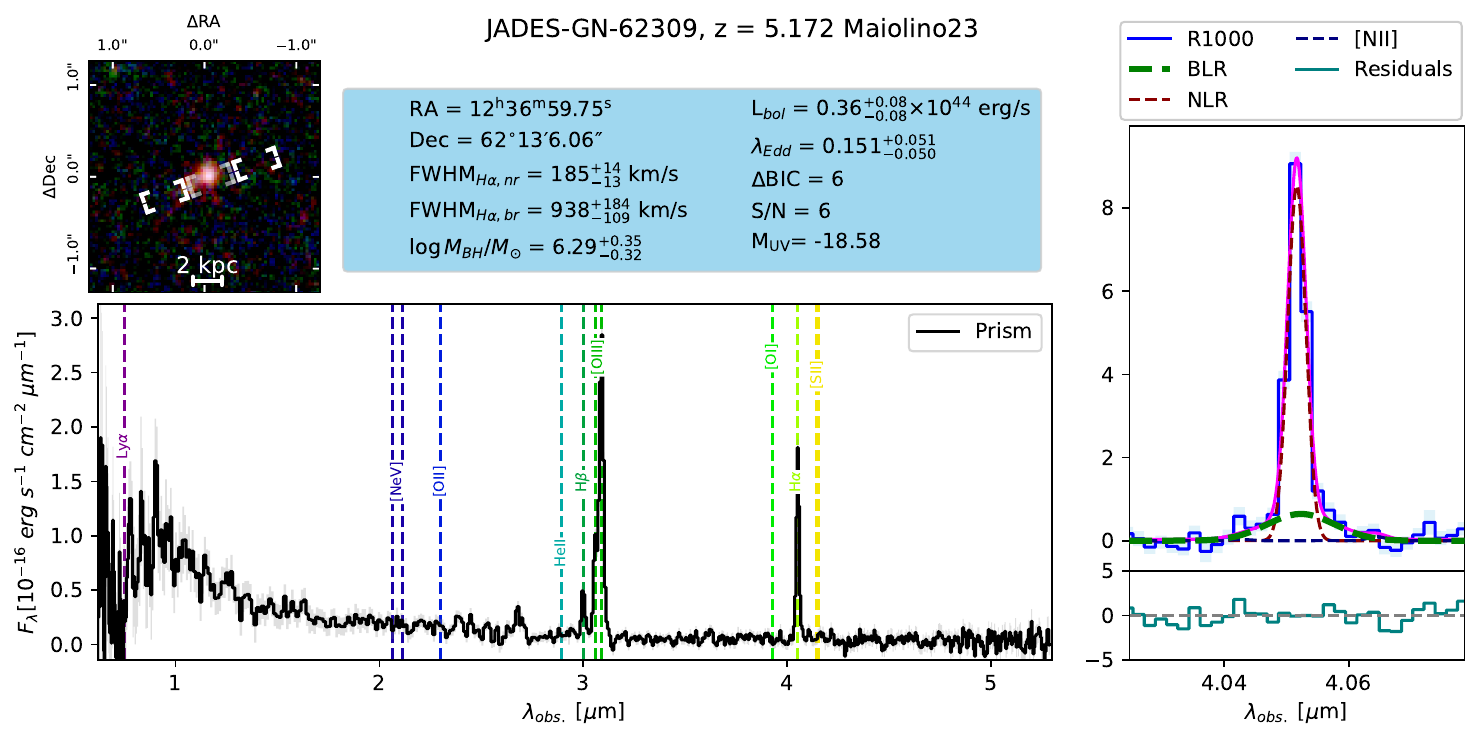}}
    \hfill
    \subfloat{\includegraphics[width=0.5\textwidth]{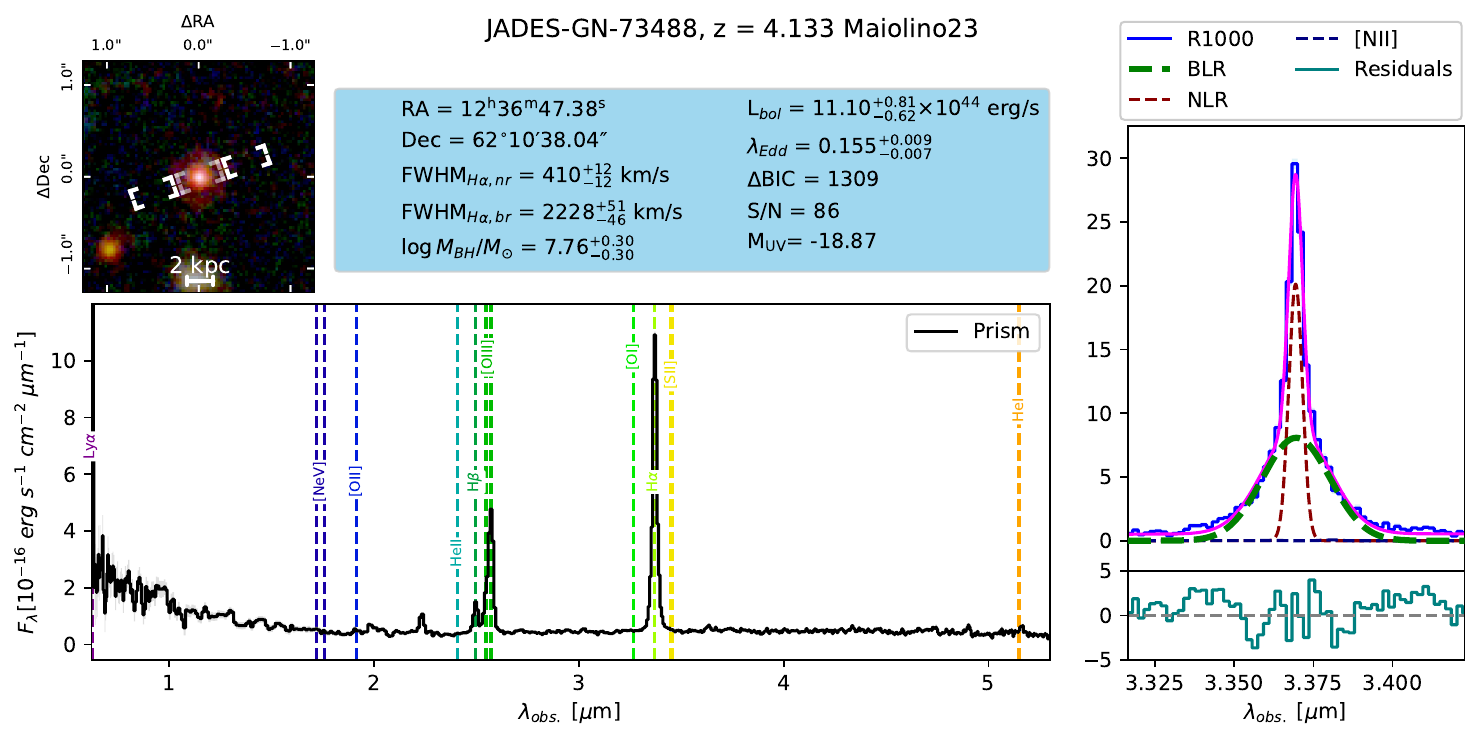}}
    \hfill
    \subfloat{\includegraphics[width=0.5\textwidth]{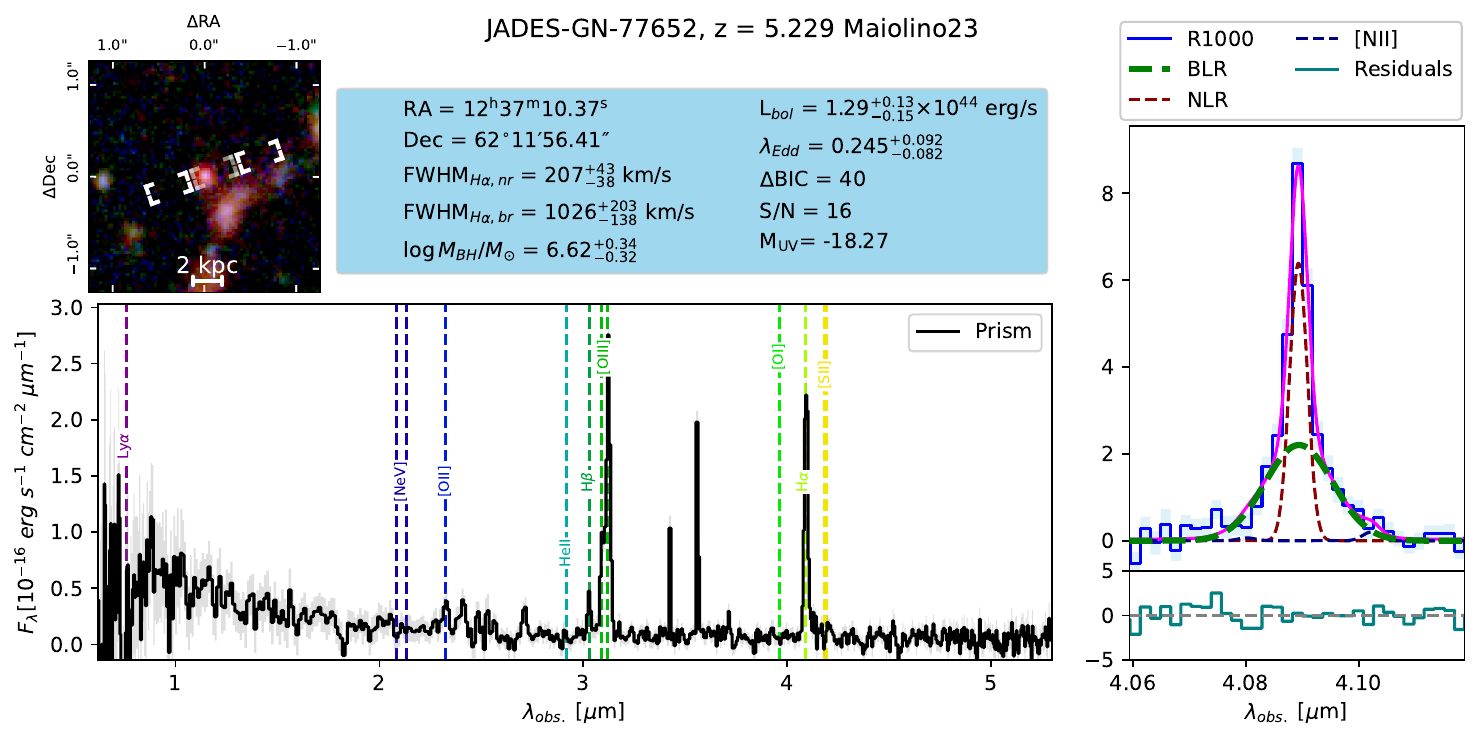}}
    \hfill
    \subfloat{\includegraphics[width=0.5\textwidth]{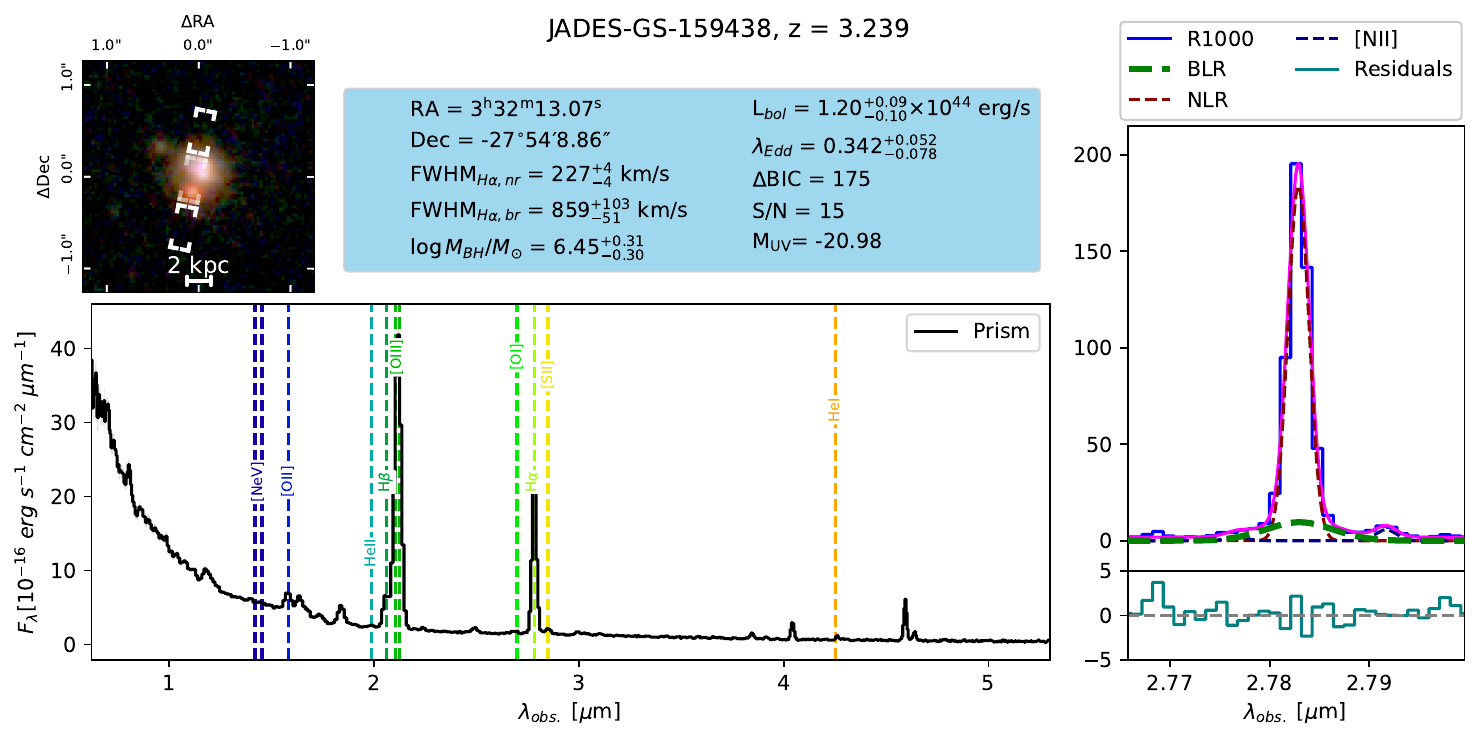}}
    \hfill
    \subfloat{\includegraphics[width=0.5\textwidth]{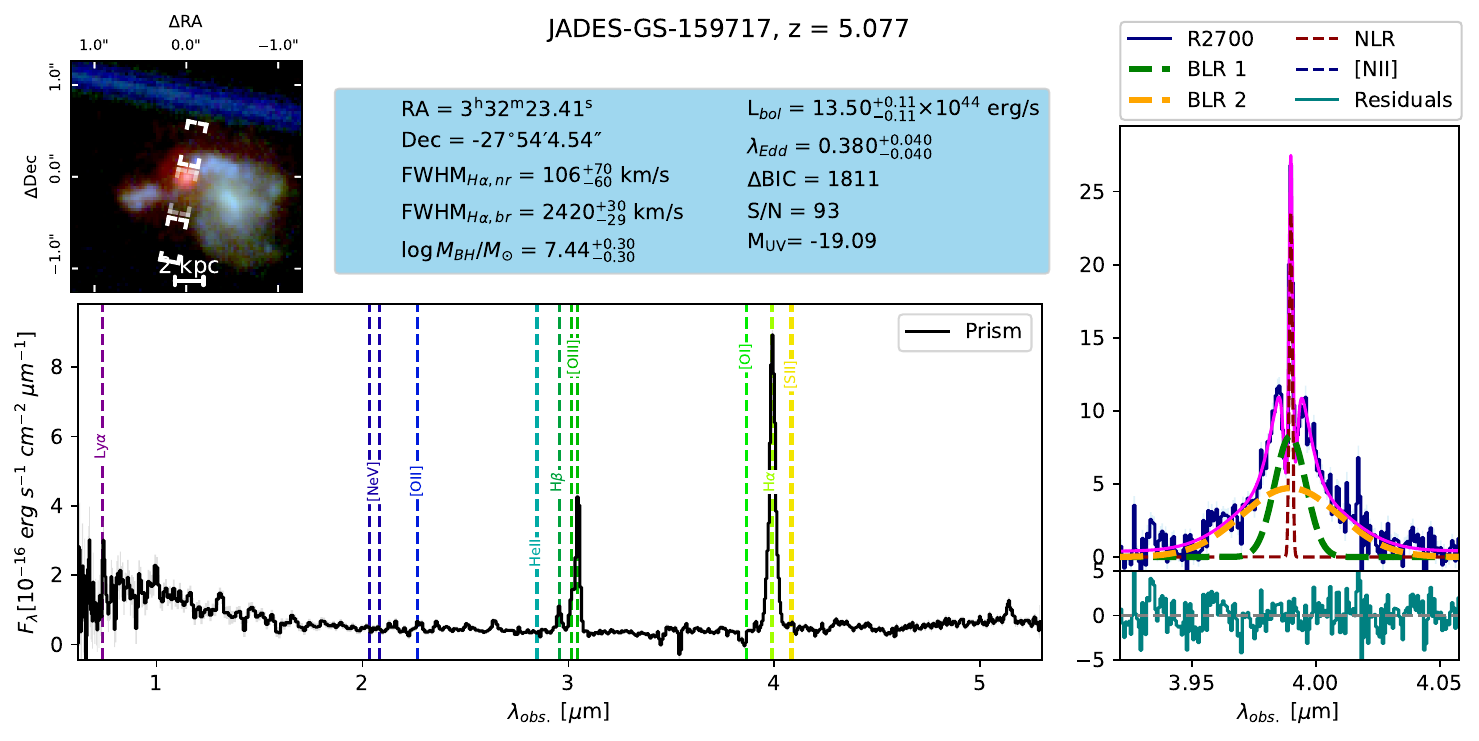}}
    \hfill
    \subfloat{\includegraphics[width=0.5\textwidth]{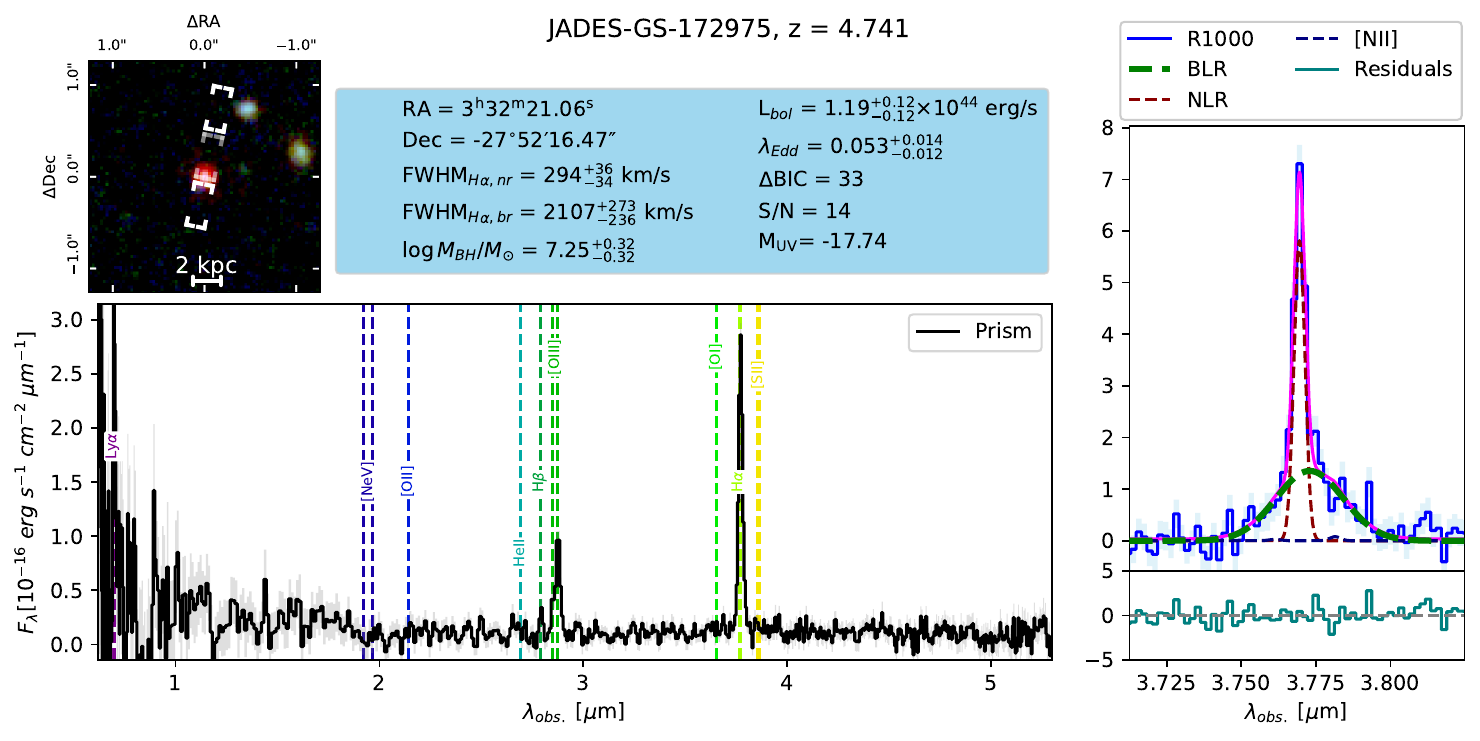}}
    \hfill
    \subfloat{\includegraphics[width=0.5\textwidth]{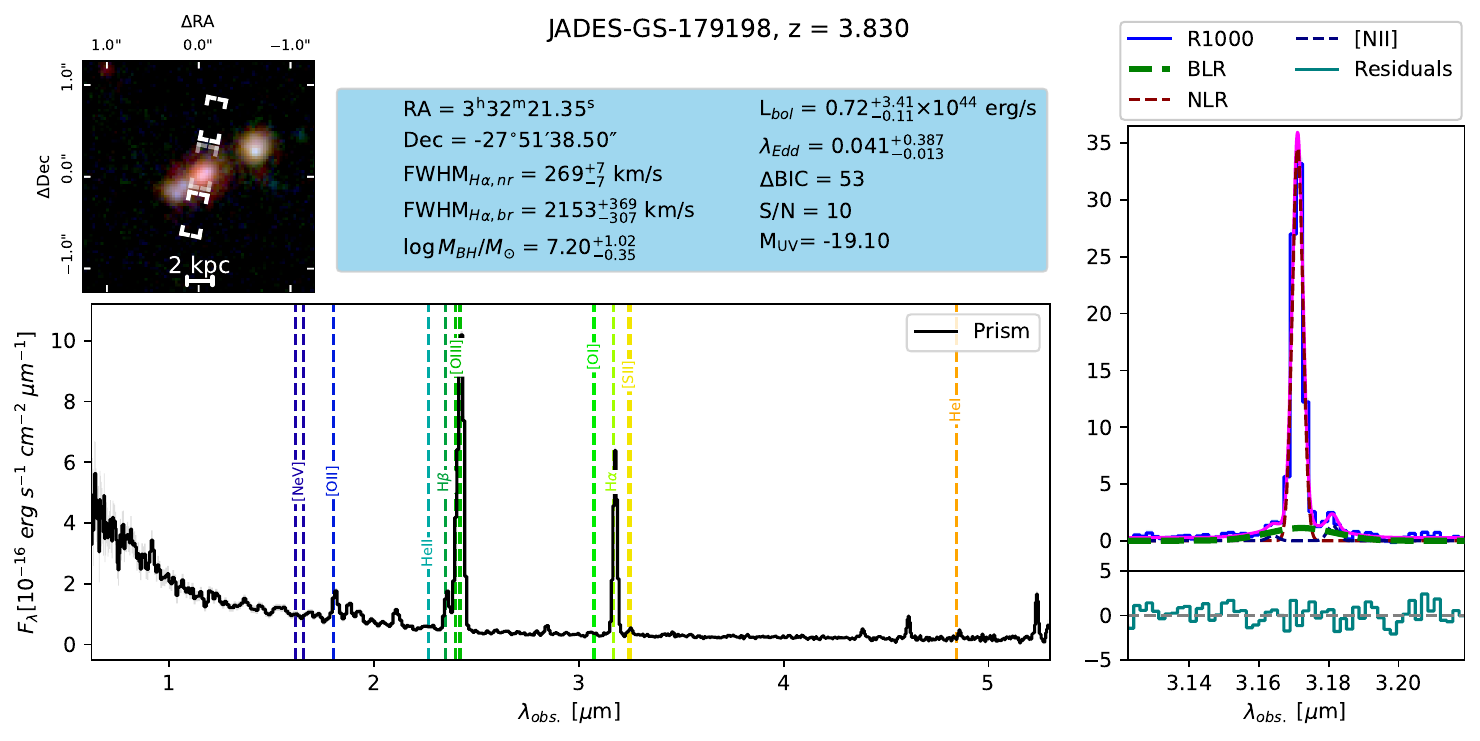}}
    \hfill
    \subfloat{\includegraphics[width=0.5\textwidth]{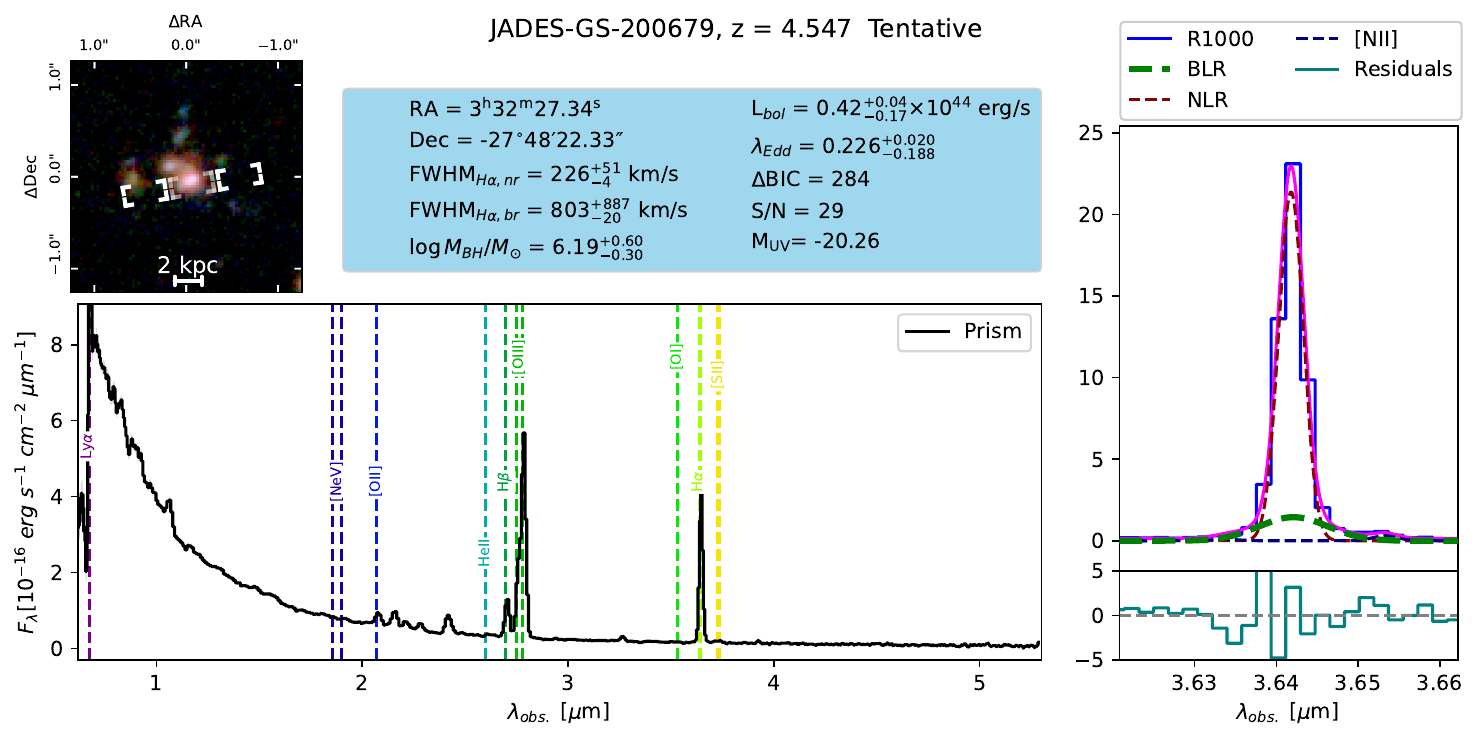}}
    \hfill
    \caption{Same structure as \autoref{fig:representative_fits} in the main text.}
    \label{fig:all_fits3}
\end{figure*}

\begin{figure*}
    \centering
    \subfloat{\includegraphics[width=0.5\textwidth]{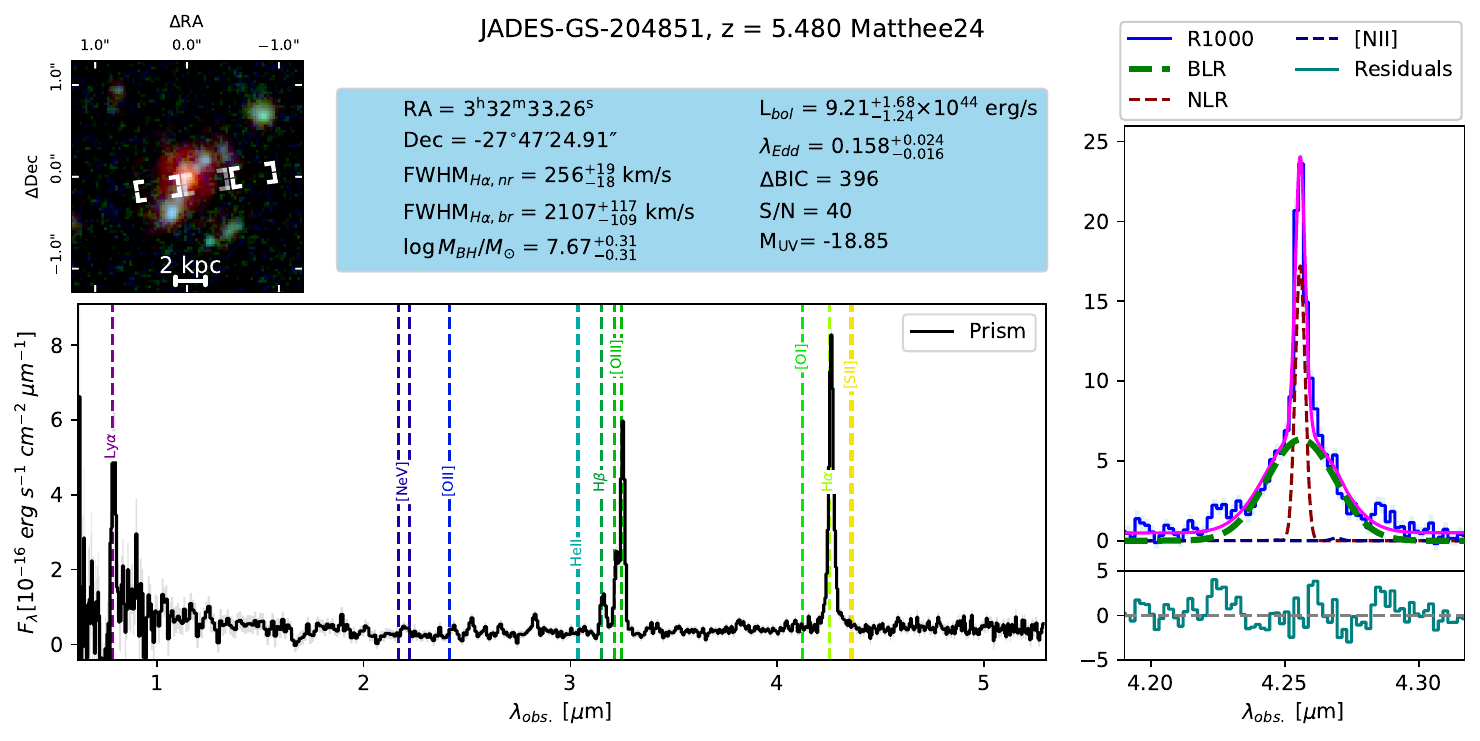}}
    \hfill
    \subfloat{\includegraphics[width=0.5\textwidth]{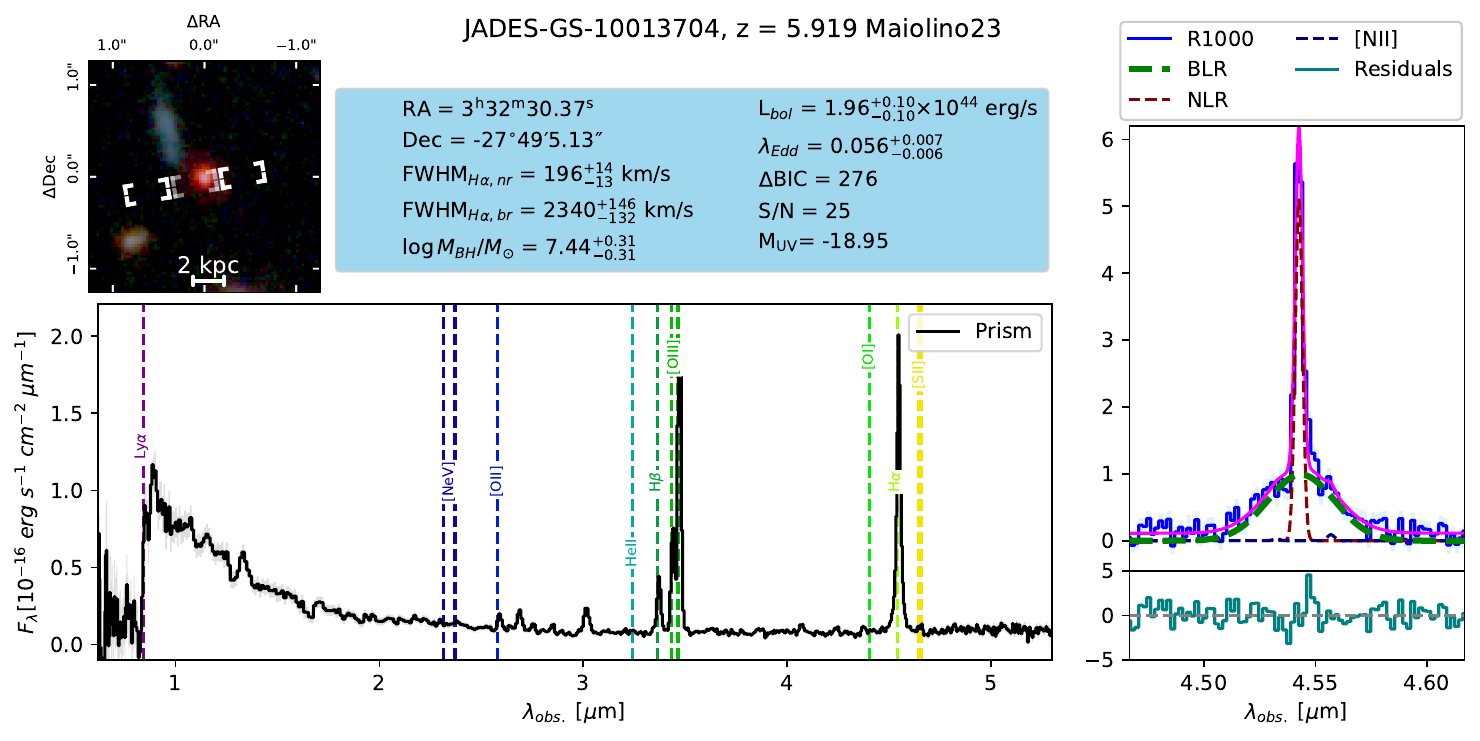}}
    \hfill
    \subfloat{\includegraphics[width=0.5\textwidth]{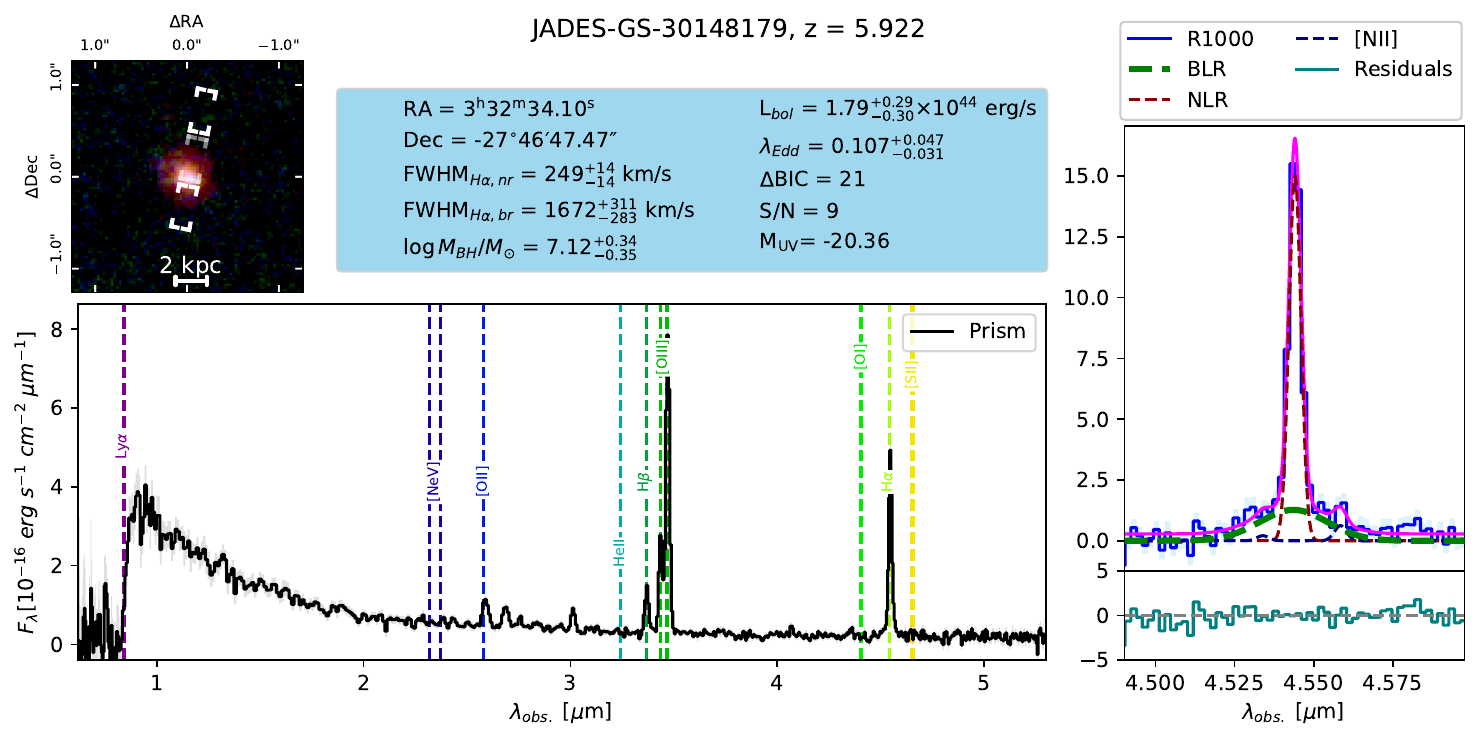}}
    \hfill
    \caption{Same structure as \autoref{fig:representative_fits} in the main text.}
    \label{fig:all_fits4}
\end{figure*}
\section{Tentative broad \Hbs emitters}
\label{sec:tent_hbeta}
\begin{figure*}
    
    \subfloat{\includegraphics[width=0.4\textwidth]{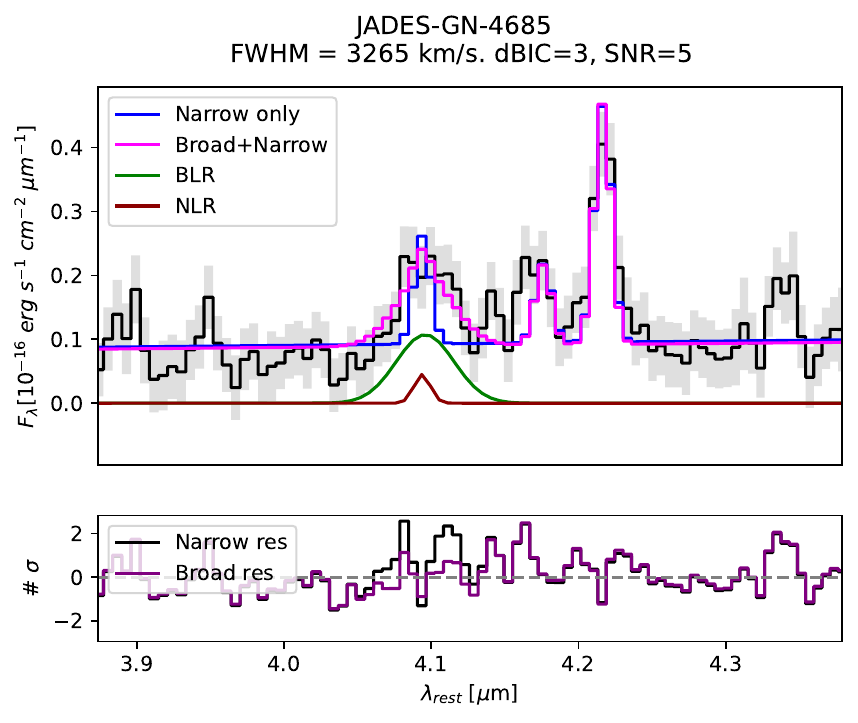}}
    \subfloat{\includegraphics[width=0.4\textwidth]{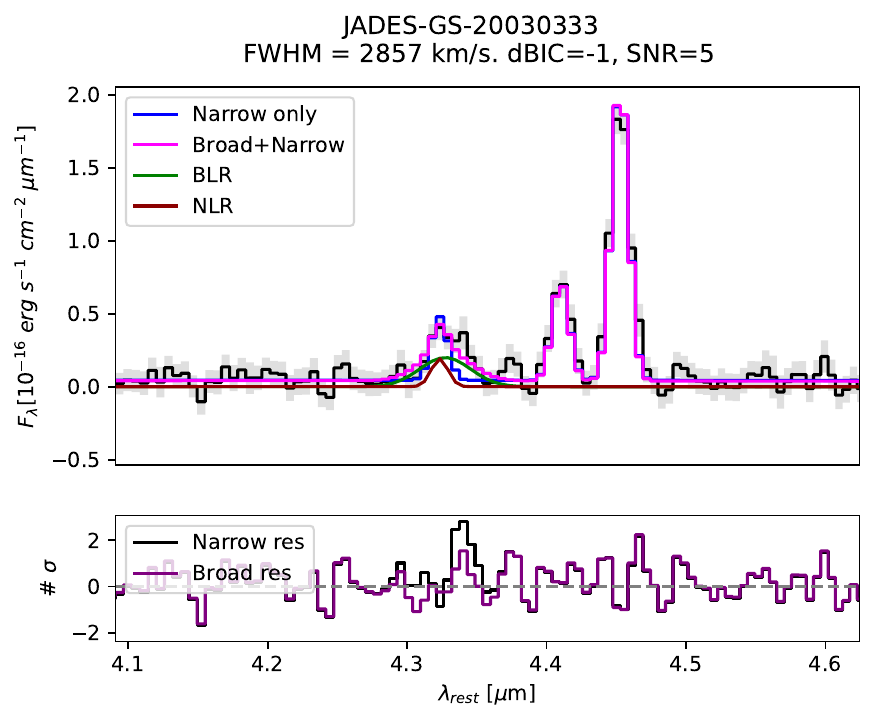}}

    \subfloat{\includegraphics[width=0.4\textwidth]{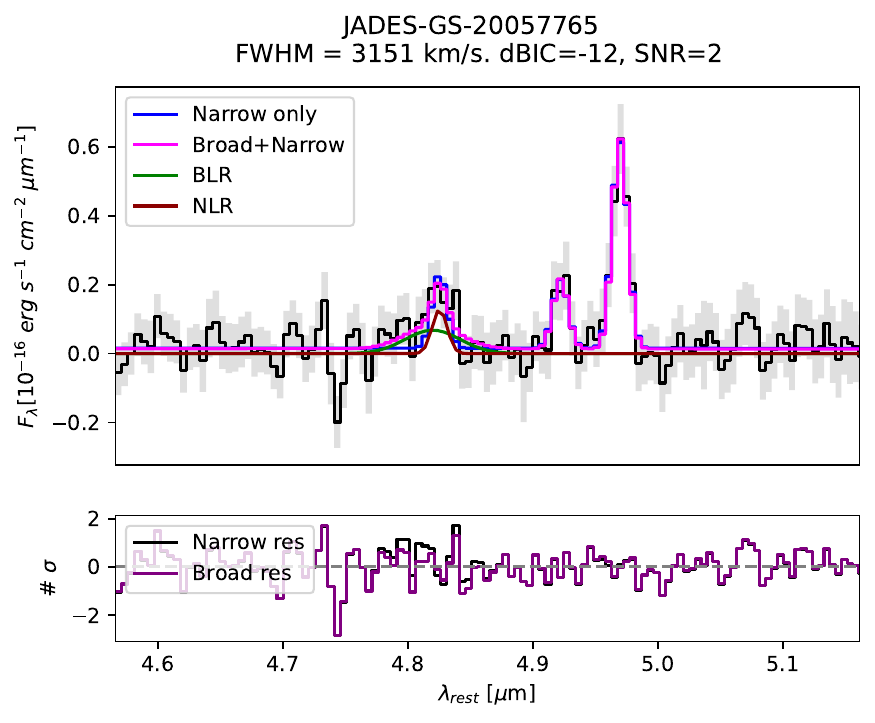}}
    \subfloat{\includegraphics[width=0.4\textwidth]{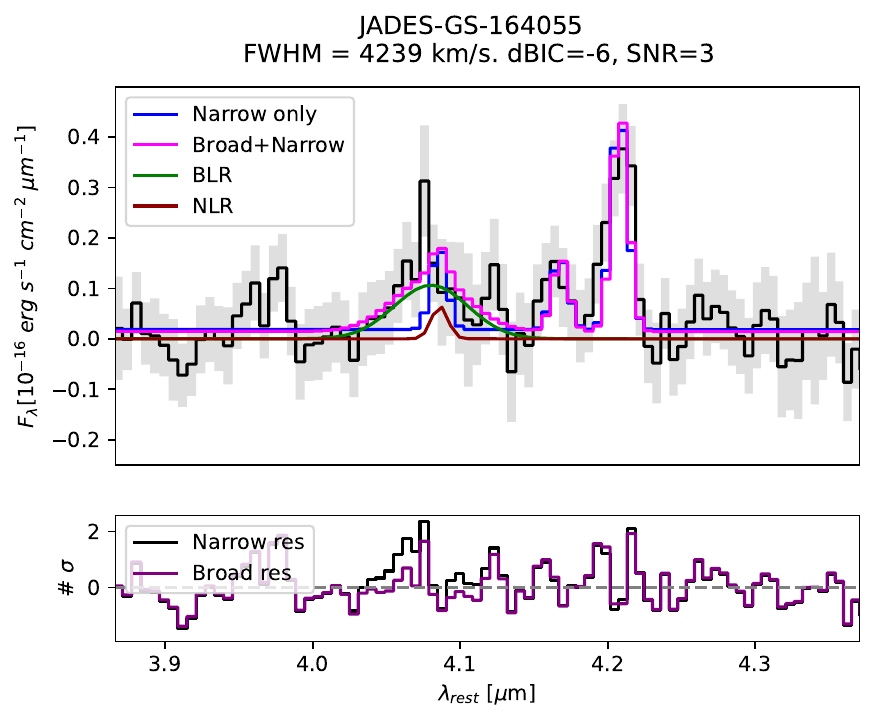}}

    \caption{Showcase of the individual spectra of the tentative broad \Hbs emitters. The narrow line only fit is shown in blue, while the magenta line shows shows the fit including a broad \Hbs component. The BLR and NLR components of the broad line fit are shown in green and dark red respectively. The lower panel beneath each spectrum shows the residuals of the two fits. It can be seen that, while the narrow line only fit does leave larger residuals, they are not significant enough to include these objects in our main sample.}
    \label{fig:individual_hb}
\end{figure*}

\begin{figure}
    \centering
    \includegraphics[width=\linewidth]{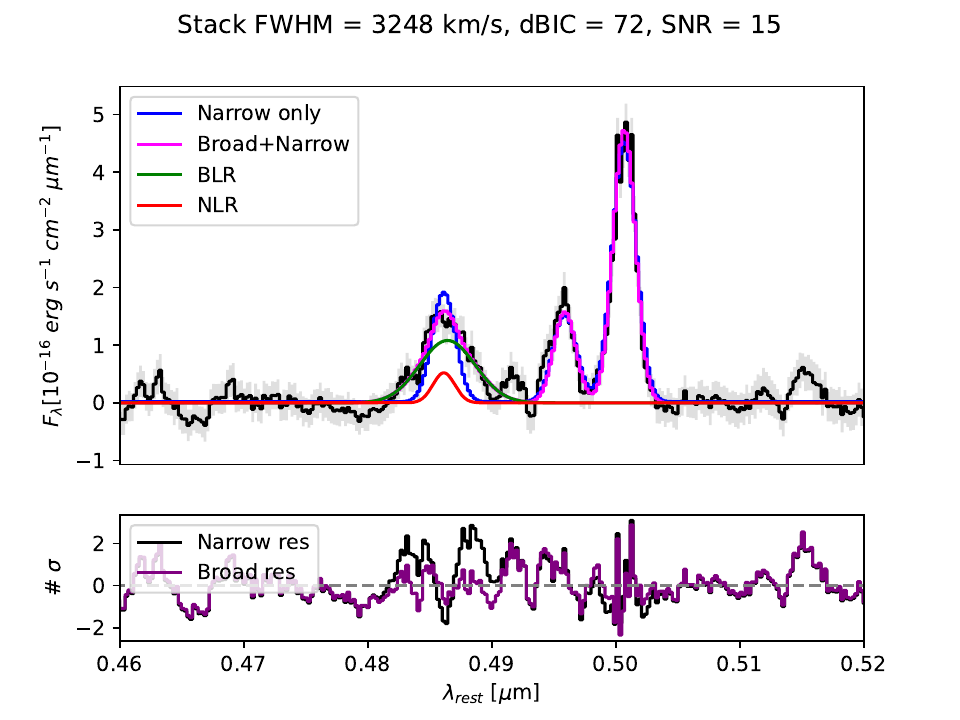}
    \caption{Stacked, continuum subtracted spectrum of the four objects visually identified to have broad \Hbs in prism. The plot is organized the same as the individual ones in \autoref{fig:individual_hb} It can be seen that not including a broad line in the fit leaves significant symmetric wings in the residual. In addition, the narrow-line fit would result in a \OIII/H$\beta$ ratio of 3, uncharacteristically low for these high-redshift objects.
    }
    \label{fig:Hb_stack}
\end{figure}

\begin{figure}
    \centering
    \includegraphics[width=\linewidth]{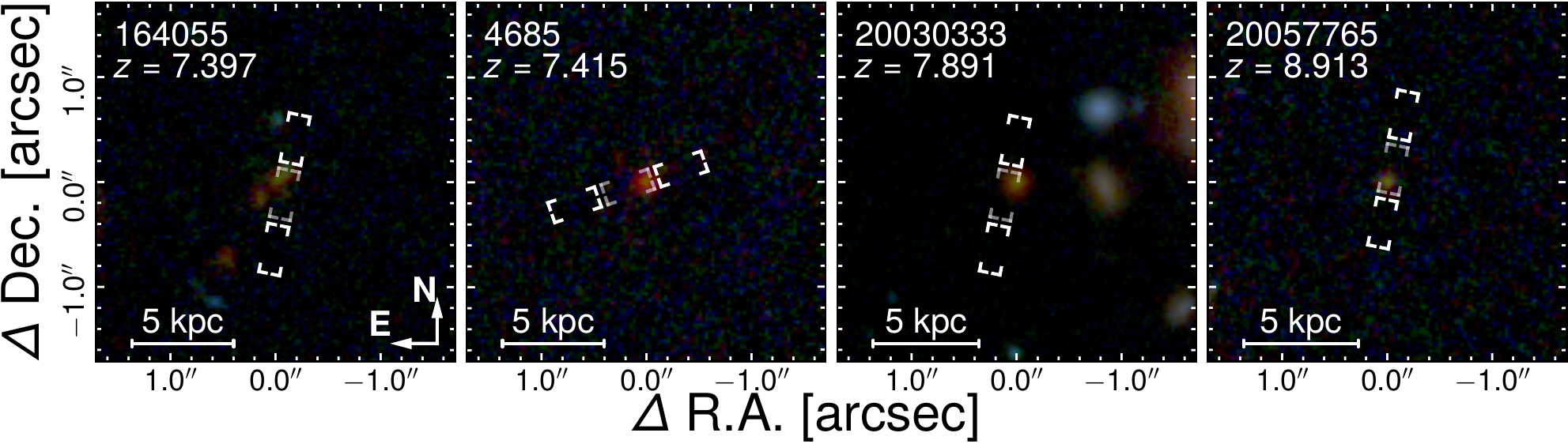}
    \caption{RGB NIRCam cutouts of the tentative broad \Hbs emitters. The shutter position is indicated in white rectangles. The images clearly showcase the compact, LRD-like nature of these sources. However, GS-164055 does appear to be an interacting system.}
    \label{fig:hb_stamps}
\end{figure}

In this section we present the individual and stacked spectra of the 4 sources with tentative broad \Hbs emission. The individual spectra are shown in \autoref{fig:individual_hb} and display visually broad, but not formally significant \Hbs profiles. A stack of these spectra (\autoref{fig:Hb_stack}), however, does show a prominent broad \Hbs profile. Lastly, the NIRCam cutouts for these sources are shown in \autoref{fig:hb_stamps} and reveal their compact nature.


\bsp	
\label{lastpage}
\end{document}